\documentclass[usenatbib,usegraphicx,usedcolumn]{mn2e}
\usepackage{times,amssymb,amsmath}
\usepackage{relsize}
\usepackage{hyperref}
\title[Constrained transport on a moving mesh]{A constrained transport scheme for MHD on unstructured static and moving meshes}
\author[P. Mocz et. al.]{Philip Mocz$^{1}$\thanks{E-mail: pmocz@cfa.harvard.edu (PM)}, Mark Vogelsberger$^{2}$, and Lars Hernquist$^{1}$ \\
$^{1}$Harvard-Smithsonian Center for Astrophysics, 60 Garden Street, Cambridge, MA 02138, USA \\
$^{2}$Department of Physics, Kavli Institute for Astrophysics and Space Research, Massachusetts Institute of Technology, Cambridge, MA 02139, USA}
\begin{document}

\date{accepted to MNRAS, May 2014}

\pagerange{\pageref{firstpage}--\pageref{lastpage}} \pubyear{2013}

\maketitle

\label{firstpage}

\begin{abstract}

Magnetic fields play an important role in many astrophysical systems and a detailed understanding of their impact on the gas dynamics requires robust numerical simulations. Here we present a new method to evolve the ideal magnetohydrodynamic (MHD) equations on unstructured static and moving meshes that preserves the magnetic field divergence-free constraint to machine precision. The method overcomes the major problems of using a cleaning scheme on the magnetic fields instead, which is non-conservative, not fully Galilean invariant, does not eliminate divergence errors completely, and may produce incorrect jumps across shocks. Our new method is a generalization of the constrained transport (CT) algorithm used to enforce the $\nabla\cdot \mathbf{B}=0$ condition on fixed Cartesian grids. Preserving $\nabla\cdot \mathbf{B}=0$ at the discretized level is necessary to maintain the orthogonality between the Lorentz force and $\mathbf{B}$. The possibility of performing CT on a moving mesh provides several advantages over static mesh methods due to the quasi-Lagrangian nature of the former (i.e., the mesh generating points move with the flow), such as making the simulation automatically adaptive and significantly reducing advection errors. Our method preserves magnetic fields and fluid quantities in pure advection exactly.


\end{abstract}


\section{Introduction}\label{sec:intro}

The equations of fluid dynamics and magnetohydrodynamics (MHD) may be evolved using a variety of numerical approaches. One can choose from a number of discretizations that are formally intended to yield a solution that is accurate to some order $n$. However, a system of partial differential equations for physical flows also gives rise to conserved quantities due to the symmetries of the problem and these quantities are not necessarily preserved with some algorithms. Such conserved quantities for the MHD equations include mass, momentum, angular momentum, energy, and the solenoidal nature of the magnetic field. It then becomes a concern to design techniques in a clever way such that the discrete representations of these conserved quantities are in fact preserved to machine precision. Failure to do this may result in instabilities and unphysical solutions; that is, the system's quantity that should be conserved may chaotically evolve away from its initial value. One well-known example of such a phenomenon occurs when evolving an $N$-body system with a non-symplectic time integrator, which does not conserve the total energy and hence may lead to energy drifts and decaying orbits.

In this work, we propose a scheme for evolving the MHD equations on a moving mesh such that conservation of mass, momentum, and energy, and, importantly, the solenoidal (divergence-free) nature of the magnetic field is preserved. On static Cartesian meshes, the constrained transport (CT) algorithm achieves this goal. It uses a finite volume formalism to evolve the density, momentum, and energy, and exploits Stokes' theorem and uses a face-averaged representation of the magnetic fields (called the `staggered-mesh' approach) to enforce $\nabla\cdot \mathbf{B}=0$ \citep{1988ApJ...332..659E}. The CT algorithm has been described in the literature as being quite difficult (if not impossible) to extend to an unstructured mesh \citep{2011ApJS..197...15D,2011MNRAS.418.1392P,2013MNRAS.432..176P} and leading to the development of alternate divergence-cleaning schemes to keep $\nabla\cdot \mathbf{B}$ small but non-zero on moving meshes based on the Powell \citep{1999JCoPh.154..284P} and Dedner \citep{2002JCoPh.175..645D} cleaning methods. Moving mesh methods for MHD with cleaning schemes have been described by \cite{2011ApJS..197...15D,2011MNRAS.418.1392P,2012ApJ...758..103G,2013MNRAS.432..176P}.
Additionally, there have been recent advances in designing robust mesh-less based methods for MHD \citep{2011MNRAS.414..129G,2013MNRAS.436.2810T}, which still, however, require cleaning schemes. These schemes alter the MHD equations and may add source terms in an attempt to control the divergence errors, which has the unwanted side-effect of making the schemes non-conservative. Additionally, such an approach loses the moving mesh's Galilean invariant properties (e.g. solving pure advection problems is diffusive). One approach to improve these methods has been to develop a {\it locally} divergence-free formulation based on the discontinuous Galerkin method in the work presented in \cite{2014MNRAS.437..397M}, where local divergence-free basis functions are used to represent the solution on each cell.  In this paper we solve the outstanding problem of designing a {\it globally} divergence-free, conservative, finite volume based algorithm on an unstructured moving mesh.

Coupling a CT algorithm with a moving mesh technique is clearly highly desirable. The moving mesh algorithm for solving the inviscid Euler equations was developed recently by \cite{2010MNRAS.401..791S} and implemented in the code {\sc arepo}. The moving mesh algorithm largely eliminates a number of well-know weaknesses of both static Cartesian/adaptive mesh refinement (AMR) approaches and pseudo-Lagrangian smoothed-particle hydrodynamics (SPH). The mesh-generating points in the algorithm can be set to move with the fluid flow, making the scheme quasi-Lagrangian. The method has automatic adaptivity, is Galilean-invariant, and significantly reduces advection errors. Contact discontinuities resulting from shocks are preserved to significantly greater accuracy. In addition, conditions on the size of the time-step are less severe than on a static mesh in the cases of flows with high-Mach number bulk motions, since fluxes across cells are always calculated in the rest-frame of the faces. A method for solving the MHD equations on a moving mesh will automatically gain these benefits of the Lagrangian nature of the scheme and, ideally, should not suffer any significant disadvantage compared to a static mesh approach. Since methods exist to preserve the divergence-free condition on structured fixed grids, it is important to extend this to unstructured meshes so that if differences are observed in MHD simulations run on static and moving meshes, then they can be attributed to the advantages of the moving mesh approach and not as an artefact produced by divergence errors.

The CT method proposed here can also be applied to static arbitrarily unstructured grids, where CT algorithms are lacking. A number of simulations use non-Cartesian static grids, dictated by the geometry/symmetry of the problem, as a way of achieving greater accuracy and by using fewer grid cells. For example, the MHD solver in \cite{2013ApJS..205...19F} was developed for a hexagonal spherical geodesic grid as a way to improve the simulation of astrophysical flows of partially ionized plasmas around a central compact object. However, such methods use cleaning schemes instead of CT to handle the magnetic field, and would benefit greatly from the CT method described in what follows. 

A robust code for solving the MHD equations as accurately as possible would have many astrophysical applications because magnetic fields are ubiquitous in the Universe. For example, the magnetorotational instability in accretion discs \citep{1998RvMP...70....1B} generates turbulence and mediates angular momentum transfer. Discs around supermassive black holes may be levitated by magnetic pressure \citep{2012ApJ...758..103G}. Magnetic fields are key in the production of relativistic jets and outflows from compact sources \citep{1977MNRAS.179..433B,1982MNRAS.199..883B}. Magnetic fields under a vertical gravitational field as in the disc of a galaxy leads to the Parker instability, which is thought to play an important role in the evolution of the interstellar medium \citep{1966ApJ...145..811P,1967ApJ...149..535P}. Interstellar turbulence and star formation are linked with magnetized plasma processes \citep{1995ApJ...438..763G}. Cosmic magnetic fields are likely to have a primordial origin which leads to imprints on the temperature and polarization anisotropies of the cosmic microwave background radiation \citep{2001PhR...348..163G}. These seed fields are amplified during the formation and evolution of galaxies. Radio lobes of galaxies may also play an important role in enhancing the magnetic fields in the intergalactic medium \citep{2001ApJ...560L.115G}. Magnetic fields may be responsible for suppressing strong isotropic turbulence and conduction in clusters and make it possible for stable, $100$ kpc-scale high-density arms to exist in the cluster environment \citep{2013Sci...341.1365S}. The strength and topology of magnetic fields are responsible for determining the propagation of cosmic rays in galaxies \citep{1998ApJ...509..212S}. Magnetic fields are also present in a wide variety of stars and play a significant role in their evolution \citep{2009ARA&A..47..333D}.  Many of these problems are suitable for study by a moving mesh approach and up until now a divergence-free MHD solver has been lacking. This paper lays the framework for implementing a divergence-preserving algorithm for robust evolution of the MHD equations in moving mesh codes such as {\sc arepo} \citep{2010MNRAS.401..791S} and {\sc tess} \citep{2011ApJS..197...15D}.

In Section~\ref{sec:methods} we describe the details of the numerical method. In Section~\ref{sec:tests} we show the results of numerical tests (with comparisons to fixed grid CT and the Powell cleaning scheme on a moving mesh) demonstrating that the method works well and has several advantages over the other two techniques. In Section~\ref{sec:disc} we discuss variations of our method and future directions. In Section~\ref{sec:conclusion} we provide concluding remarks.

\section{Numerical Method}\label{sec:methods}

\begin{figure*}
\centering
\includegraphics[width=\textwidth]{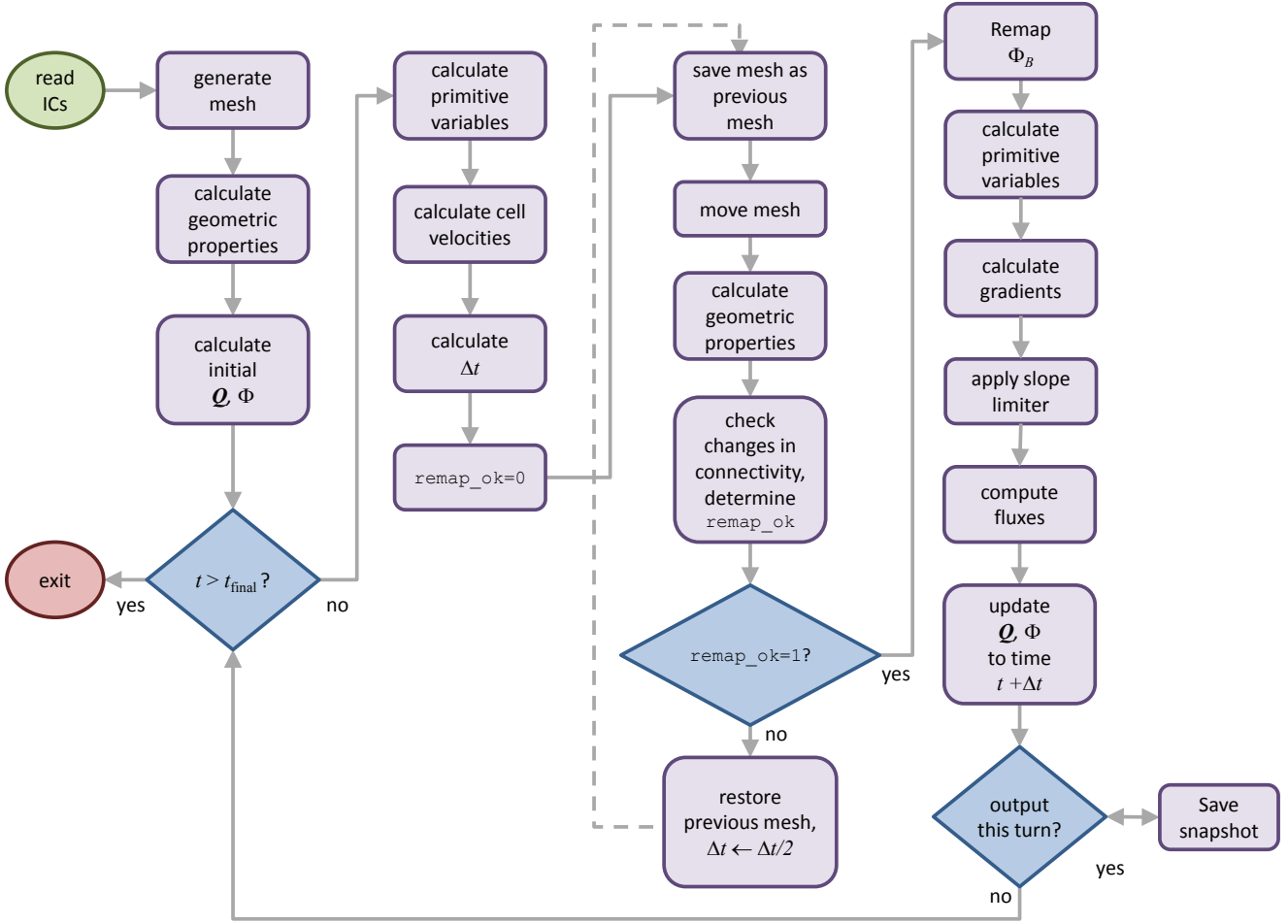}
\caption{A flow chart of the CT algorithm for a moving mesh. After initialization, the code enters the main loop. In each iteration of the loop, the system advances by a time interval $\Delta t$ as long as the mesh connectivity does not change too drastically (in which case the flag {\tt remap\_ok} is set to 1), else the timestep $\Delta t$ is halved and the system attempts to update itself according to this new timestep.}
\label{fig:flow}
\end{figure*}

This section is dedicated to describing the second-order numerical method in detail. Our method is implemented in 2D in {\sc matlab}. The method can be generalized to 3D (see section~\ref{sec:3d}). For reference, a flow chart of our numerical algorithm is presented in Fig.~\ref{fig:flow}.

\subsection{The magnetohydrodynamic equations}\label{sec:mhd}

The ideal MHD equations can be written as a system of conservation laws:
\begin{equation}
\frac{\partial \mathbf{U}}{\partial t} + \nabla \cdot \mathbf{F} = 0
\end{equation}
where $\mathbf{U}$ is the vector of the conserved variables and $\mathbf{F}(\mathbf{U})$ is the flux, namely:
\begin{equation}
\mathbf{U} = \begin{pmatrix} \rho \\ \rho\mathbf{v} \\ \rho e \\ \mathbf{B} \end{pmatrix},
\,\,\,\,\,\,
\mathbf{F}(\mathbf{U}) = 
\begin{pmatrix} \rho\mathbf{v} \\ \rho\mathbf{v}\mathbf{v}^T + p -\mathbf{B}\mathbf{B}^T \\ \rho e \mathbf{v} + p\mathbf{v} -\mathbf{B}(\mathbf{v}\cdot \mathbf{B}) \\ \mathbf{B}\mathbf{v}^T-\mathbf{v}\mathbf{B}^T \end{pmatrix}
\end{equation}
where $p=p_{\rm gas}+\frac{1}{2}\mathbf{B}^2$ is the total gas pressure, $e=u+\frac{1}{2}\mathbf{v}^2+\frac{1}{2\rho}\mathbf{B}^2$ is the total energy per unit mass, $u$ is the thermal energy per unit mass. The equation of state for the fluid is given by $p=(\gamma-1)\rho u$. We describe how to discretize and solve these equations on an unstructured moving mesh with CT.

\subsection{Finite volume approach on a moving mesh}\label{sec:fv}

A finite volume strategy is used to update the density, momentum, and energy of the cells (but the magnetic fields require a different approach). Our method follows that of \cite{2010MNRAS.401..791S} closely, except for minor modifications which we point out, which are needed to link the method with our CT scheme to update the magnetic fields.

The domain is discretized into cells created by a Voronoi tessellation from mesh generating points. For each cell, we define volume averaged quantities (`conservative variables'): the total mass $m_i$, momentum $\mathbf{p}_i$, and energy $E_i$, given by:
\begin{equation}
\mathbf{Q}_i =  \begin{pmatrix} m_i \\ \mathbf{p}_i \\ E_i \end{pmatrix}
= \int_{V_i} \tilde{\mathbf{U}}  \,dV
\end{equation}
where $V_i$ is the volume of cell $i$, and
\begin{equation}
\tilde{\mathbf{U}} = \begin{pmatrix} \rho \\ \rho\mathbf{v} \\ \rho e \end{pmatrix}
\end{equation}

By Gauss' theorem, the conservation laws for a moving cell can be rewritten as:
\begin{equation}
\frac{d\mathbf{Q}_i}{dt} = -\int_{\partial V_i} \left(  \mathbf{F}(\tilde{\mathbf{U}}) - \tilde{\mathbf{U}}\mathbf{w}^T \right)\,d\mathbf{n}=0
\label{eqn:evo1}
\end{equation}
where $\mathbf{w}$ is the velocity of each point of the boundary of the cell and $\mathbf{n}$ is the outward normal of the cell surface.

Voronoi cells are polyhedra (polygons in 2D) with $A_{ij}$ denoting the area of the face between cells $i$ and $j$. Equation~\ref{eqn:evo1} is discretized as follows:
\begin{equation}
\mathbf{Q}_i^{(n+1)} = \mathbf{Q}_i^{(n)} - \Delta t \sum_j A_{ij}\hat{\mathbf{F}}_{ij}^{n+1/2}=0
\label{eqn:evo2}
\end{equation}
where $\mathbf{Q}_i^{(n)}$ is the state of the cell at timestep $n$, $\Delta t$ is the timestep that evolves the system to the next point in time, and $\hat{\mathbf{F}}_{ij}^{n+1/2}$ is a time-averaged approximation to the flux through the cell over the duration of the timestep. $\hat{\mathbf{F}}_{ij}$ is anti-symmetric, making the method conservative. All that remains to update mass, density, and energy is to find an appropriate second-order estimate of the numerical flux $\hat{\mathbf{F}}_{ij}$, which we describe next.

The flux computation for the mass, momentum, and energy for the moving mesh algorithm is calculated in the rest-frame of each of the faces. An important difference with the approach taken in \cite{2010MNRAS.401..791S} is that we first move the mesh generating points over a time interval $\Delta t$ according to their velocities, reconstruct the Voronoi mesh (this is the mesh at the end of the timestep), and then extrapolate fluid quantities to the face centroid of the mesh using the geometry at the end of the timestep rather than the beginning of the timestep for the flux computations. Using the cell geometry at the end of the timestep rather than at the beginning is an equally accurate reconstruction technique, but it turns out to be easier to account for changes in mesh connectivity for the CT algorithm for divergence-free evolution of magnetic fields, described in Section~\ref{sec:ct}.

In the scheme, the primitive variables of each cell are estimated from the conserved variables and used to determine the wave-speeds local to the cell and the timesteps. The Courant-Friedrichs-Levy (CFL) timestep criterion is used to set the timestep:
\begin{equation}
\Delta t_i = C_{\rm CFL} \frac{R_i}{c_{{\rm f},i}+|\mathbf{v}_i-\mathbf{w}_i|}
\end{equation}
where we choose $C_{\rm CFL}\lesssim 0.4$ for stable evolution of the MHD equations on a moving mesh. Here, $c_{{\rm f},i}$ is the fast magnetosonic sound-speed in cell $i$.

The mesh is then evolved to the geometry of the next timestep. The mesh-generating points move closely with the fluid flow with some small regularization corrections to keep the cells `round'. The cell velocity $\mathbf{w}_i$ of each cell $i$ is calculated at the beginning of each timestep as:
\begin{equation}
\mathbf{w}_i = \mathbf{v}_i + \chi \begin{cases}
0, & d_i/(\eta R_i) < 0.9 \\
c_i\frac{\mathbf{s}_i-\mathbf{r}_i}{d_i} \frac{d_i-0.9\eta R_i}{0.2\eta R_i}, & 0.9 \leq d_i/(\eta R_i) < 1.1 \\
c_i\frac{\mathbf{s_i}-\mathbf{r}_i}{d_i}, & 1.1 \leq  d_i/(\eta R_i)
\end{cases}
\end{equation}
where $R_i$ is the effective radius of a cell (calculated from the volume), $c_i$ is the local sound-speed of the cell, and $d_i=|\mathbf{r}_i-\mathbf{s}_i|$ is the distance between the cell's mesh generating point and its centre of mass. The quantities $\eta$, and $\chi$ are free parameters defining how aggressively to apply mesh regularization. For our simulations, we use $\eta=0.05$, $\chi=1.0$. The full geometry of the Voronoi mesh is then calculated from the positions of the mesh generating points.

The primitive variables are re-estimated once the mesh is moved from the conserved variables. This makes the reconstruction step fully conservative.

To obtain second-order accurate estimates of the flux, we estimate gradients of the primitive fluid variables ($\rho$, $\mathbf{v}$, $p_{\rm gas}$, $\mathbf{B}$) in each cell. We use Equation~22 of \cite{2010MNRAS.401..791S}, namely, the gradient in cell $i$ of a primitive fluid variable $\phi$ is given by:
\begin{equation}
\langle \phi \rangle_i = \frac{1}{V_i} \sum_{j\neq i} A_{ij} \left( [\phi_j-\phi_i]\frac{\mathbf{c}_{ij}}{r_{ij}} - \frac{\phi_i+\phi_j}{2}\frac{\mathbf{r}_{ij}}{r_{ij}} \right)
\end{equation}
where $\mathbf{r}_{ij}=\mathbf{r}_i-\mathbf{r}_j$ with $\mathbf{r}_i$ being the mesh generating point of cell $i$, and $\mathbf{c}_{ij}$ is the vector from the midpoint between $i$ and $j$ to the centre-of-mass of the face between $i$ and $j$.

Now, discontinuities in the fluid variables due, for example, to the presence of shocks, may cause spurious oscillations in the solutions. To suppress these, we are required to limit the gradients of the fluid variables, using the rule that linearly reconstructed quantities on face centroids may not exceed the maxima or minima among all cell neighbours. The slope limiting rule is expressed in mathematical terms as:
\begin{equation}
\langle \nabla\phi\rangle^\prime_i = \alpha_i \langle \nabla\phi\rangle_i
\end{equation} 
where $0\leq \alpha_i\leq 1$ is the slope limiter computed as:
\begin{equation}
\alpha_i = {\rm min}(1,\psi_{ij})
\end{equation} 
with the minimum being taken over all neighbouring cells $j$ of cell $i$, with $\psi_{ij}$ defined as:
\begin{equation}
\phi_{ij} = \begin{cases}
(\phi_i^{\rm max} - \phi_i)/\Delta \phi_{ij},  & \Delta \phi_{ij} > 0 \\
(\phi_i^{\rm min} - \phi_i)/\Delta \phi_{ij},  & \Delta \phi_{ij} < 0 \\
1, & \Delta \phi_{ij} = 0 
\end{cases}
\end{equation}
where $\nabla \phi_{ij} = \langle \phi \rangle_i \cdot (\mathbf{f}_{ij} - \mathbf{s}_i)$ is the estimated change between the centroid $\mathbf{f}_{ij}$ of the face and the centre-of-mass $\mathbf{s}_i$ of the cell $i$, and $\phi_i^{\rm max} = {\rm max}(\phi_j)$, $\phi_i^{\rm min} = {\rm min}(\phi_j)$ are the maximum/minimum values of $\phi$ among all neighbouring cells of cell $i$, including cell $i$.

Supposing we have neighbouring cells $i$ and $j$, the estimate of the face velocity $\mathbf{w}_{ij}$ can be computed as:
\begin{equation}
\mathbf{w}_{ij} = \frac{\mathbf{w}_i+\mathbf{w}_j}{2} + \frac{(\mathbf{w}_i-\mathbf{w}_j)\cdot[\mathbf{f}_{ij}-(\mathbf{r}_i+\mathbf{r}_j)/2]}{|\mathbf{r}_j-\mathbf{r}_i|}\cdot\frac{\mathbf{r}_j-\mathbf{r}_i}{|\mathbf{r}_j-\mathbf{r}_i|}
\end{equation}
We then change the primitive variables $\mathbf{W}^T=(\rho,\mathbf{v},p_{\rm gas},\mathbf{B})$ of cells $i$ and $j$ from the lab-frame to the rest-frame of the face:
\begin{equation}
\mathbf{W}^\prime_{i,j} =  \mathbf{W}_{i,j} - \begin{pmatrix} 0 \\ \mathbf{w}_{ij} \\ 0 \\ 0 \\ 0 \end{pmatrix}
\end{equation}
We then linearly predict the states on both sides of the centroid of the face, also predicting them forward by half a timestep:
\begin{equation}
\mathbf{W}^{\prime\prime}_{i,j} =  \mathbf{W}^\prime_{i,j} 
+ \frac{\partial \mathbf{W}^\prime}{\partial \mathbf{r}}\bigg|_{i,j} (\mathbf{f}_{ij} - \mathbf{s}_{i,j}) 
+ \frac{\partial \mathbf{W}^\prime}{\partial t}\bigg|_{i,j} \frac{\Delta t}{2} 
\end{equation}
where the spatial derivatives $\frac{\partial \mathbf{W}^\prime}{\partial \mathbf{r}}$ are the slope-limited gradients of the primitive fluid variables. Note again that we are extrapolating to the centroids of the cell faces of the mesh geometry at the end of the timestep in this step. The time derivatives $\frac{\partial \mathbf{W}^\prime}{\partial t}$ can be obtained from the primitive form of the MHD equations:
\begin{equation}
\frac{\partial \mathbf{W}^\prime}{\partial t} + \mathbf{A}(\mathbf{W}) \frac{\partial \mathbf{W}^\prime}{\partial \mathbf{r}} = 0
\end{equation}
where
\begin{equation}
\mathbf{A}_x(\mathbf{W}) = \mathsmaller{\mathsmaller{\mathsmaller{\mathsmaller{\mathsmaller{
\begin{pmatrix}
v_x & \rho & 0 & 0 & 0 & 0\\
0 & v_x & 0 & 1/\rho & -B_x/\rho & B_y/\rho \\
0 & 0 & v_x & 0 & -B_y/\rho & -B_x/\rho \\
0 & \gamma p_{\rm gas} & 0 & v_x & (\gamma-1)\mathbf{B}\cdot\mathbf{v} & 0 \\
0 & 0 & 0 & 0 & 0 & 0 \\
0 & B_y & -B_x & 0 & -v_y & v_x \\
\end{pmatrix}
}}}}}
\end{equation}
and
\begin{equation}
\mathbf{A}_y(\mathbf{W}) = \mathsmaller{\mathsmaller{\mathsmaller{\mathsmaller{\mathsmaller{
\begin{pmatrix}
v_y & 0 & \rho & 0 & 0 & 0\\
0 & v_y & 0 & 0 & -B_y/\rho & -B_x/\rho \\
0 & 0 & v_y & 1/\rho & B_x/\rho & -B_y/\rho \\
0 & 0 & \gamma p_{\rm gas} & v_y & 0 & (\gamma-1)\mathbf{B}\cdot\mathbf{v} \\
0 & -B_y & B_x & 0 & v_y & -v_x \\
0 & 0 & 0 & 0 & 0 & 0 \\
\end{pmatrix}
}}}}}
\end{equation}

The state is rotated into a new coordinate frame such that the $x$-axis is parallel to the normal vector of the face, pointing from cell $i$ to cell $j$:
\begin{equation}
\mathbf{W}^{\prime\prime\prime}_{i,j} = 
\begin{pmatrix}
1 & 0 & 0 & 0 \\
0 & \Lambda_{2{\rm D}}  & 0 & 0 \\
0 & 0 & 1 & 0 \\
0 & 0 & 0 & \Lambda_{2{\rm D}} 
\end{pmatrix}
\mathbf{W}^{\prime\prime}_{i,j}
\end{equation}
where $\Lambda_{2{\rm D}}$ rotates the velocity and magnetic field components appropriately. In this new frame, the normal components of the magnetic field are averaged to guarantee their continuity across the cell (which is required by the divergence-free condition of MHD for 1D):
\begin{equation}
B_{x,(i,j)}^{\prime\prime\prime} \leftarrow  \frac{B_{x,i}^{\prime\prime\prime}+B_{x,j}^{\prime\prime\prime}}{2}
\label{eqn:Bavg}
\end{equation}

At the interface between the two cells, we now must approximately solve the Riemann problem to obtain the flux. The primitive variables $\mathbf{W}^{\prime\prime\prime}_{i,j}$ are converted to conservative variables $\mathbf{U}^{\prime\prime\prime}_{i,j}$, and the HLLD flux \citep{2005JCoPh.208..315M} is computed:
\begin{equation}
\mathbf{F}^{\prime\prime\prime}_{ij}= \mathcal{F}_{\rm HLLD}(\mathbf{U}^{\prime\prime\prime}_{i},\mathbf{U}^{\prime\prime\prime}_{j})
\end{equation}
The HLLD flux is a highly-accurate, widely used, Riemann solver for the MHD equations which approximates the Riemann fan by 5 waves. In some instances, the solver may result in an unphysical negative pressure, in which case the Riemann solver defaults to calculating the Rusanov flux (also called local Lax-Friedrichs flux) which is more diffusive but always stable.

De-rotation is then used to obtain the fluxes in the rest-frame of the face but the $x$ and $y$ axes restored to their directions in the lab frame:
\begin{equation}
\mathbf{F}^{\prime\prime}_{ij}=
\begin{pmatrix}
1 & 0 & 0 & 0 \\
0 & \Lambda_{2{\rm D}}^{-1}  & 0 & 0 \\
0 & 0 & 1 & 0 \\
0 & 0 & 0 & \Lambda_{2{\rm D}}^{-1} 
\end{pmatrix}
\mathbf{F}^{\prime\prime\prime}_{ij}
\end{equation}
where we will introduce notation for the components of this flux: $\mathbf{F}^{\prime\prime}_{ij} = (F_1,F_2,F_3,F_4,F_5,F_6)^T$. We can now obtain the numerical flux required to update Equation~\ref{eqn:evo2} from the flux in the rest frame of the face with a correction term for the movement of the face which retains an upwind character, derived in \cite{2011MNRAS.418.1392P}:
\begin{equation}
\hat{\mathbf{F}}_{ij}^{n+1/2} = 
\mathbf{F}^{\prime\prime}_{ij} + 
\begin{pmatrix}
0 \\ w_{x,ij} F_1 \\ w_{y,ij} F_1 \\ w_{x,ij} F_2 + w_{y,ij} F_3 + \frac{1}{2}F_1\mathbf{w}_{ij}^2 \\
-w_{x,ij} B_{x,(i,j)}^{\prime\prime\prime} \\
-w_{y,ij} B_{x,(i,j)}^{\prime\prime\prime} \\
\end{pmatrix}
\end{equation}

\subsection{Constrained transport on a moving mesh}\label{sec:ct}

The CT algorithm relies on evolving the average normal component of the flux of the magnetic field through each face of each cell using Faraday's induction equation:
\begin{equation}
\frac{\partial \mathbf{B}}{ \partial t } + \nabla \times \mathbf{E} = 0
\label{eqn:evo3}
\end{equation}
where $\mathbf{E} = -\mathbf{v}\times\mathbf{B}$ is the electric field. The normal component of the magnetic flux through some face $A$ is defined as:
\begin{equation}
\Phi = \int_A \mathbf{B}\cdot d\mathbf{A}
\end{equation}
Suppose that the surface $A=A(t)$ moves in time. Rewriting Equation~\ref{eqn:evo3} in integral form and applying Stokes' theorem yields:
\begin{equation}
\frac{d\Phi}{dt} + \int_{\partial A(t)} (\mathbf{E}+\mathbf{w}\times\mathbf{B})\cdot d\mathbf{\ell} = 0
\label{eqn:evo4}
\end{equation}
where $\mathbf{w}$ is the velocity of the motion of the surface, which can be rewritten as:
\begin{equation}
\frac{d\Phi}{dt} - \int_{\partial A(t)} ((\mathbf{v}-\mathbf{w})\times\mathbf{B})\cdot d\mathbf{\ell} = 0
\label{eqn:evo5}
\end{equation}
We see that $-(\mathbf{v}-\mathbf{w})\times\mathbf{B}$ is just the electric field in the rest frame of a point on the boundary of the surface. 

Let $\Phi_{ij}$ be the outward normal magnetic flux from cell $i$ through the face between cell $i$ and cell $j$. Then, the average magnetic field outward normal to the surface is $B_{ij} = \Phi_{ij}/A_{ij}$. Equation~\ref{eqn:evo5} may then be discretized as follows:
\begin{equation}
\Phi_{ij}^{(n+1)} = \Phi_{ij}^{(n)} - \Delta t (E^{(n+1/2)}_{L,ij} - E^{(n+1/2)}_{R,ij})
\label{eqn:evo6}
\end{equation}
where $E^{(n+1/2)}_{L,ij}$ and $E^{(n+1/2)}_{R,ij}$ are estimates for the $z$-component of the electric fields at the two edges ($e_{L,ij}$ and $e_{R,ij}$) of the face in the middle of the timestep. The points $r_i$ (the mesh generating point), $e_{R,ij}$, $e_{L,ij}$ are oriented counter-clockwise.

The divergence of $\mathbf{B}$ can be estimated using Stokes' theorem as:
\begin{equation}
\nabla\cdot \mathbf{B}_i = \frac{1}{V_i}\sum_j \Phi_{ij} 
\label{eqn:div}
\end{equation}
which is kept zero at the level of machine precision as long as the initial conditions have zero divergence, due to the opposite signs of $E^{(n+1/2)}_{L,ij}$ and $E^{(n+1/2)}_{R,ij}$ and the exact cancellation that is obtained in looping around the border of the cell and summing the changes to the fluxes through the faces. Note that the CT approach only keeps this one particular discretization of the divergence to zero to machine precision.

One now just needs to accurately estimate $E^{(n+1/2)}$ to be able to update the magnetic flux. We do so using a flux-interpolated approach as follows. In this Section we will assume that the mesh connectivity does not change between timesteps, and in Section~\ref{sec:remapping} we describe the additional remapping technique that needs to be applied on magnetic fluxes through faces whenever the mesh connectivity changes between timesteps (which corresponds to the appearance and disappearance of faces).

We define a map from the magnetic field fluxes through the faces of a cell and the magnetic field at the centre of the cell $\mathbf{B}_i$ by solving the linear least-squares problem:
\begin{equation}
\begin{pmatrix}
w_{ij_1} n_{x,ij}  & w_{ij_1} n_{y,ij}  \\
\vdots & \vdots \\
w_{ij_n} n_{x,ij}  & w_{ij_n} n_{y,ij}  \\
\end{pmatrix}
\begin{pmatrix}
B_{x,i} \\
B_{y,i}
\end{pmatrix}
=
\begin{pmatrix}
w_{ij_1} B_{ij_1} \\
\vdots \\
w_{ij_n} B_{ij_n} \\
\end{pmatrix}
\end{equation}
where the $j_1,\ldots,j_n$ are the neighbours of cell $i$, $\mathbf{n}_{ij}$ is the outward normal at the face, and $w_{ij}$ is a weight, where we use the reciprocal of the distance from the centre of mass of the cell to the centroid of the face between cell $i$ and cell $j$. This map is used to obtain the conservative variable$\int B \,dV_i$ of each cell at the beginning of the timestep. The choice of $w_{ij}$ is not unique. We adopt a form for $w_{ij}$ that minimizes the difference of the magnetic field on the face centroids compared to the volume average magnetic field, weighting closer faces as the inverse of the displacement (i.e., the error obtained by extending the cell-average value to the edges is assumed to be dominated by a first order term). One may choose to additionally multiply the weight by the area of faces, or the area of the triangle determined by the centre of mass and the face. We explored these alternate options and found that they made no significant difference for regularized meshes.

The mesh is moved its location at the end of the timestep, and gradient information for $\mathbf{B}$ is calculated, as in Section~\ref{sec:fv} for the other primitive variables. The normal component of the magnetic field across each face is obtained by averaging the two predictions obtained by extrapolating to the face from the cell centre of mass at either side of the face (Equation~\ref{eqn:Bavg}). The Riemann solver requires a constant magnetic field across the shock (which is a consequence of $\nabla\cdot\mathbf{B}=0$ in 1D). Note that in the CT method the divergence of the magnetic field in a cell is not necessarily $0$ when we extrapolate to estimate the flux at the middle of the timestep, but at the end of the timestep it will always evolve to a state with the discretized divergence (equation~\ref{eqn:div}) kept at zero to machine-precision.

The next step is to estimate the electric fields in the rest-frames of each edge. This is done by using the fluxes for the magnetic fields obtained from the Riemann solver and applying a correction to change the frame into the rest frame of the edge. There are multiple estimates for the electric field at each edge, obtained from using the flux of any of the faces that include that edge (3 faces in the case of a non-degenerate 2D Voronoi mesh). These estimates are averaged with barycentric weights of a triangle determined by the centroids of the faces. This choice of interpolation gives the best-possible accuracy for the case of smooth fields. One may design variants of this scheme 
in which the weights are adjusted in the vicinity of magnetosonic shocks to account for the upwind direction, as described in \cite{1999JCoPh.149..270B}, although these modifications were not found to be necessary for our simulations (see discussions in \cite{1999JCoPh.149..270B} and \cite{Toth:2000:DBC:349920.349997} for cases where this variant may be useful). The electric field estimate at an edge $e_{L,ij}$ obtained from the flux across cell $i$ and $j$ is calculated as:
\begin{equation}
\begin{matrix}
E^{\prime\prime}_{L,ij} &=& -(F_{6,ij}\hat{n}_{\perp,x,ij}-F_{5,ij}\hat{n}_{\perp,y,ij}) \\
&\qquad& + (v_{x,L,ij}-w_{x,ij})B_{y,ij} \\
&\qquad& - (v_{y,L,ij}-w_{y,ij})B_{x,ij}
\end{matrix}
\label{eqn:eupdate}
\end{equation}
where $F_{5,ij}$ and $F_{6,ij}$ are the magnetic field fluxes of the Riemann solver in the rest-frame of the faces, $\hat{\mathbf{n}}_{\perp,ij}$ is the outward normal of the face between cells $i$ and $j$, pointing towards $j$, $\mathbf{v}_{L,ij}$ is the edge velocity, calculated exactly from the mesh geometry, and $\mathbf{B}_{ij}$ is the magnetic field (in the lab frame) returned by the 1D Riemann solver sampled at $x=0$.

An alternate approach to the above is to use a field-interpolated approach; i.e., to extrapolate the magnetic and velocity fields to each edge of each face in the rest-frame of the edge. However, flux-interpolated approaches have slightly better performance and desirable properties due to the consistency of coupling the electric field estimate to the Riemann solver \citep{Toth:2000:DBC:349920.349997}. Equation~\ref{eqn:eupdate}, without the last two terms that account for the change of frame, is the familiar way to estimate the electric field in flux-interpolated CT schemes on fixed grids (see, e.g. \cite{1999JCoPh.149..270B}).

The above describes the basics of a CT algorithm for moving unstructured meshes, which preserves $\nabla\cdot \mathbf{B}=0$. It is important that the initial conditions must be set so that $\nabla\cdot \mathbf{B}=0$ at the beginning. This is accomplished using the vector potential $\mathbf{A}$, evaluating it on the face edges, and using $\mathbf{B} = \nabla \times \mathbf{A}$ to obtain the face-centred magnetic fields, which is the usual strategy taken by CT algorithms on fixed grids. In particular, in our 2D simulations we set the magnetic flux on each of the faces to be:
\begin{equation}
\Phi_{ij} = A_z(e_{L,ij}) - A_z(e_{R,ij})
\end{equation}
If the divergence error is not set to zero in the initial conditions, the scheme will preserve any glitch in the initial conditions (effectively conserving a magnetic monopole).

\subsubsection{Remapping}\label{sec:remapping}

During one timestep to the next, the moving Voronoi tessellation may change connectivity, resulting in faces appearing and disappearing. A face that appears at the end of a timestep poses no large issue for preserving $\nabla\cdot\mathbf{B}=0$ because it can be treated as coming from a degenerate face of area and magnetic flux $0$ during the previous step. The nice cancellation property of the update terms for the magnetic fluxes of the faces of a cell in the calculation of the divergence still holds. A bigger concern is the disappearance of a face. In the continuous limit, the flux through the face should go to $0$ as the face disappears, since its area goes to zero. But in our discretization, a vanishing face would have some small amount of residual flux that would not be properly accounted after the face disappears in the next timestep, and this would result in a breakdown of the CT method. However, this issue can be redressed by using a remapping technique, described here.

\begin{figure}
\centering
\includegraphics[width=0.47\textwidth]{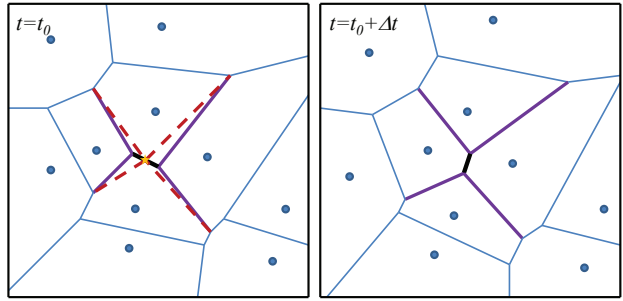}
\caption{During one timestep (left) and the next (right), the connectivity between cells in the moving Voronoi mesh may change. In such a case, a face disappears (thick black line). The geometry at the beginning of the timestep is remapped by adding a degenerate vertex (yellow star) at the midpoint of the face that is to disappear and connecting the surrounding faces (red-dashed lines) to it. The magnetic flux through the vanishing face is redistributed to the faces shown in red-dashed lines. The connectivity of the mesh is not permitted to change too drastically from one time-step to the next: if a face disappears, there must be remaining neighbouring faces that do not change connectivity to which the flux is redistributed. Voronoi diagrams change continuously, so drastic changes can always be avoided by taking the timestep small enough.}
\label{fig:remapping}
\end{figure}

The approach taken in this work is to remap the geometry slightly at the beginning of each timestep in the cases where mesh connectivity changes. A face that is identified to disappear is mapped to a degenerate point at the centroid of the face (i.e., one can think of shrinking the face to a degenerate point at the centroid and remapping the magnetic fluxes through the faces). The remapping is illustrated in Fig.~\ref{fig:remapping}. The vanishing face touches two cells. For each touching cell, the flux through the vanishing face is equally distributed to the two faces that touch this face. This technique preserves $\nabla\cdot\mathbf{B}=0$ exactly to machine-precision. The method does require that the connectivity between time-steps does not change drastically; that is, if a face disappears then its surrounding neighbouring faces are not allowed to change connectivity. This is always possible in the limit of small timesteps since the Voronoi mesh evolves continuously. After each timestep, we are required to check that the connectivity has not changed too drastically, and if it did, then the current timestep is halved until the mesh evolves in a satisfactory way. This is typically a rare occurrence in our test problems because the CFL condition does limit the size of the timestep and typically $0$ or $1$ faces appear/disappear in a cell. But more sophisticated techniques can be employed to improve the efficiency of the algorithm and avoid this halving of timesteps (see Section~\ref{sec:var}). 

\subsubsection{Extension to 3D}\label{sec:3d}

The method described here is generalizable to three dimensions. In the 3D case, the boundaries of a face are a polygon rather than two points. Thus when updating the magnetic fluxes one needs to perform a loop integral over the boundary, just as in the regular CT method. The key difference is to estimate the electric fields in the rest frames of the centroid of each linear segment of the loop. An edge at the end of the timestep may not have been present at the beginning of the timestep, in which case in order to estimate its velocity it can be considered as a `degenerate face' (a single point) in the previous Voronoi tessellation. In the remapping step, the magnetic flux through a face that disappears may be equally redistributed to all the faces that touch the vanishing face, and the face that disappears is shrunk down to its centroid.

In short, the key ideas necessary for extending CT to an arbitrary moving mesh are to always calculate electric fields in the moving frame of the edge and to remap the magnetic fluxes of faces that disappear to the neighbouring faces.

\begin{figure*}
\centering
\begin{tabular}{ccccc}
 & $\rho, \,\,\, t=0.2$  & $p_{\rm gas}, \,\,\, t=0.2$ &  $\mathbf{B}^2/2, \,\,\, t=0.2$ 
& $\log_{10}\left(  \frac{|\nabla\cdot\mathbf{B}_i|\cdot |R_i|}{\sqrt{2p_i}} \right), \,\,\,  t=0.2$ \\
  &
\includegraphics[width=0.22\textwidth]{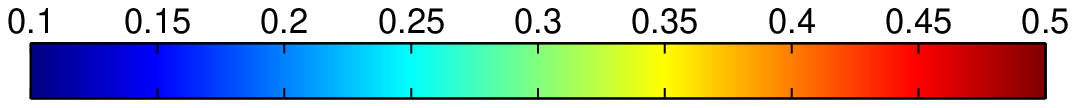} &
\includegraphics[width=0.22\textwidth]{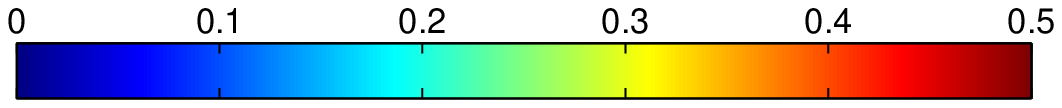} &
\includegraphics[width=0.22\textwidth]{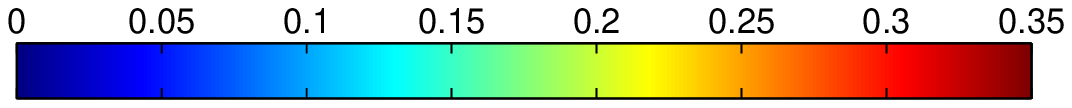} &
\includegraphics[width=0.22\textwidth]{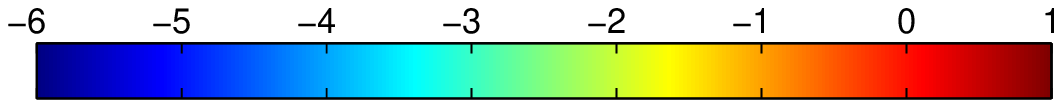} \\
\rotatebox{90}{\hspace{8 mm} moving CT (Rusanov)} &
\includegraphics[width=0.22\textwidth]{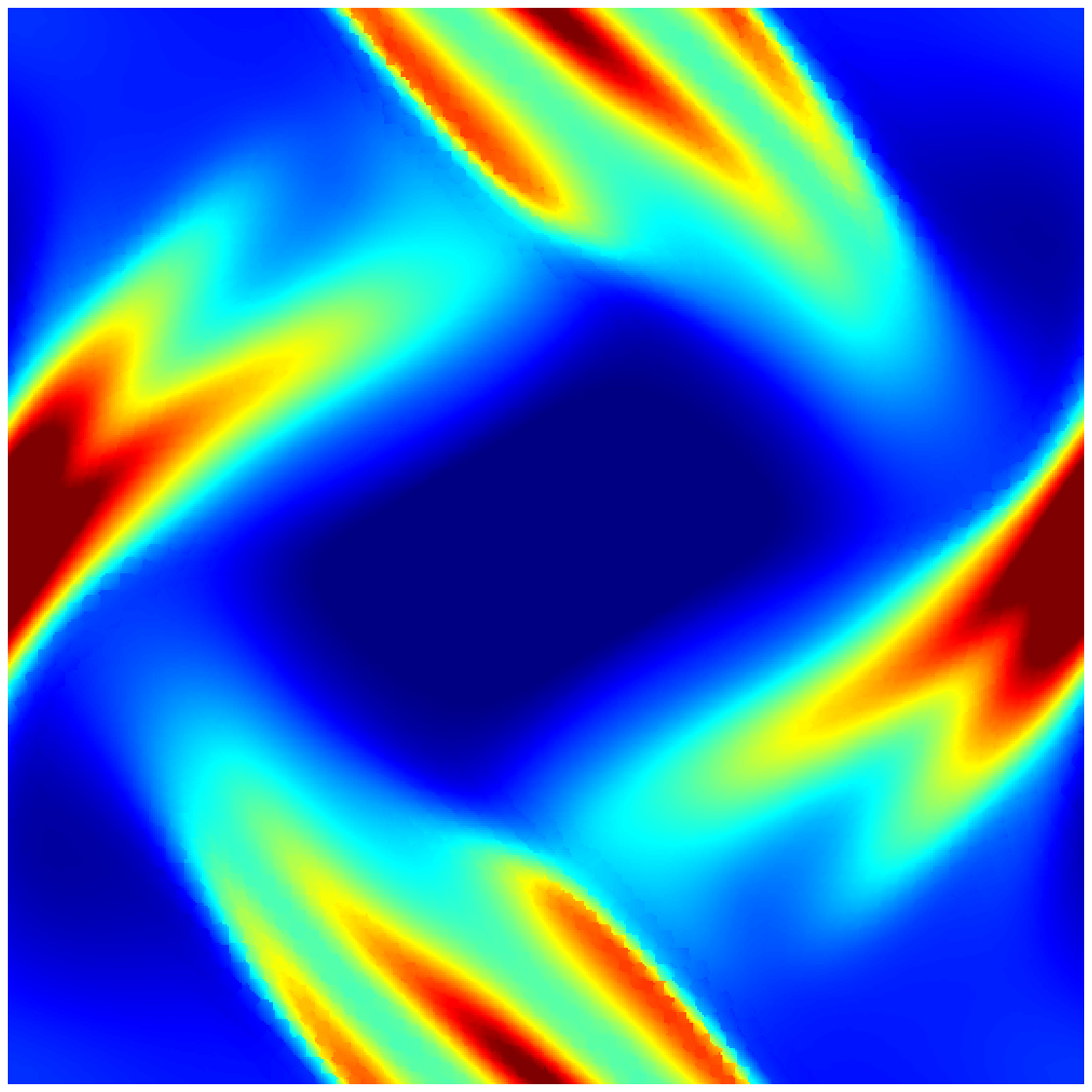} &
\includegraphics[width=0.22\textwidth]{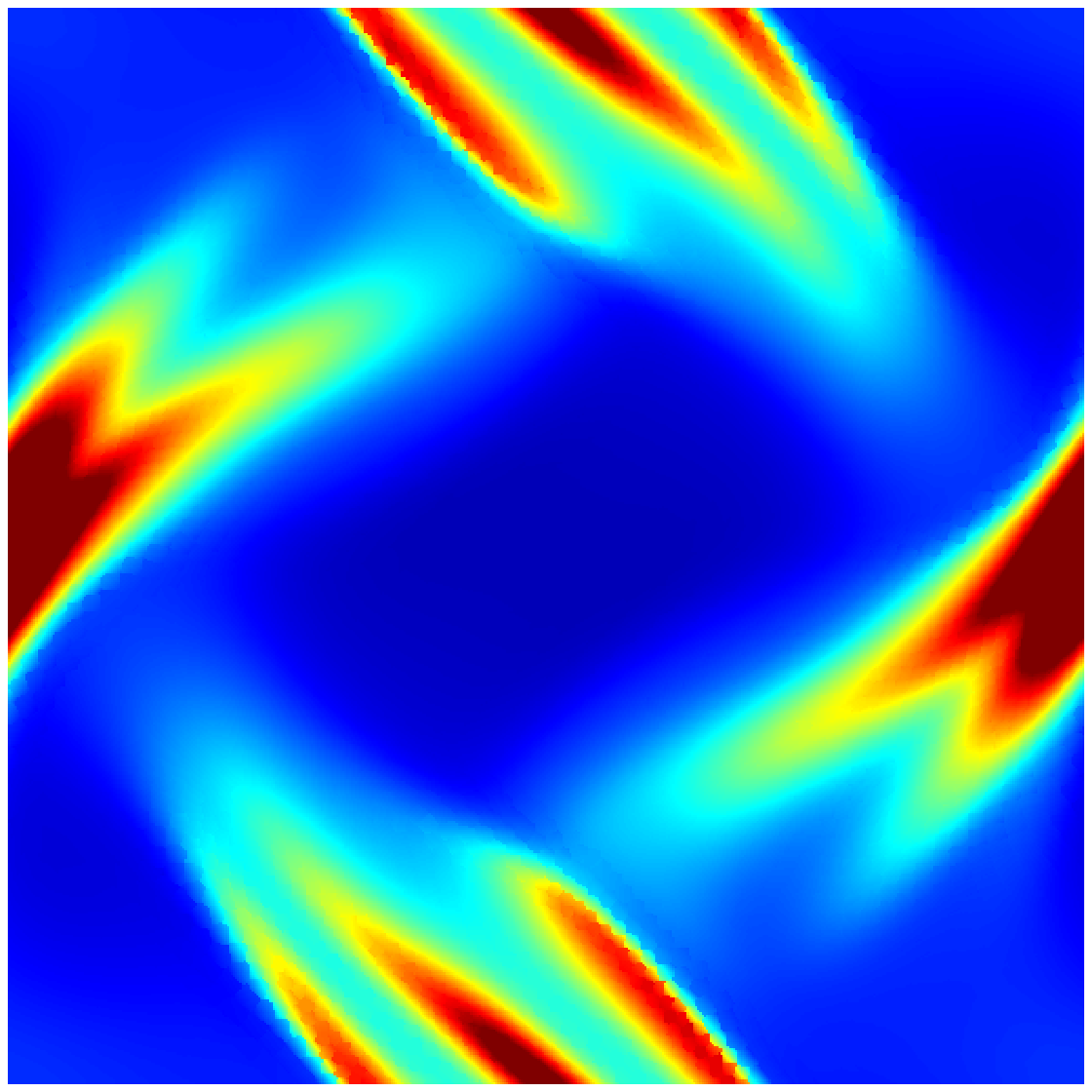} &
\includegraphics[width=0.22\textwidth]{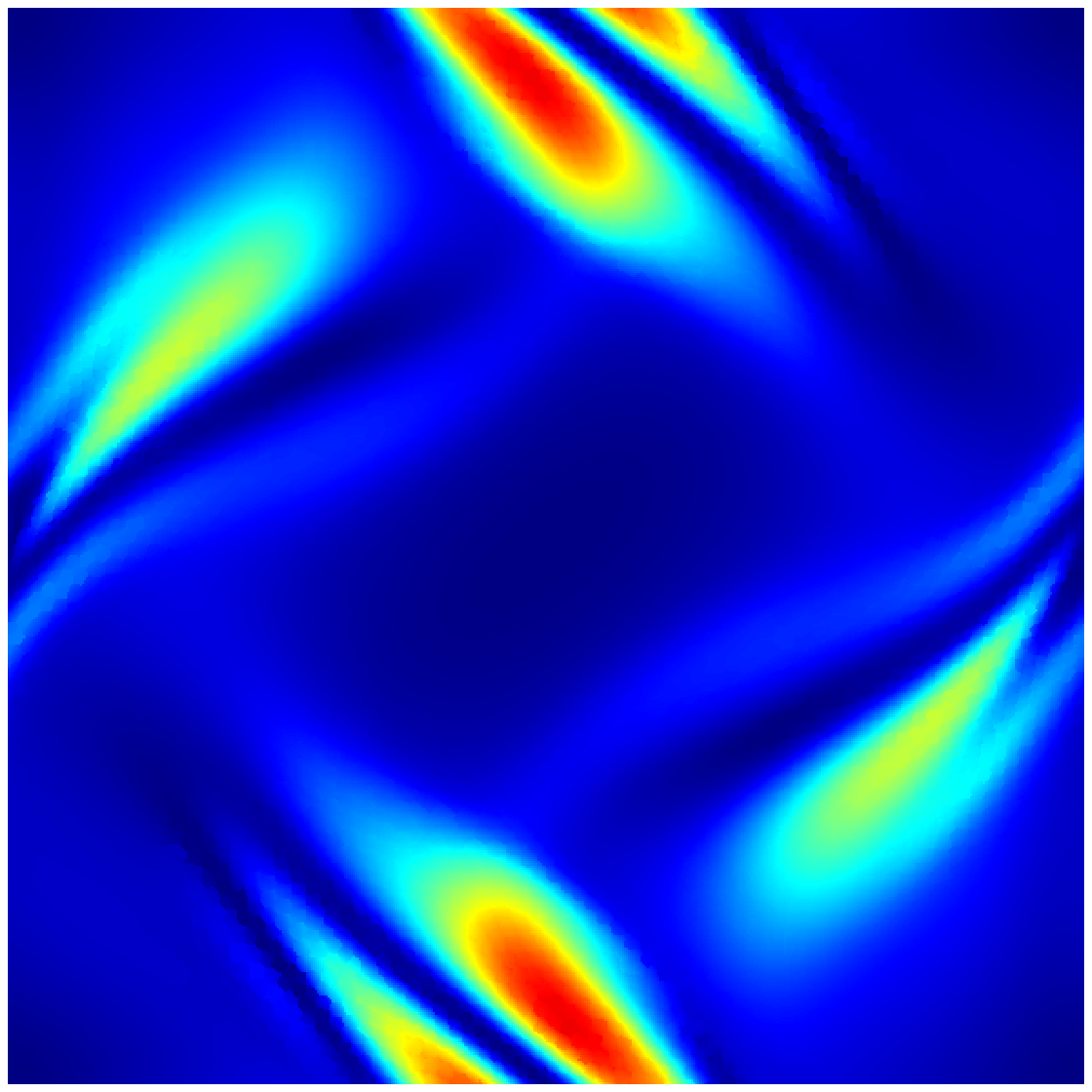} &
\includegraphics[width=0.22\textwidth]{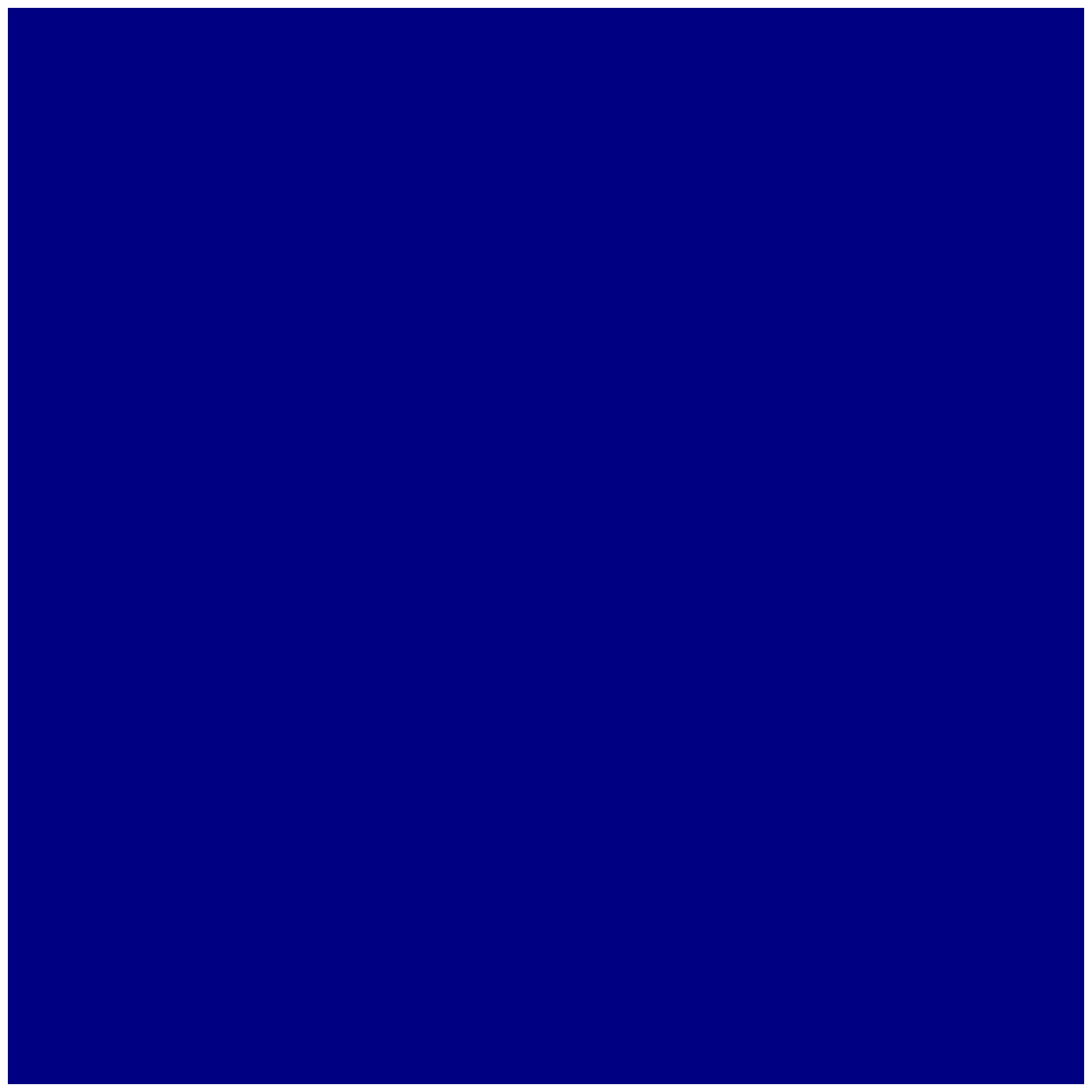} \\
\rotatebox{90}{\hspace{9 mm} moving CT (HLLD)} &
\includegraphics[width=0.22\textwidth]{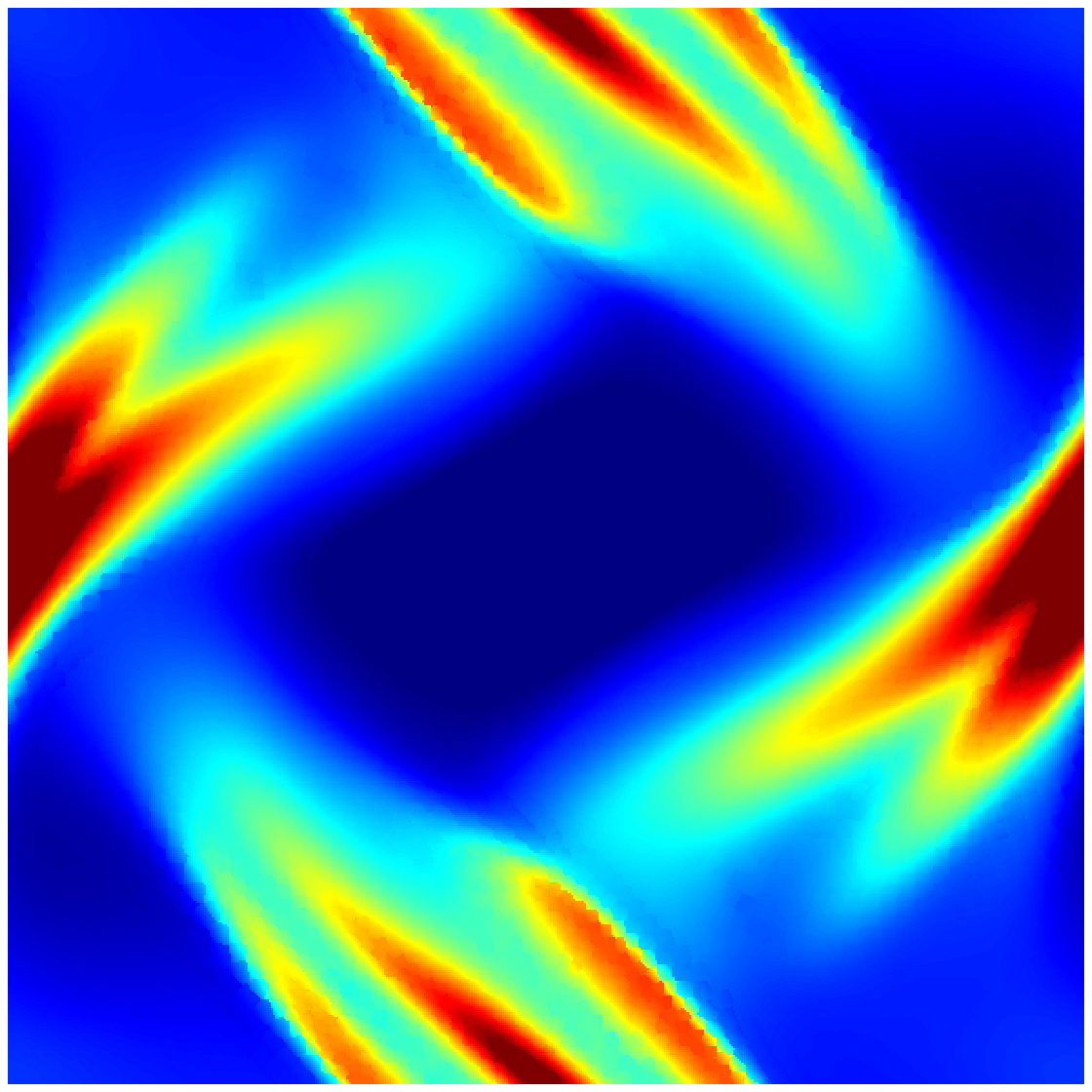} &
\includegraphics[width=0.22\textwidth]{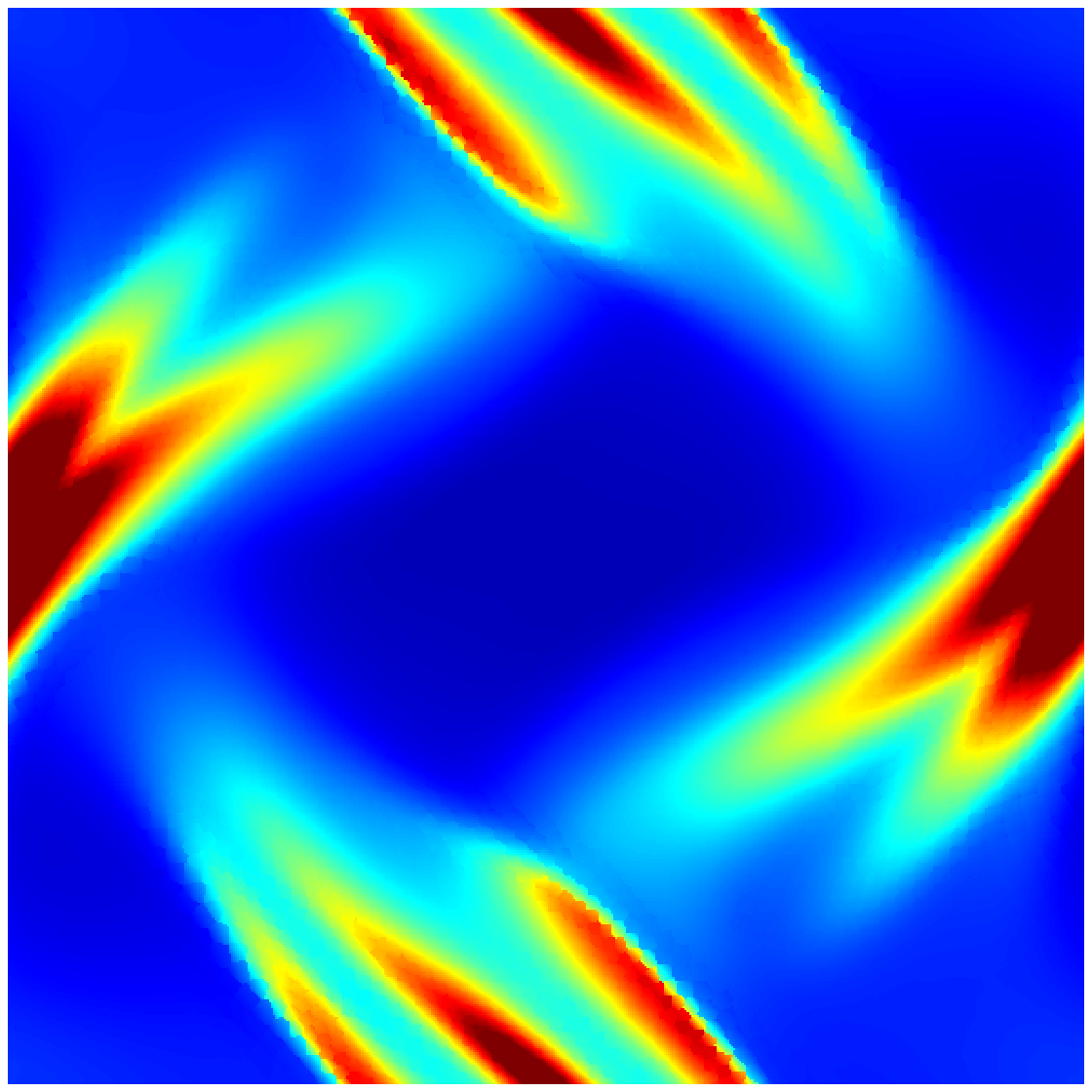} &
\includegraphics[width=0.22\textwidth]{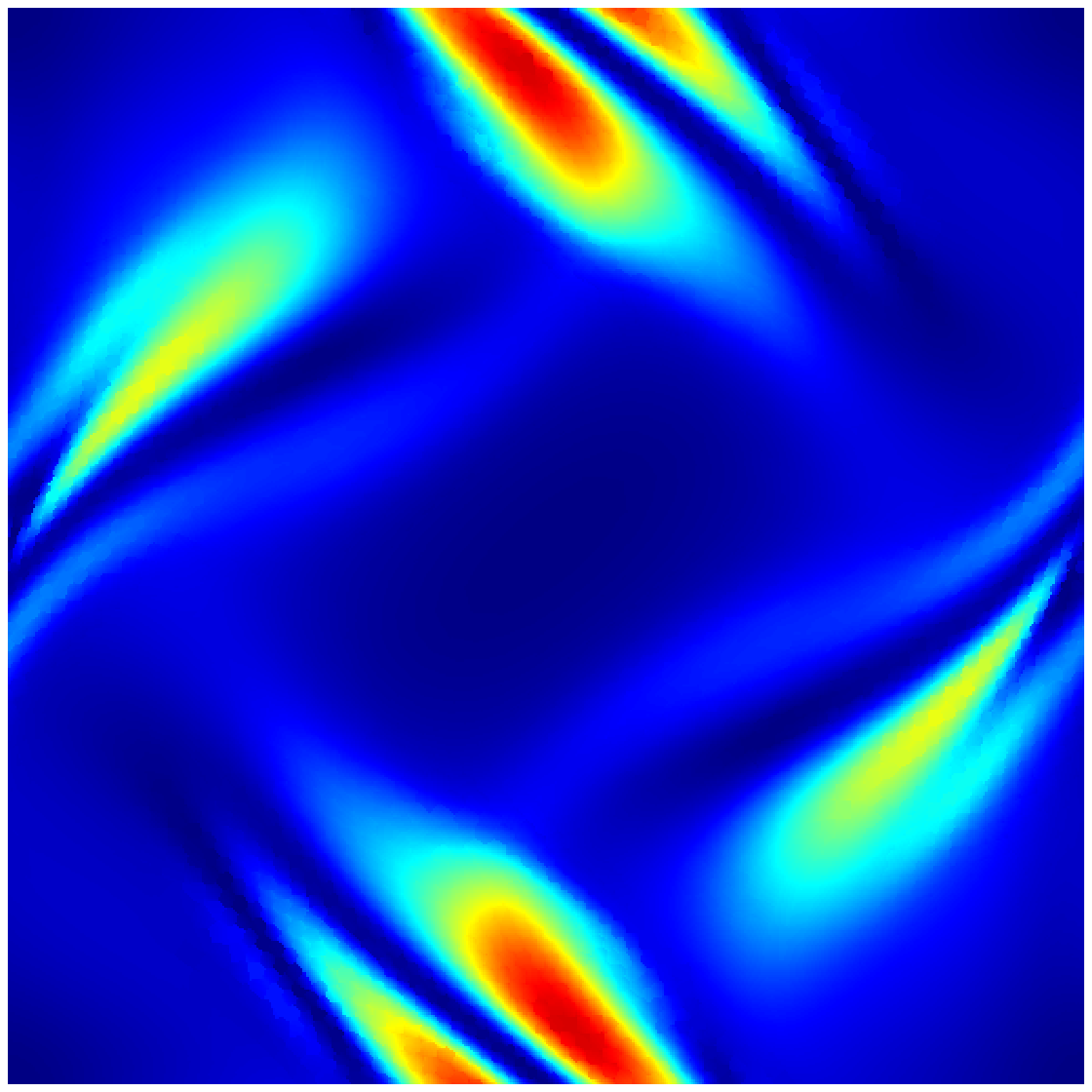} &
\includegraphics[width=0.22\textwidth]{onull.eps} \\
\rotatebox{90}{\hspace{4 mm} static CT (hexagonal,HLLD)} &
\includegraphics[width=0.22\textwidth]{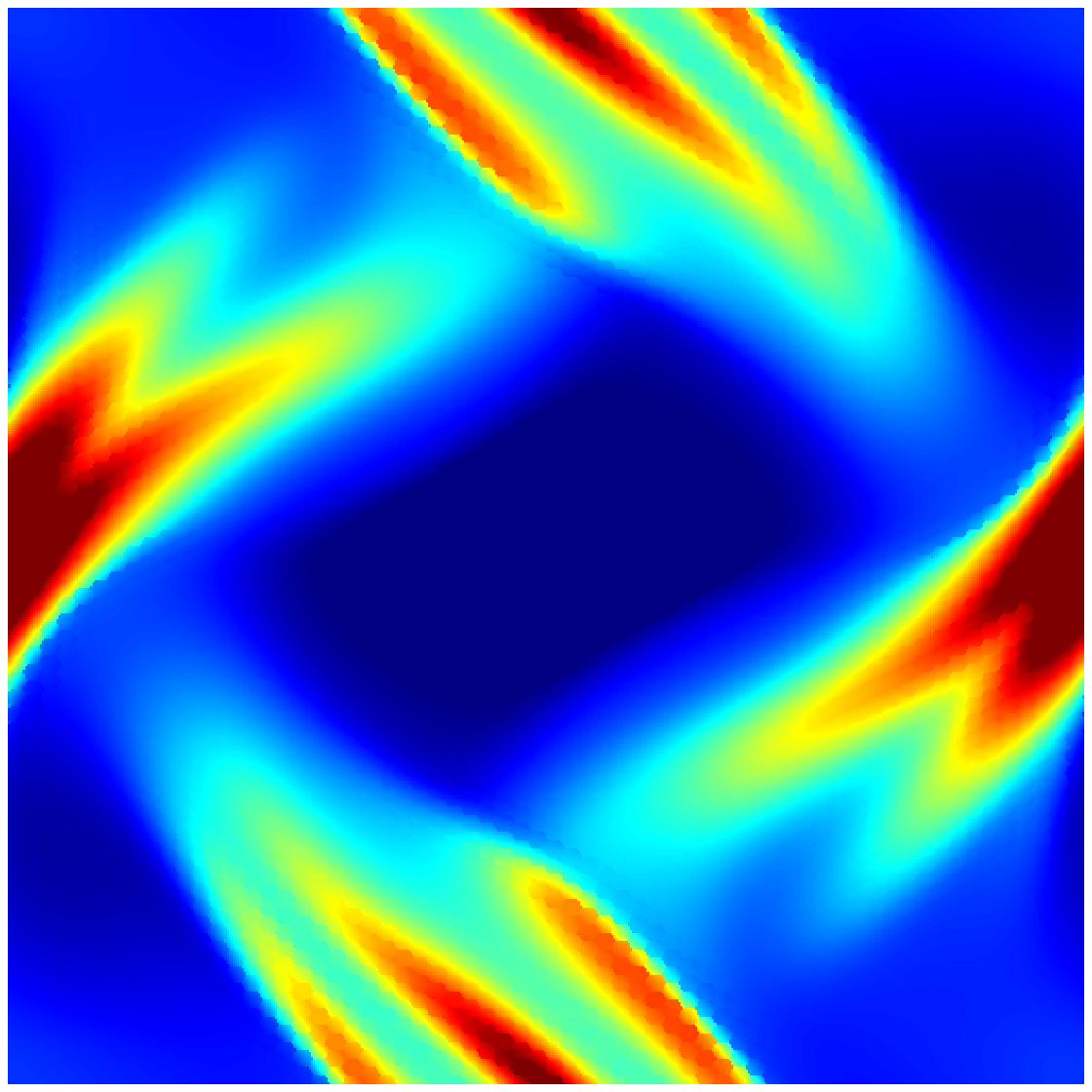} &
\includegraphics[width=0.22\textwidth]{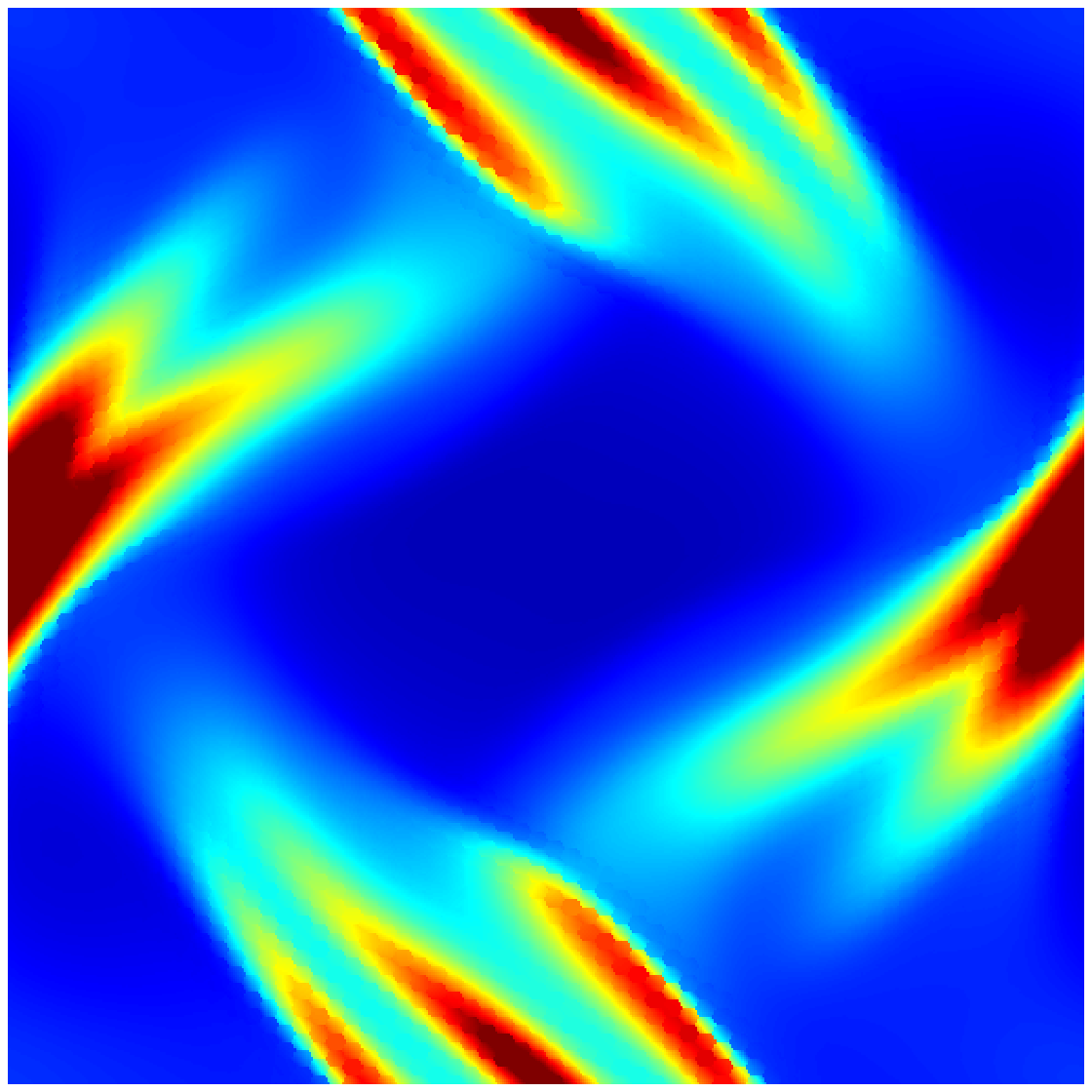} &
\includegraphics[width=0.22\textwidth]{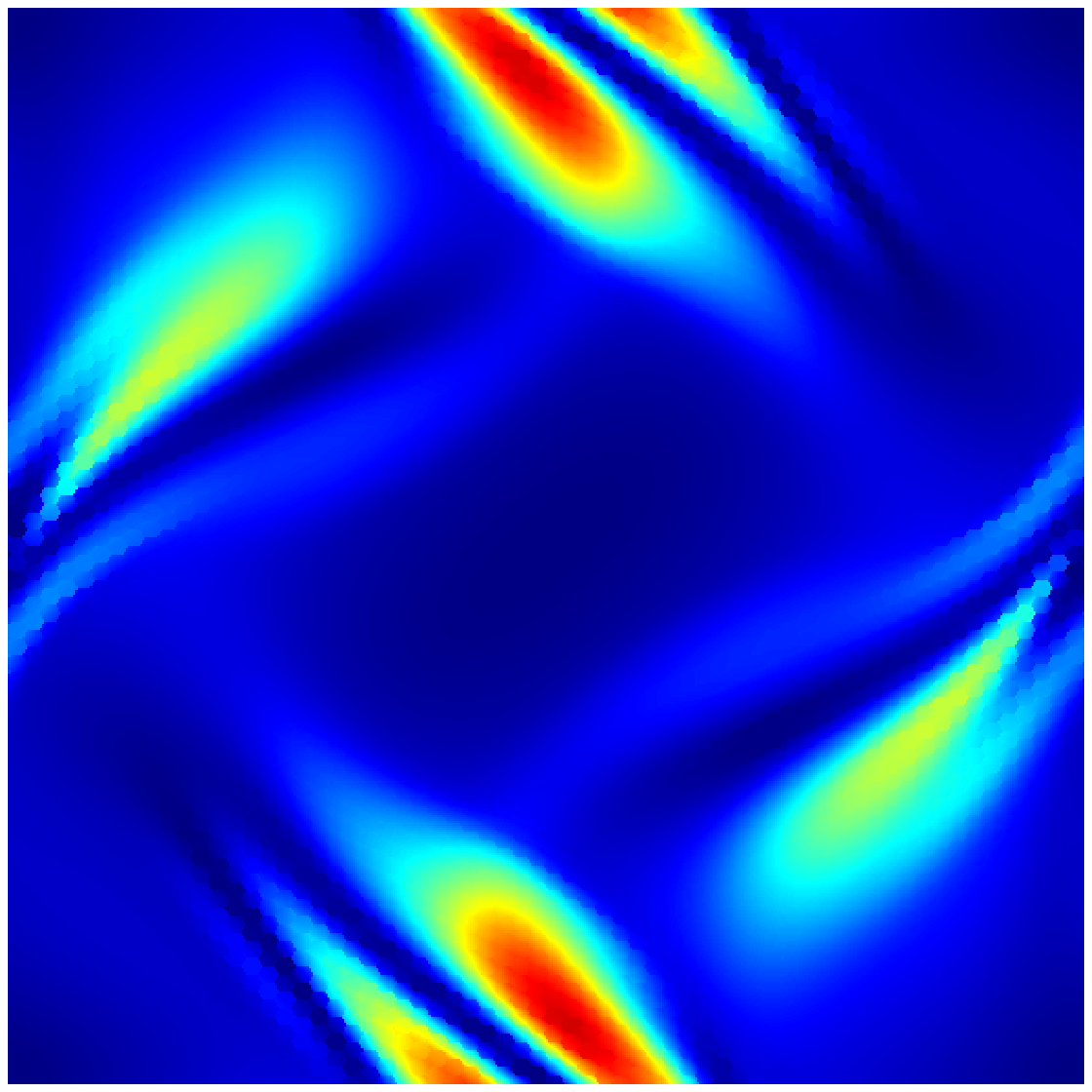} &
\includegraphics[width=0.22\textwidth]{onull.eps} \\
\rotatebox{90}{\hspace{5 mm} moving Powell (Rusanov)} &
\includegraphics[width=0.22\textwidth]{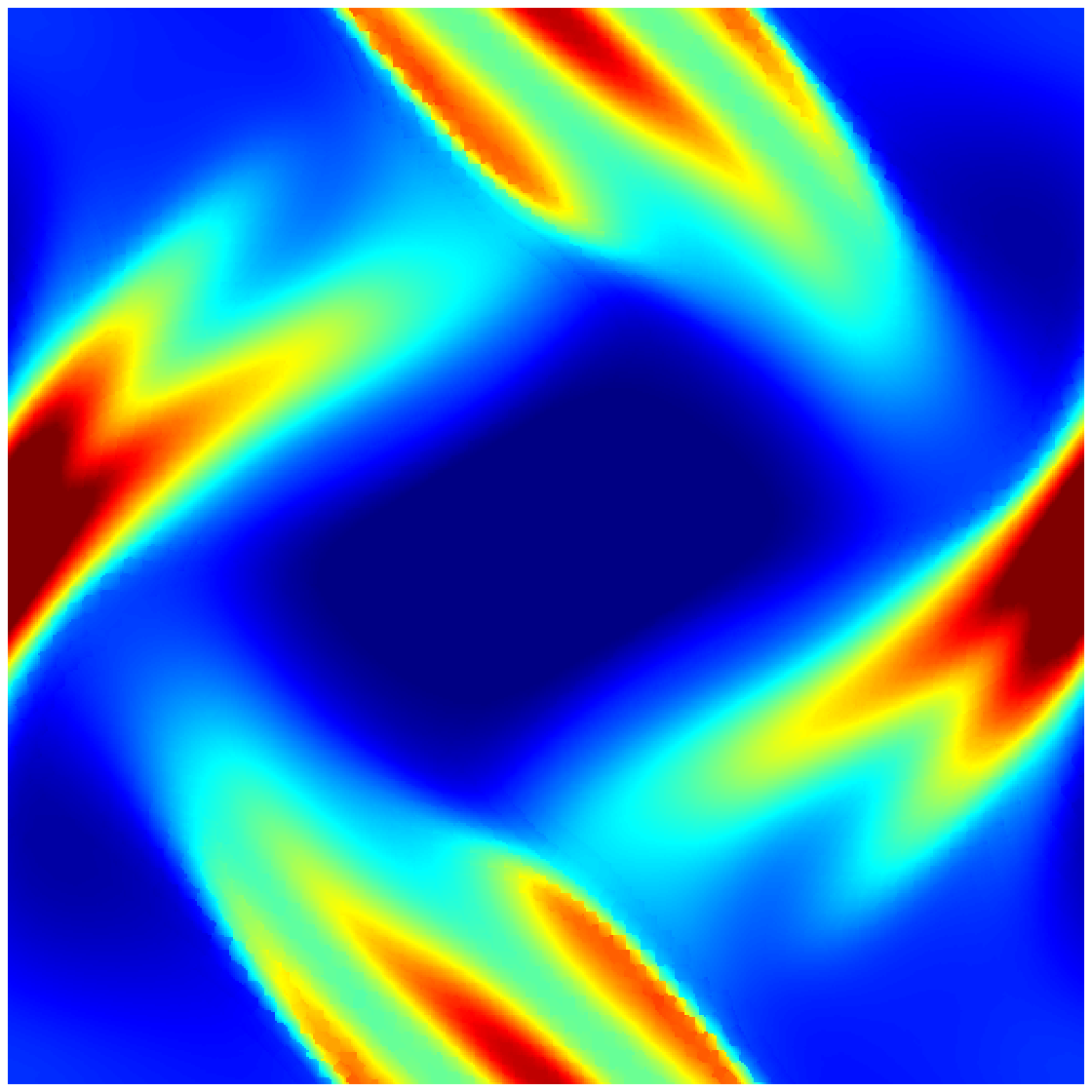} &
\includegraphics[width=0.22\textwidth]{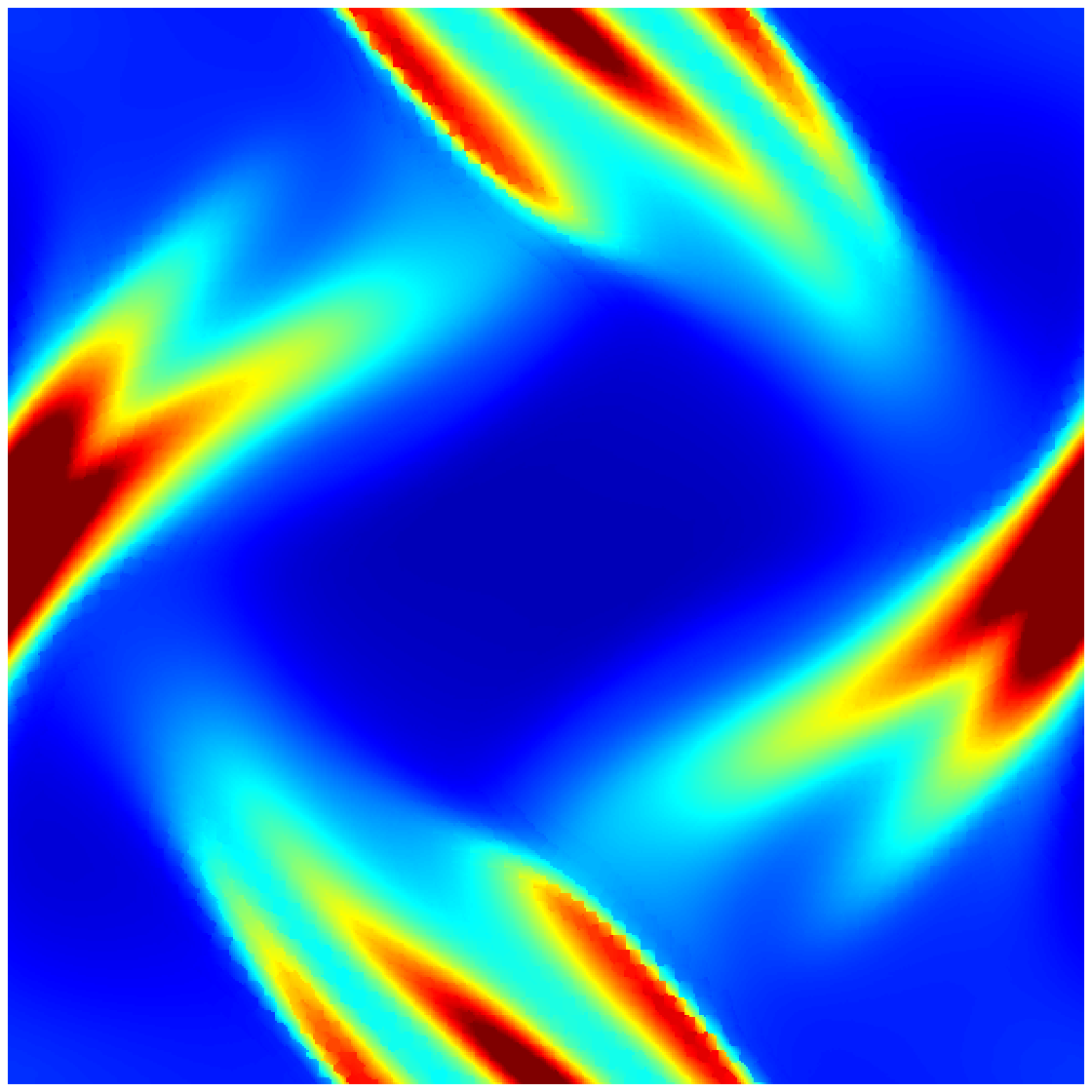} &
\includegraphics[width=0.22\textwidth]{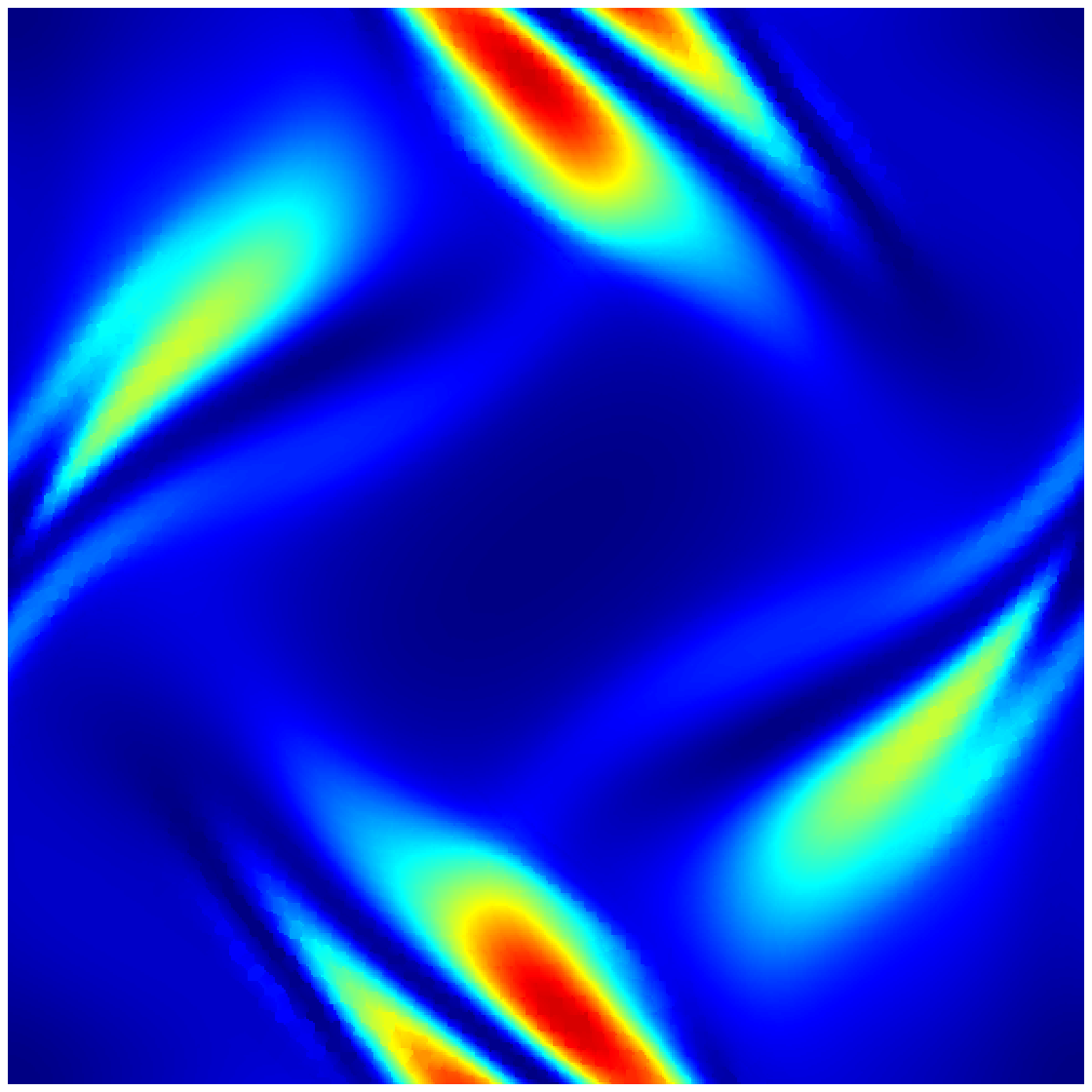} &
\includegraphics[width=0.22\textwidth]{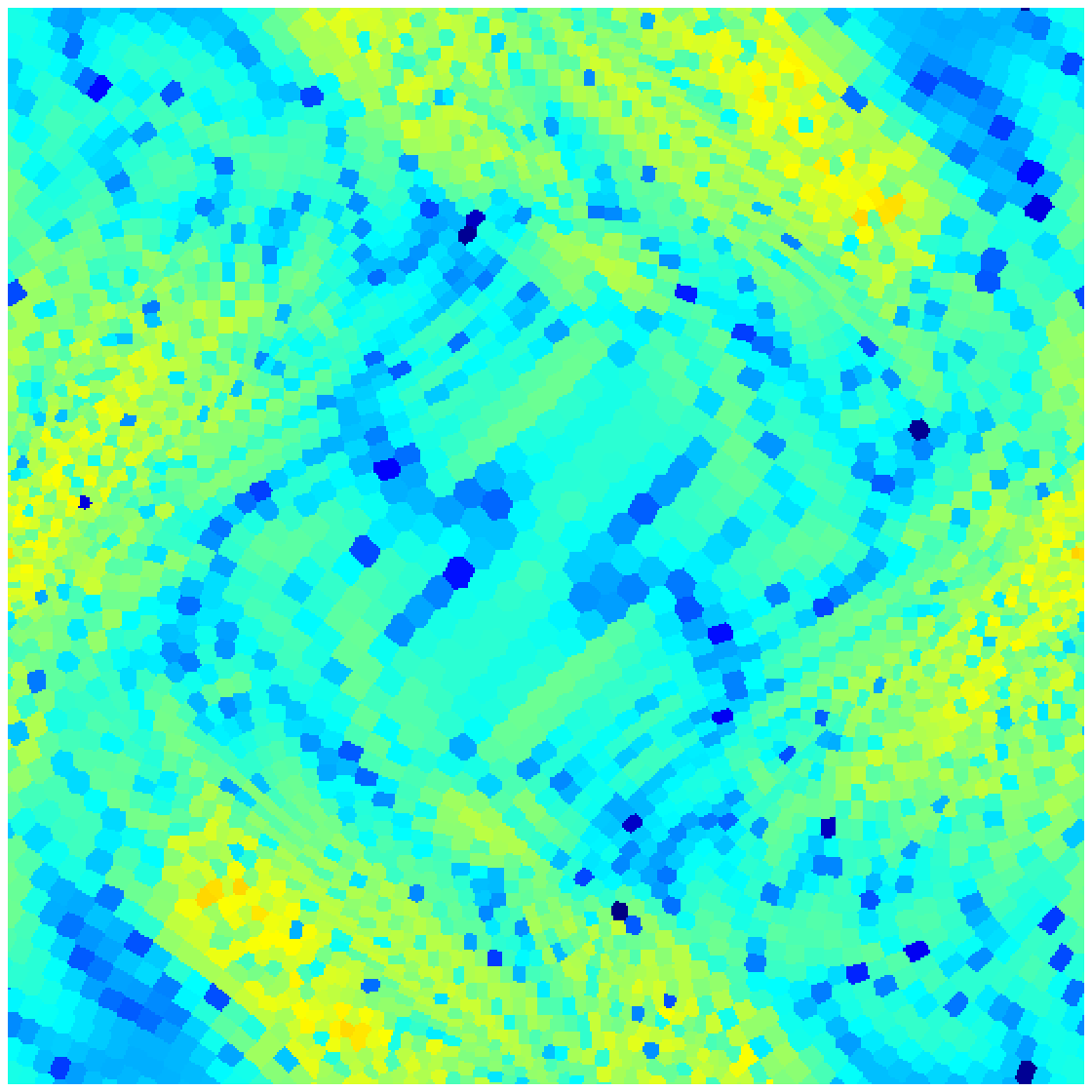} \\
\rotatebox{90}{\hspace{6 mm} moving Powell (HLLD)} &
\includegraphics[width=0.22\textwidth]{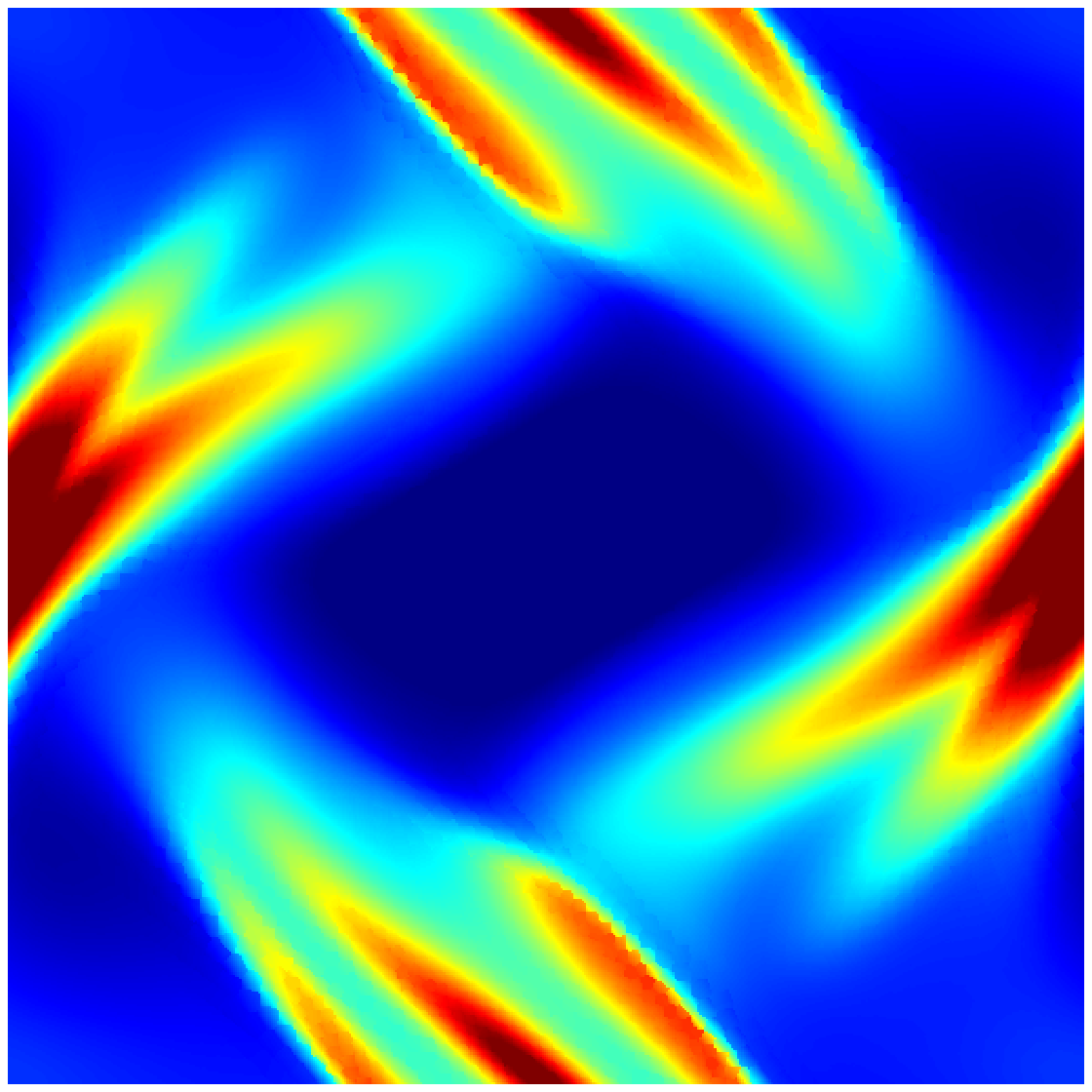} &
\includegraphics[width=0.22\textwidth]{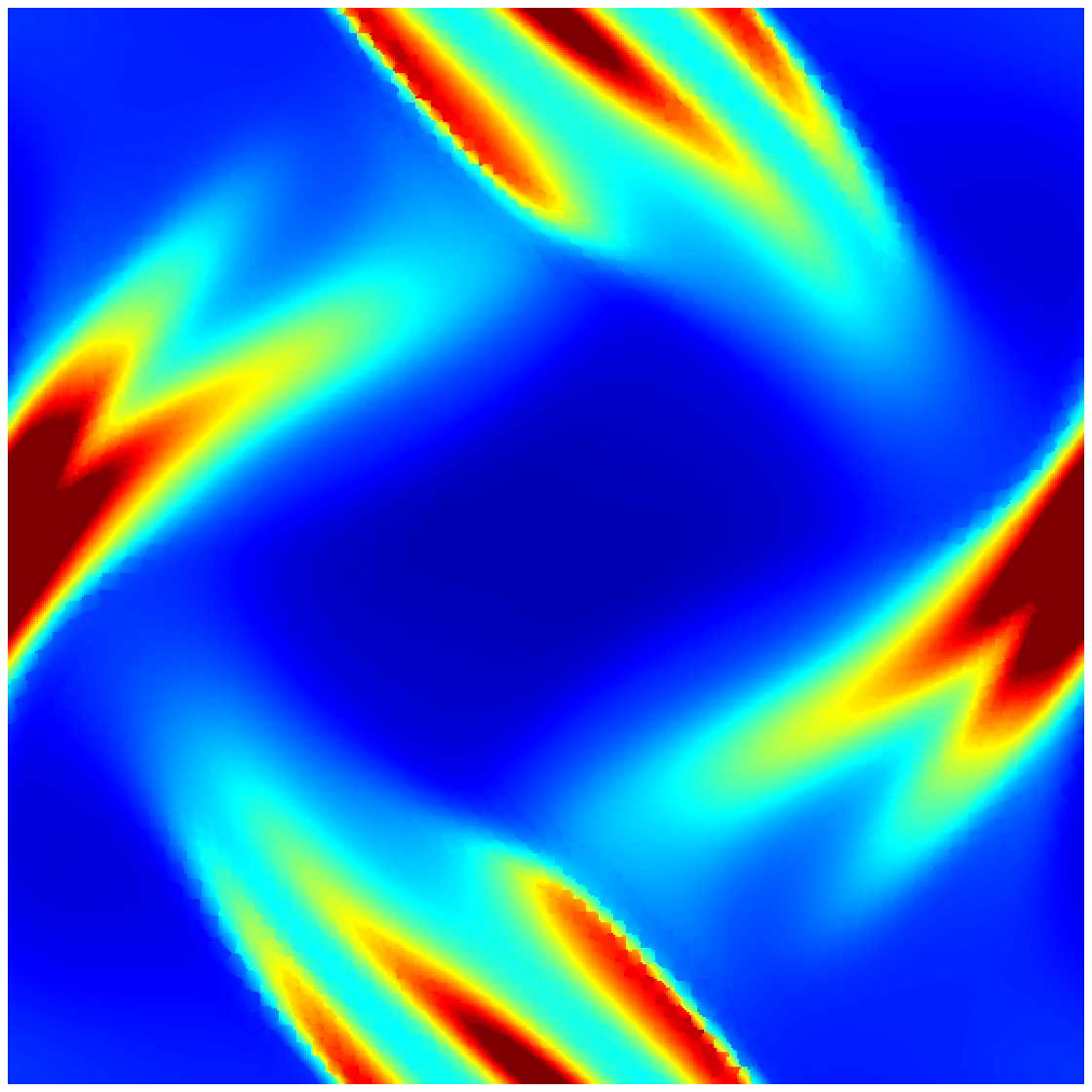} &
\includegraphics[width=0.22\textwidth]{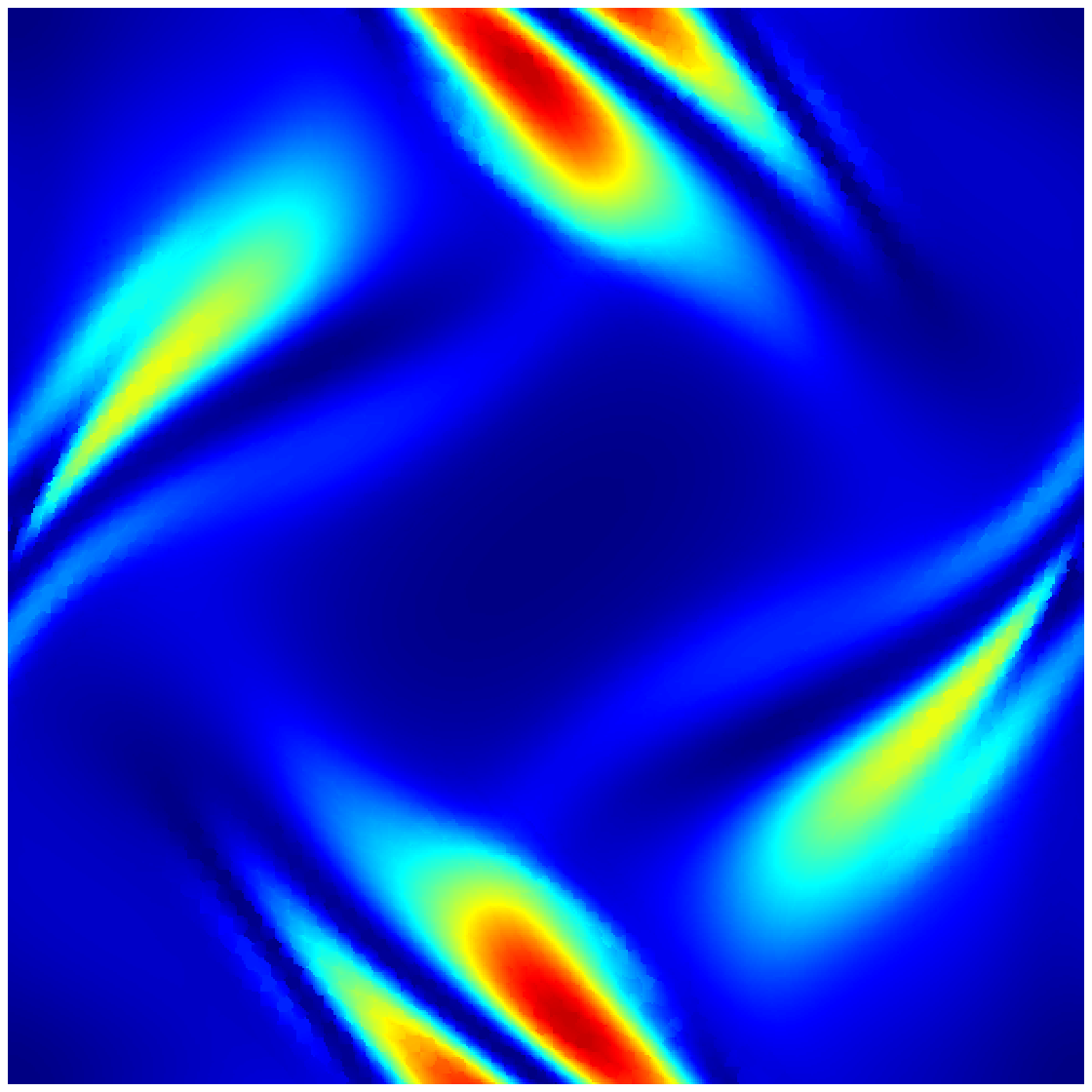} &
\includegraphics[width=0.22\textwidth]{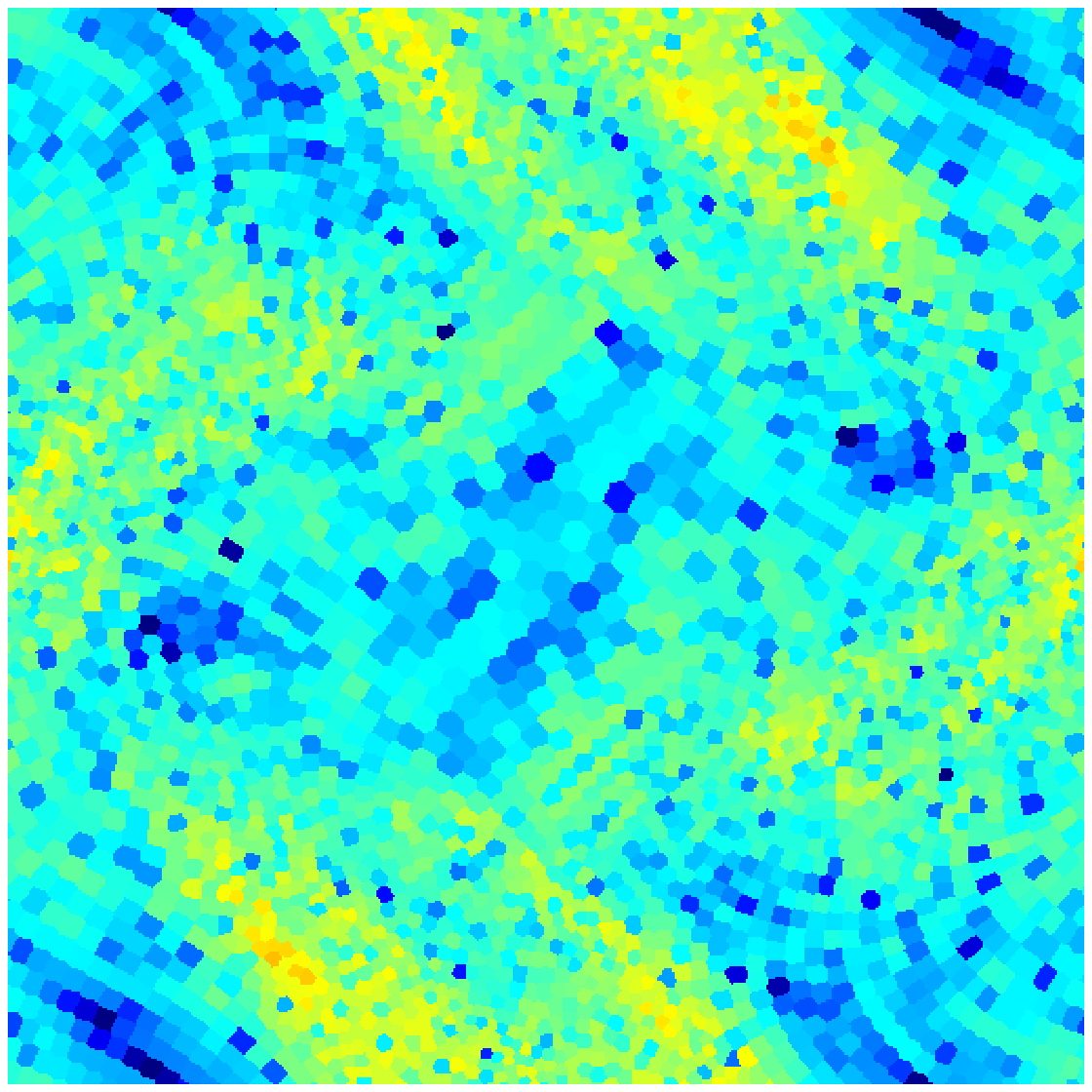} \\
\end{tabular}
\caption{A comparison of various numerical methods used to evolve the Orszag-Tang test at $t=0.2$. Shown are the moving CT method (with Rusanov and HLLD flux solvers), the static CT method (on a hexagonal grid with an HLLD solver), and the moving Powell method (with Rusanov and HLLD flux solvers). Plotted are the density, pressure, magnetic energy density, and relative divergence error of the magnetic fields (compared to fluid pressure). For the CT schemes, the divergence errors are of the order of machine precision $\sim 10^{-15}$, so they are much smaller than the minimum colour value indicated ($10^{-6}$) by the colour bar.}
\label{fig:otmain}
\end{figure*}

\begin{figure*}
\centering
\begin{tabular}{ccccc}
 & $\rho, \,\,\, t=0.5$  & $p_{\rm gas}, \,\,\, t=0.5$ &  $\mathbf{B}^2/2, \,\,\, t=0.5$
& $\log_{10}\left(  \frac{|\nabla\cdot\mathbf{B}_i|\cdot |R_i|}{|\sqrt{2p_i}|} \right), \,\,\,  t=0.5$ \\
  &
\includegraphics[width=0.22\textwidth]{cbd.eps} &
\includegraphics[width=0.22\textwidth]{cbp.eps} &
\includegraphics[width=0.22\textwidth]{cbb.eps} &
\includegraphics[width=0.22\textwidth]{cbdivb.eps} \\
\rotatebox{90}{\hspace{8 mm} moving CT (Rusanov)} &
\includegraphics[width=0.22\textwidth]{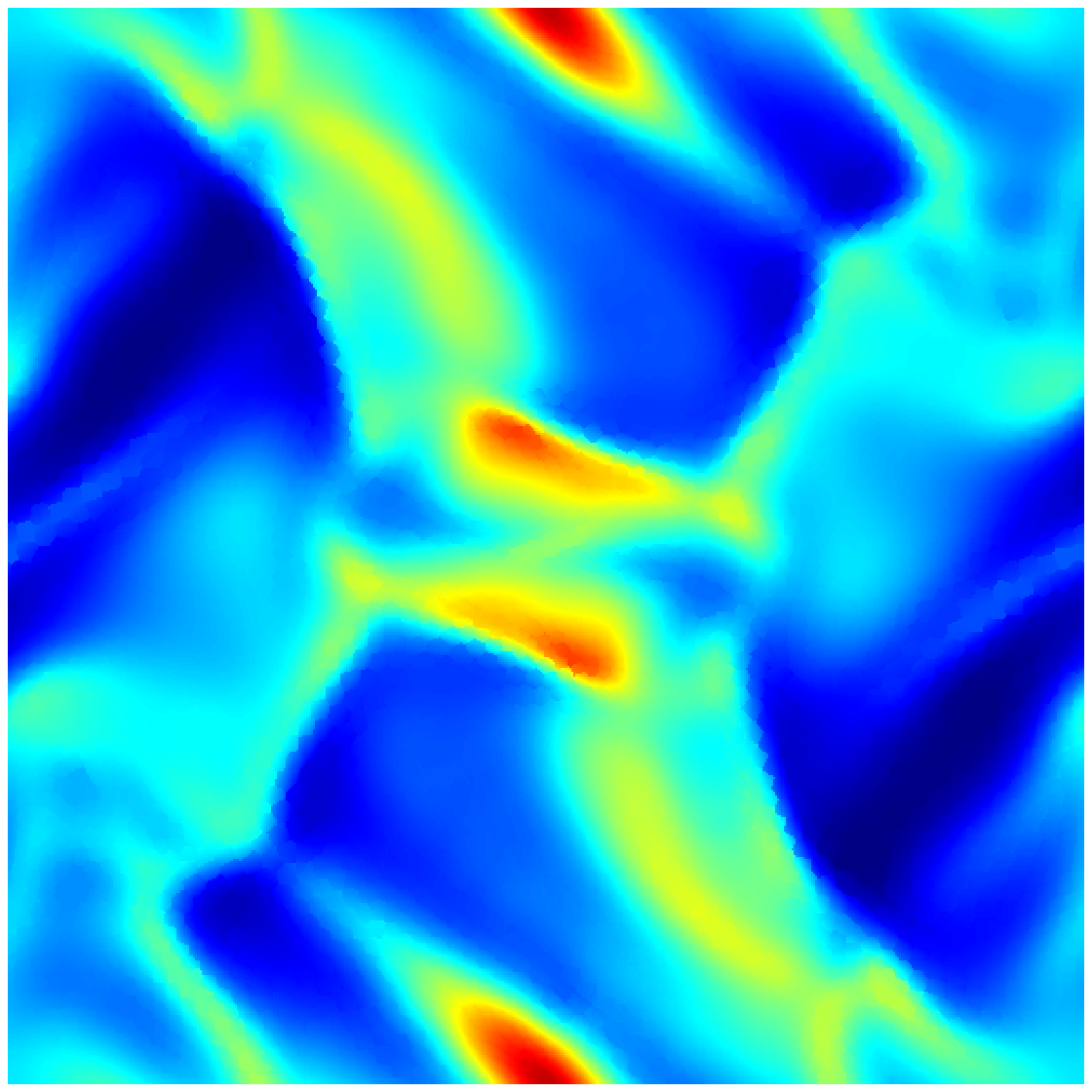} &
\includegraphics[width=0.22\textwidth]{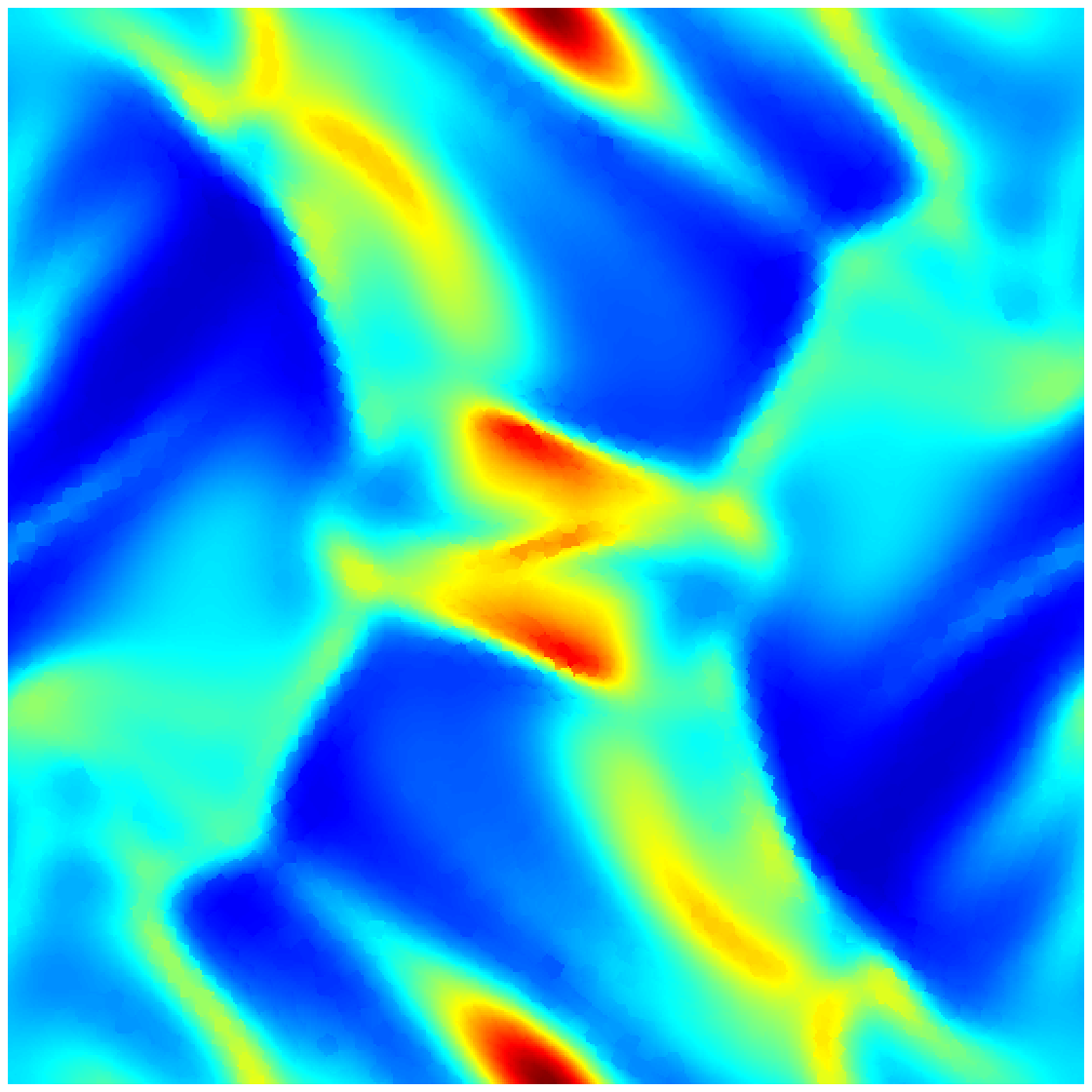} &
\includegraphics[width=0.22\textwidth]{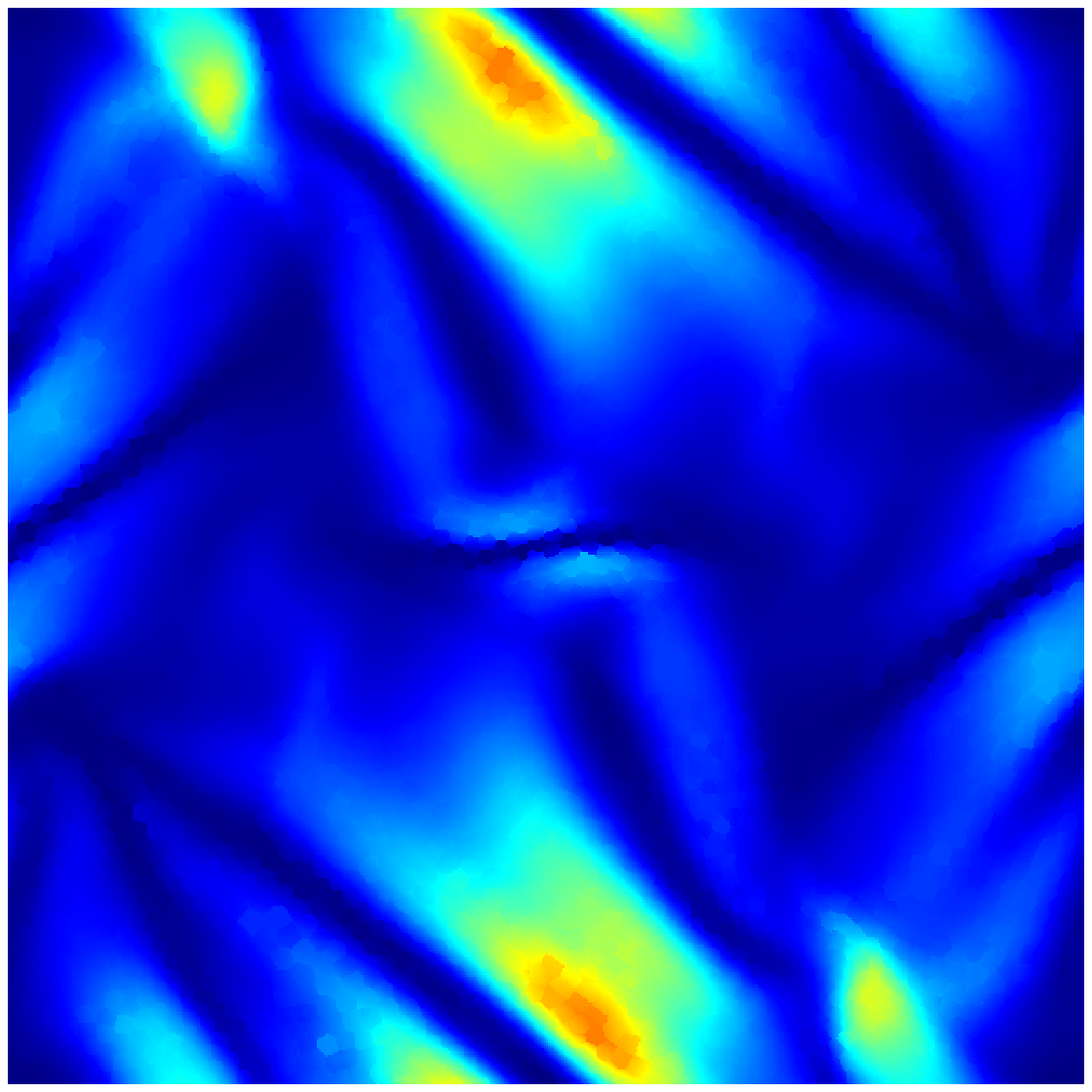} &
\includegraphics[width=0.22\textwidth]{onull.eps} \\
\rotatebox{90}{\hspace{9 mm} moving CT (HLLD)} &
\includegraphics[width=0.22\textwidth]{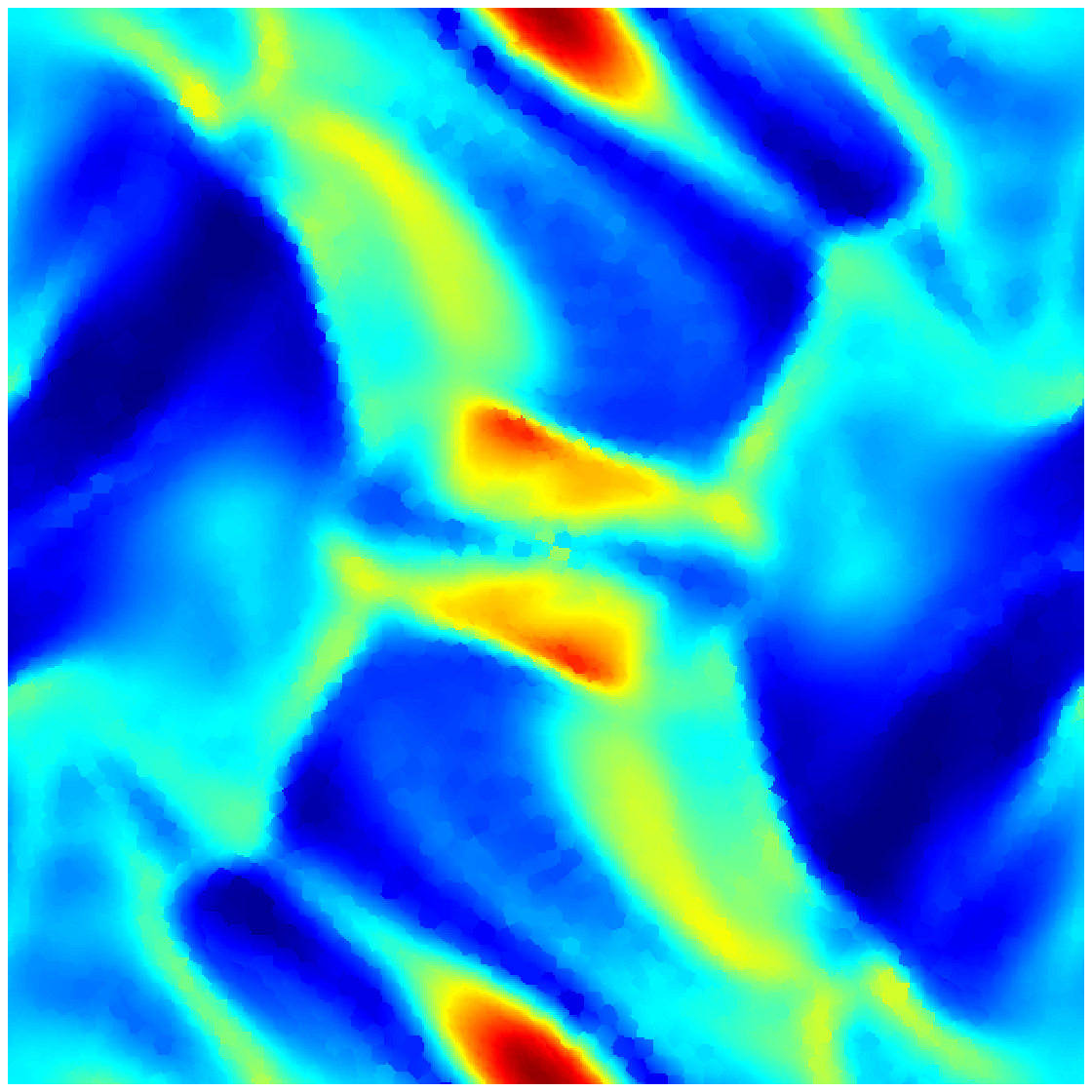} &
\includegraphics[width=0.22\textwidth]{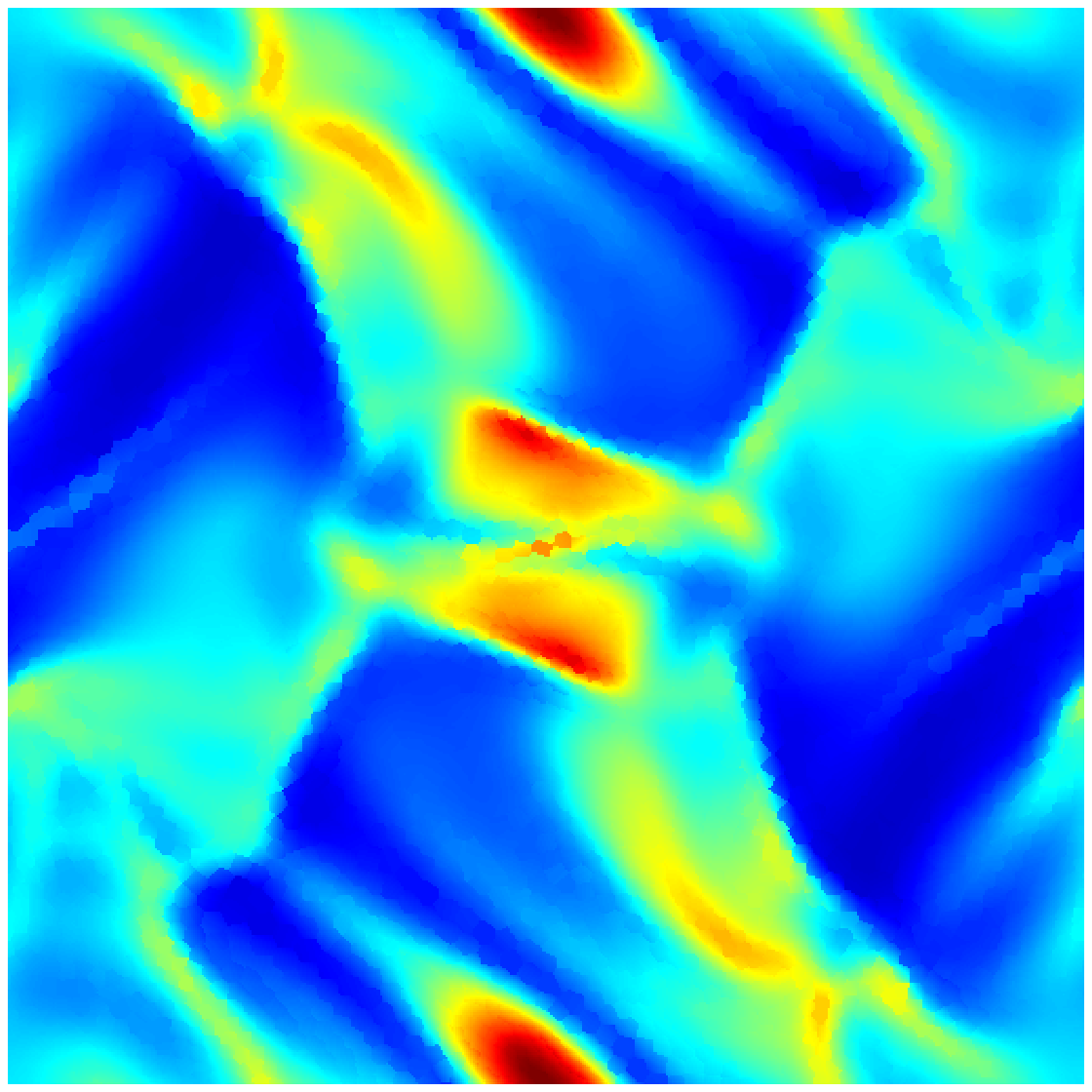} &
\includegraphics[width=0.22\textwidth]{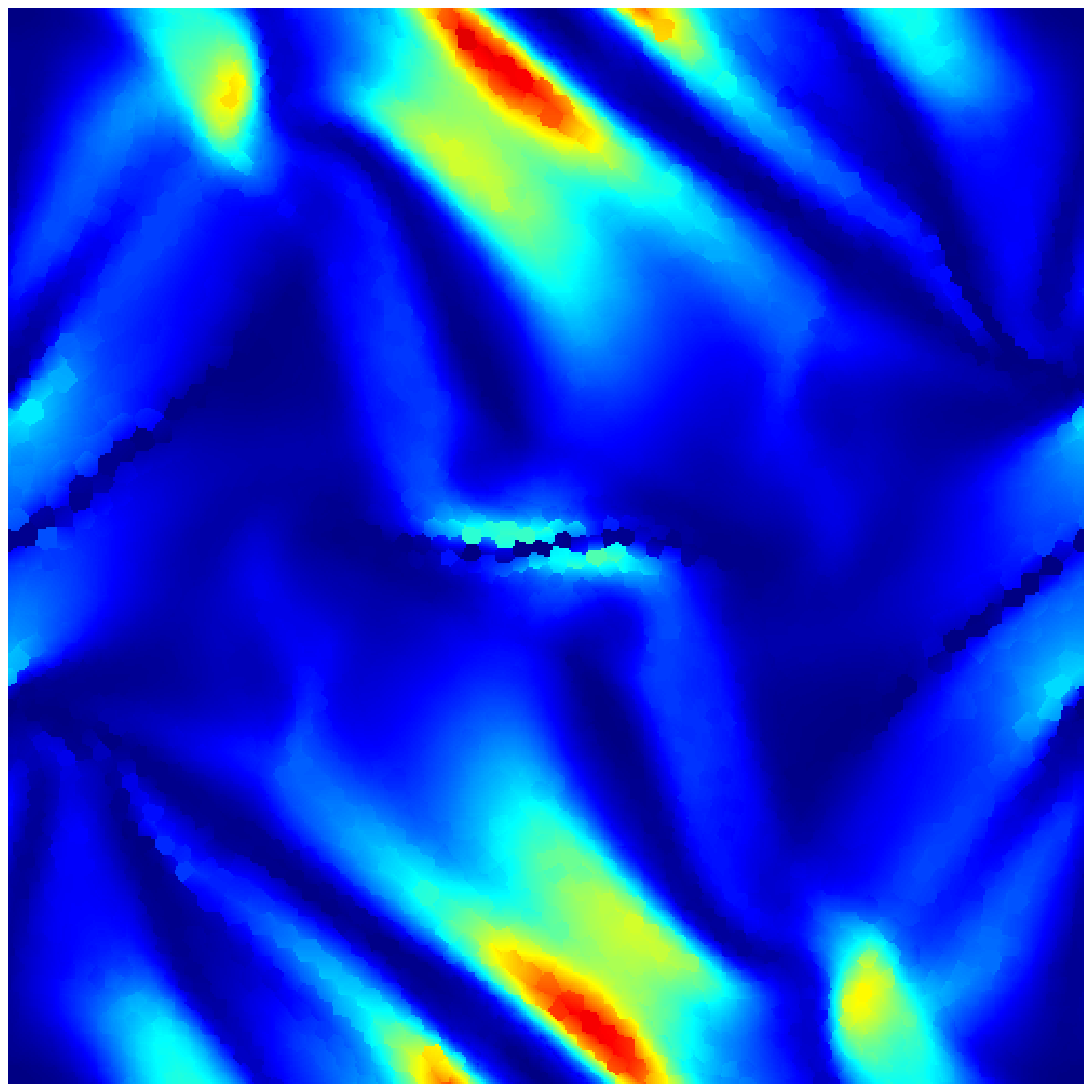} &
\includegraphics[width=0.22\textwidth]{onull.eps} \\
\rotatebox{90}{\hspace{4 mm} static CT (hexagonal,HLLD)} &
\includegraphics[width=0.22\textwidth]{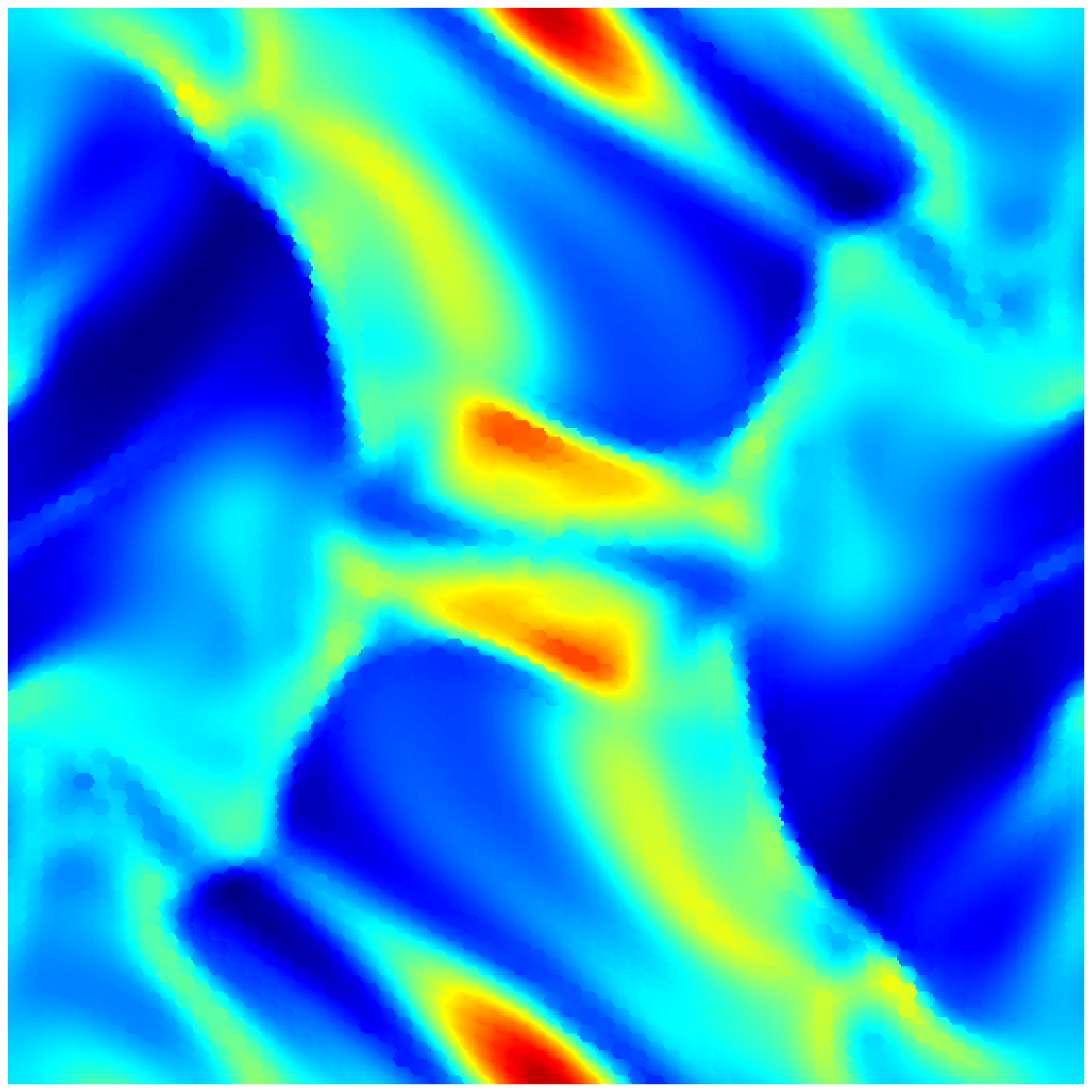} &
\includegraphics[width=0.22\textwidth]{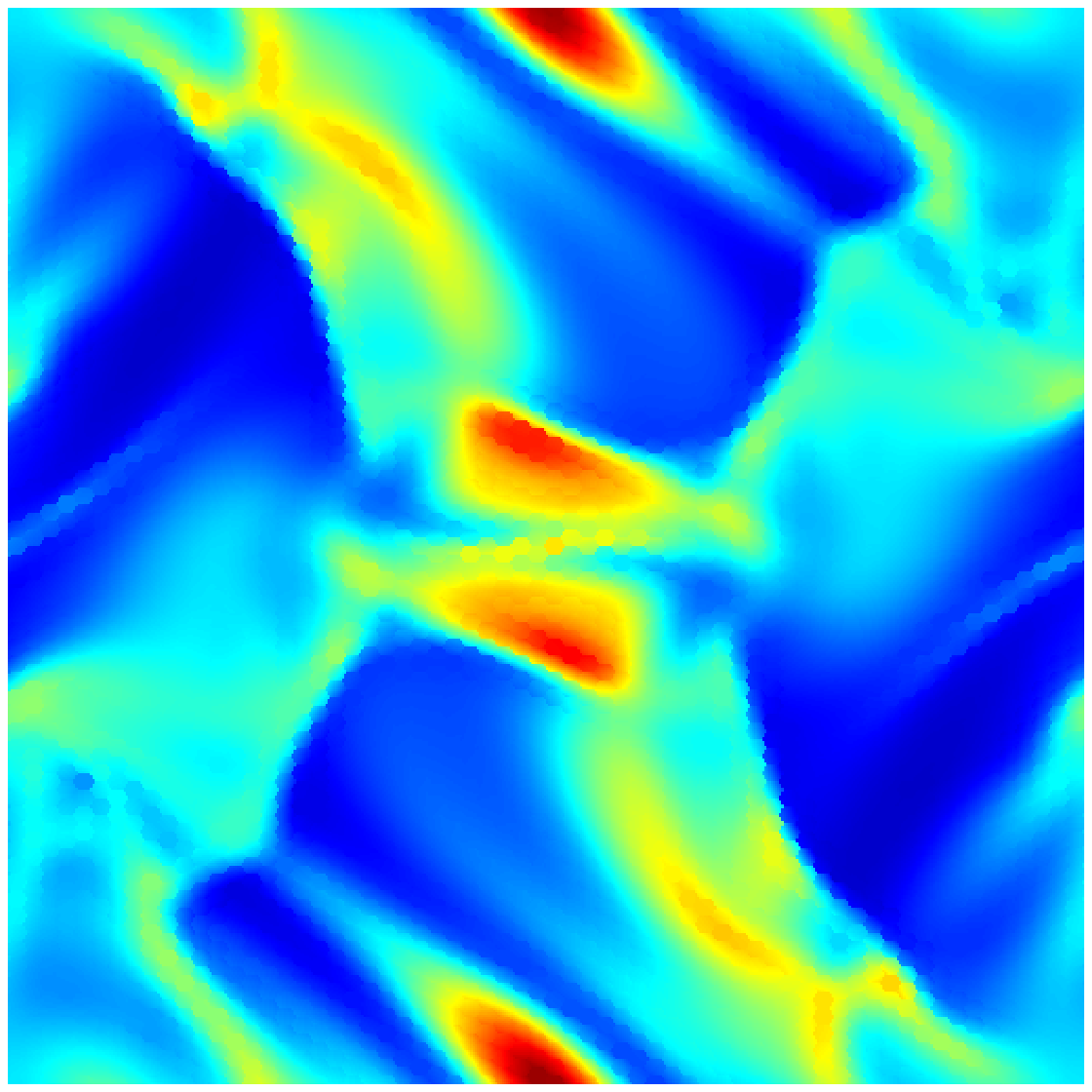} &
\includegraphics[width=0.22\textwidth]{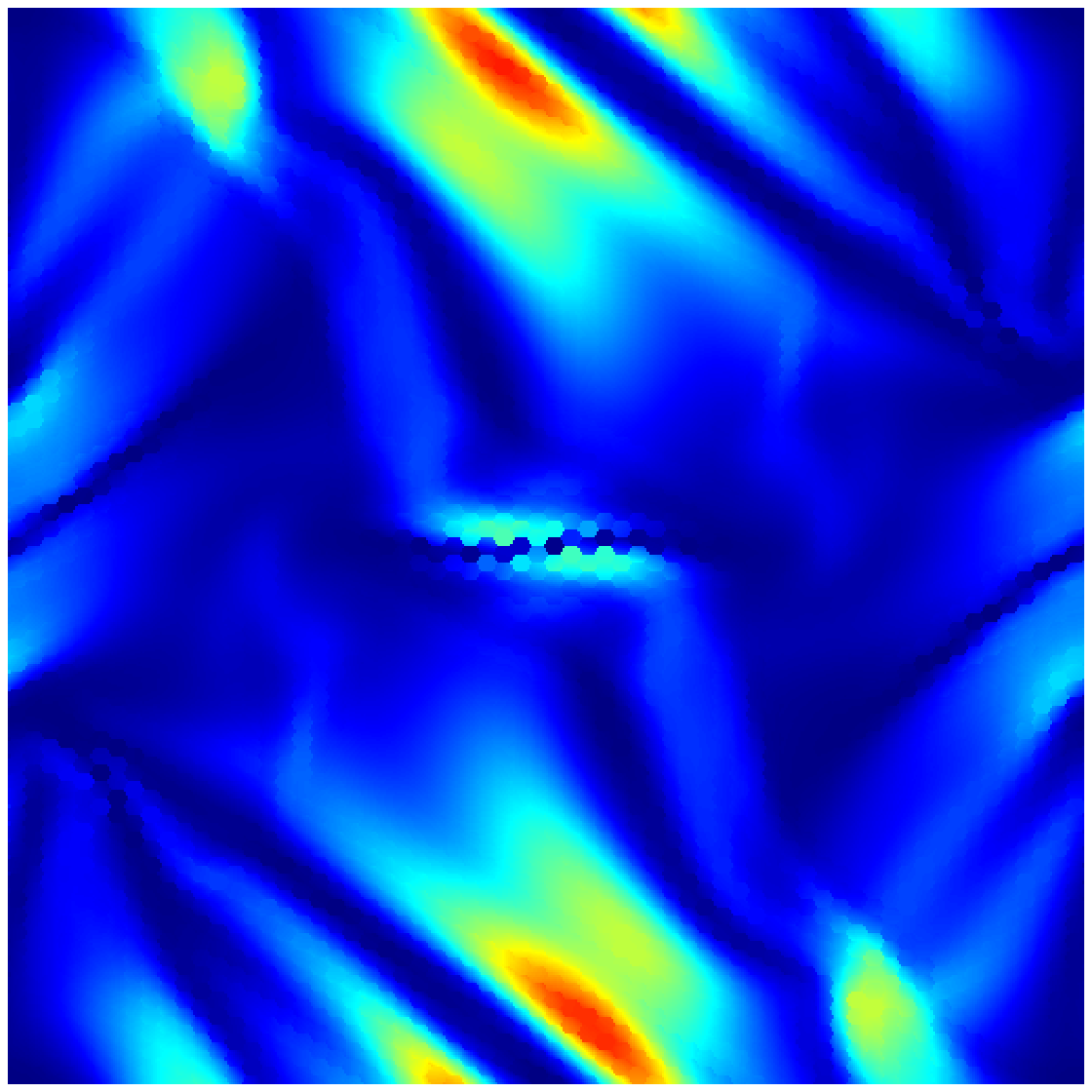} &
\includegraphics[width=0.22\textwidth]{onull.eps} \\
\rotatebox{90}{\hspace{5 mm} moving Powell (Rusanov)} &
\includegraphics[width=0.22\textwidth]{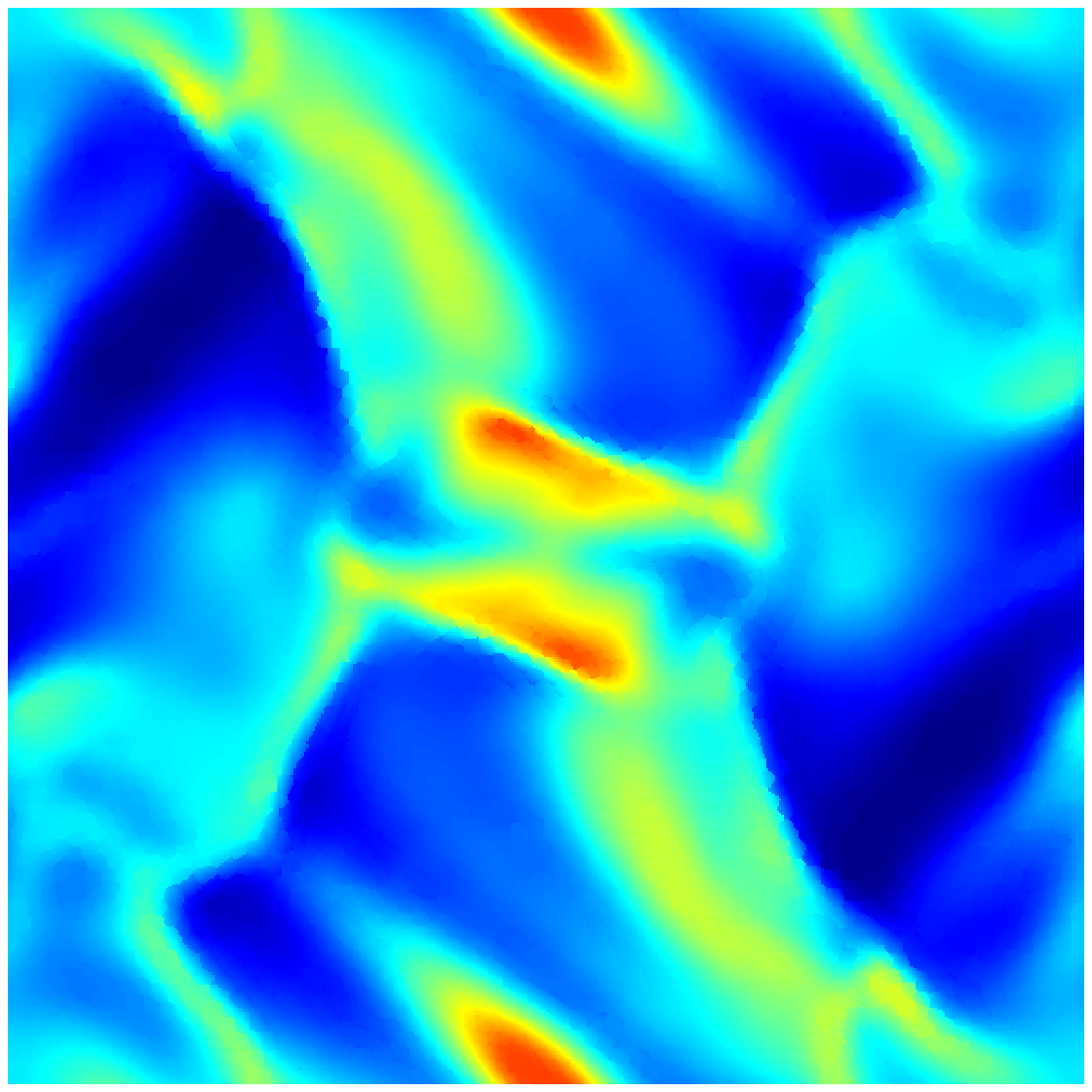} &
\includegraphics[width=0.22\textwidth]{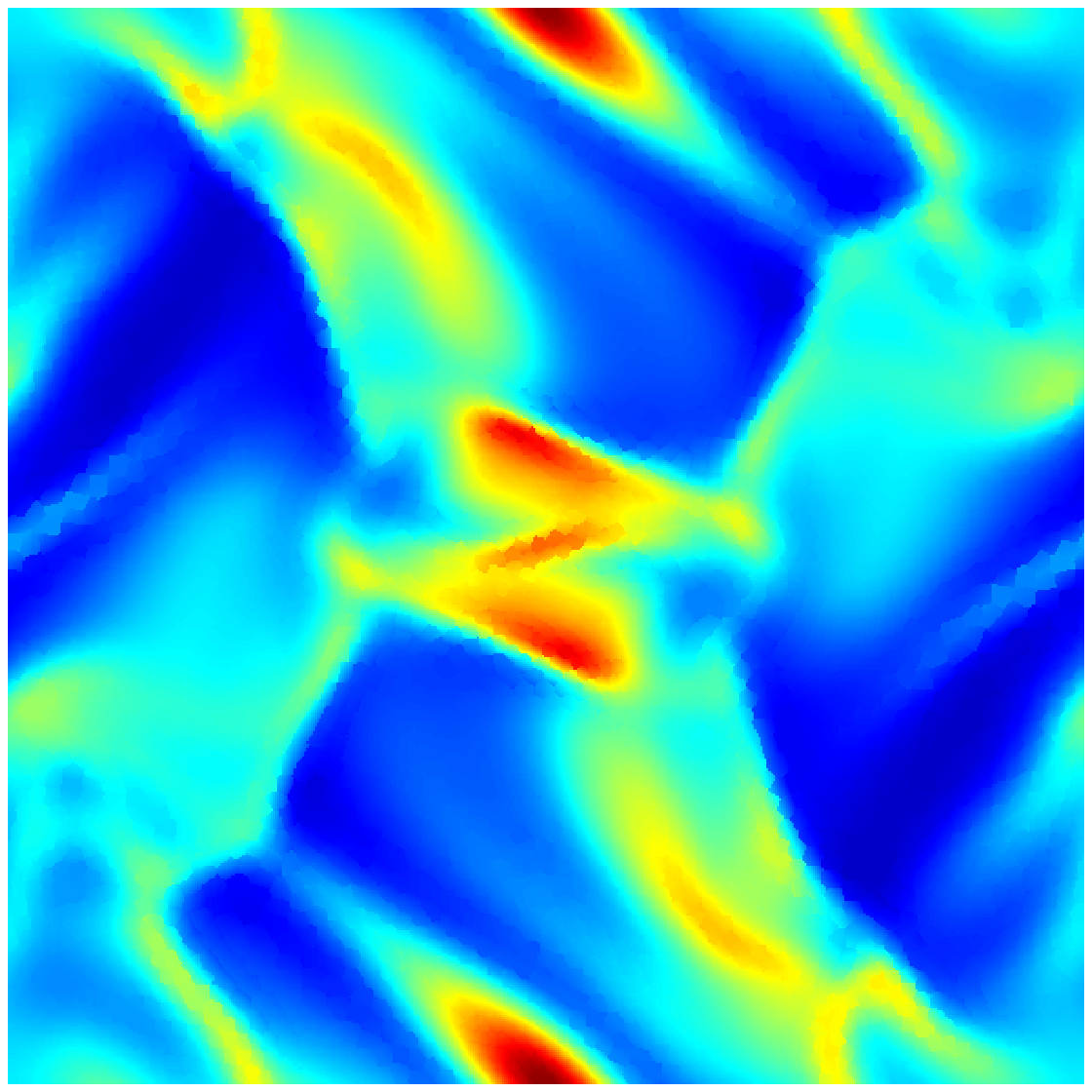} &
\includegraphics[width=0.22\textwidth]{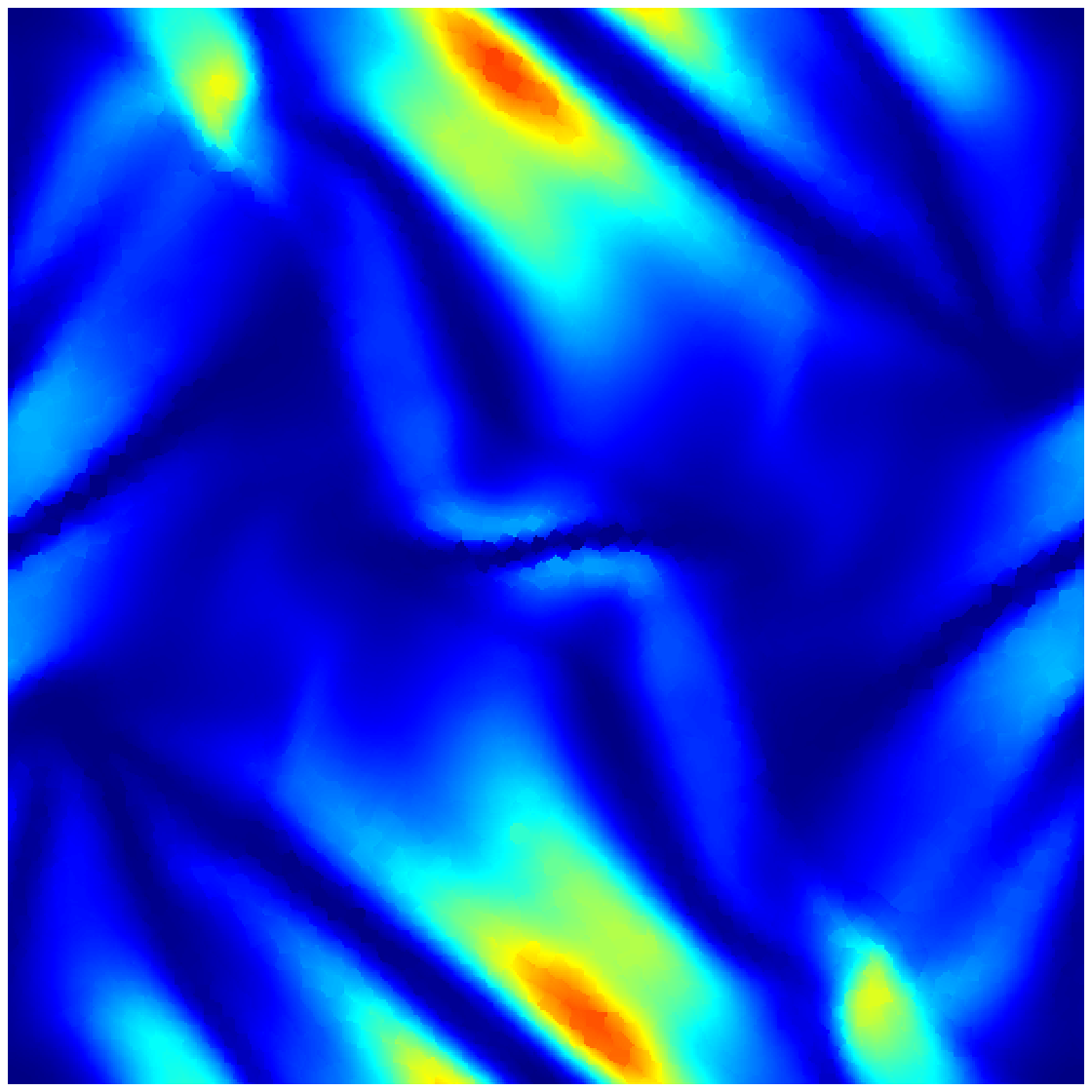} &
\includegraphics[width=0.22\textwidth]{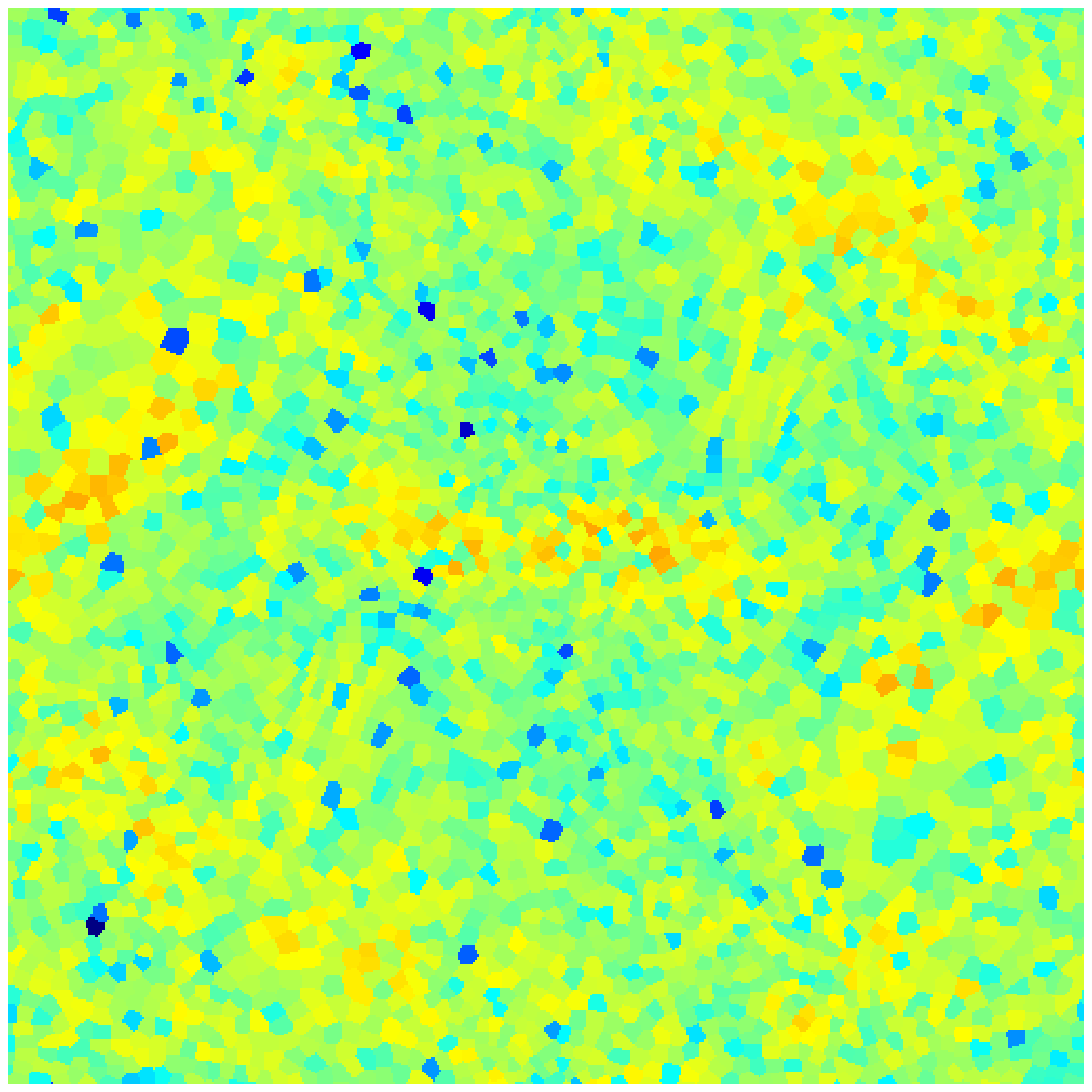} \\
\rotatebox{90}{\hspace{6 mm} moving Powell (HLLD)} &
\includegraphics[width=0.22\textwidth]{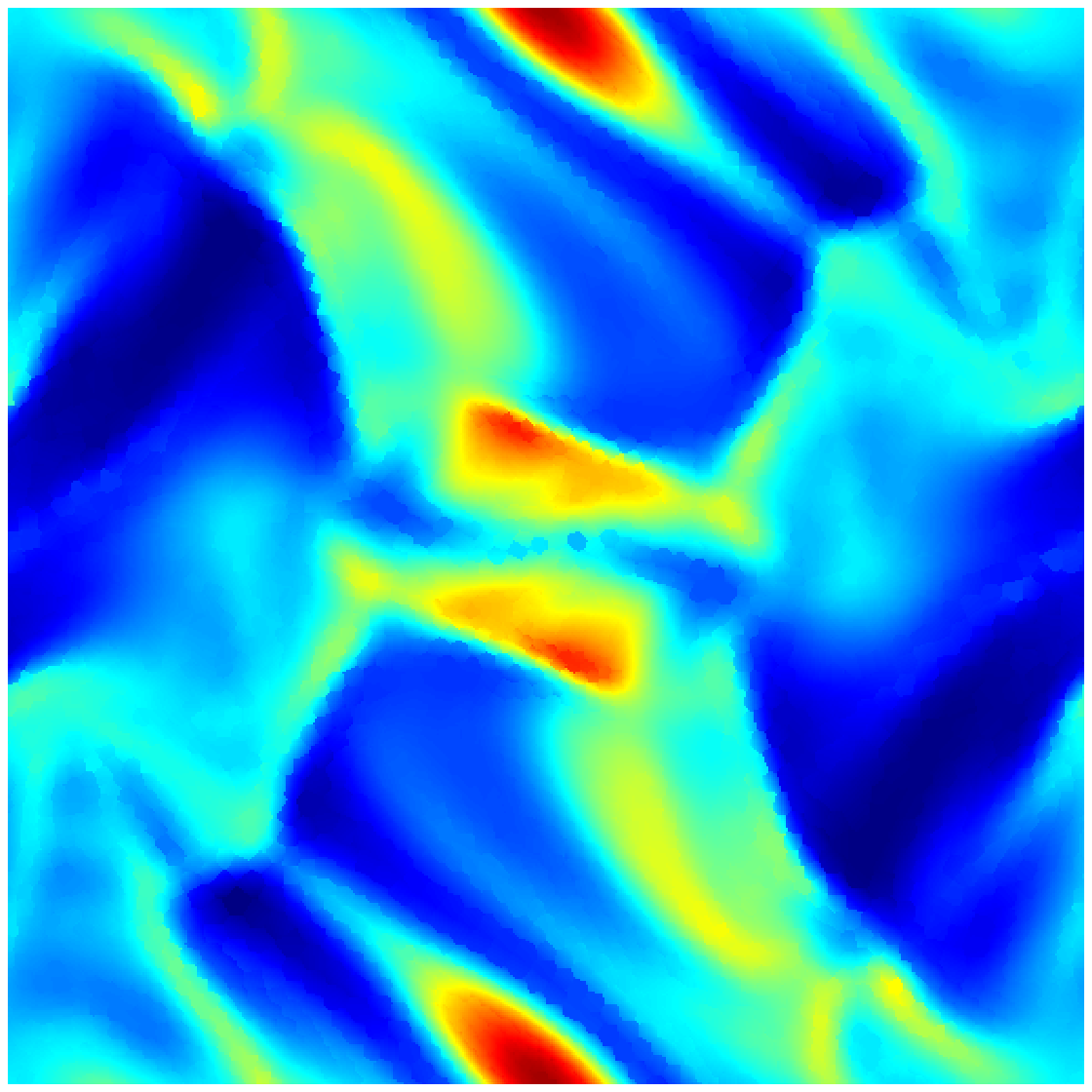} &
\includegraphics[width=0.22\textwidth]{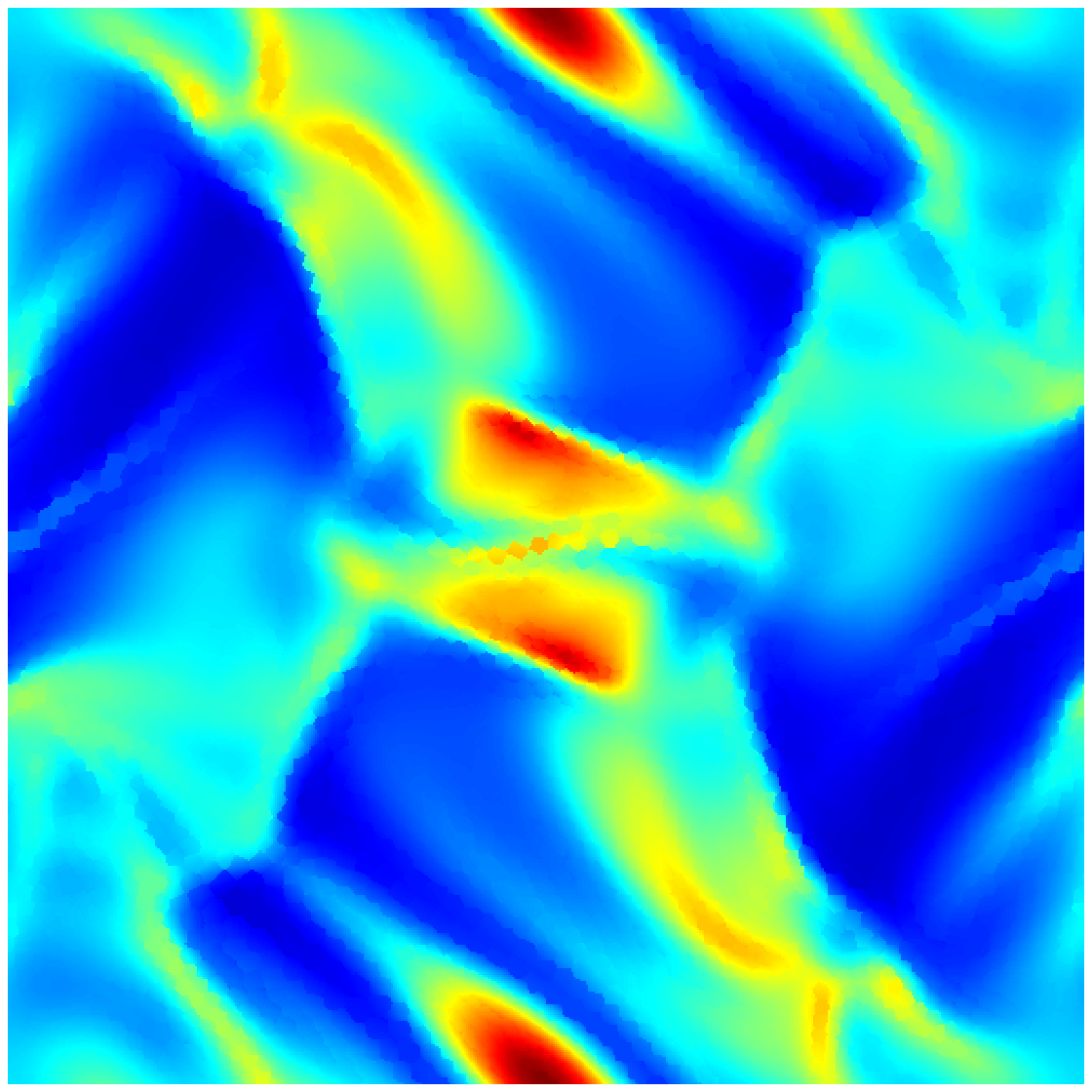} &
\includegraphics[width=0.22\textwidth]{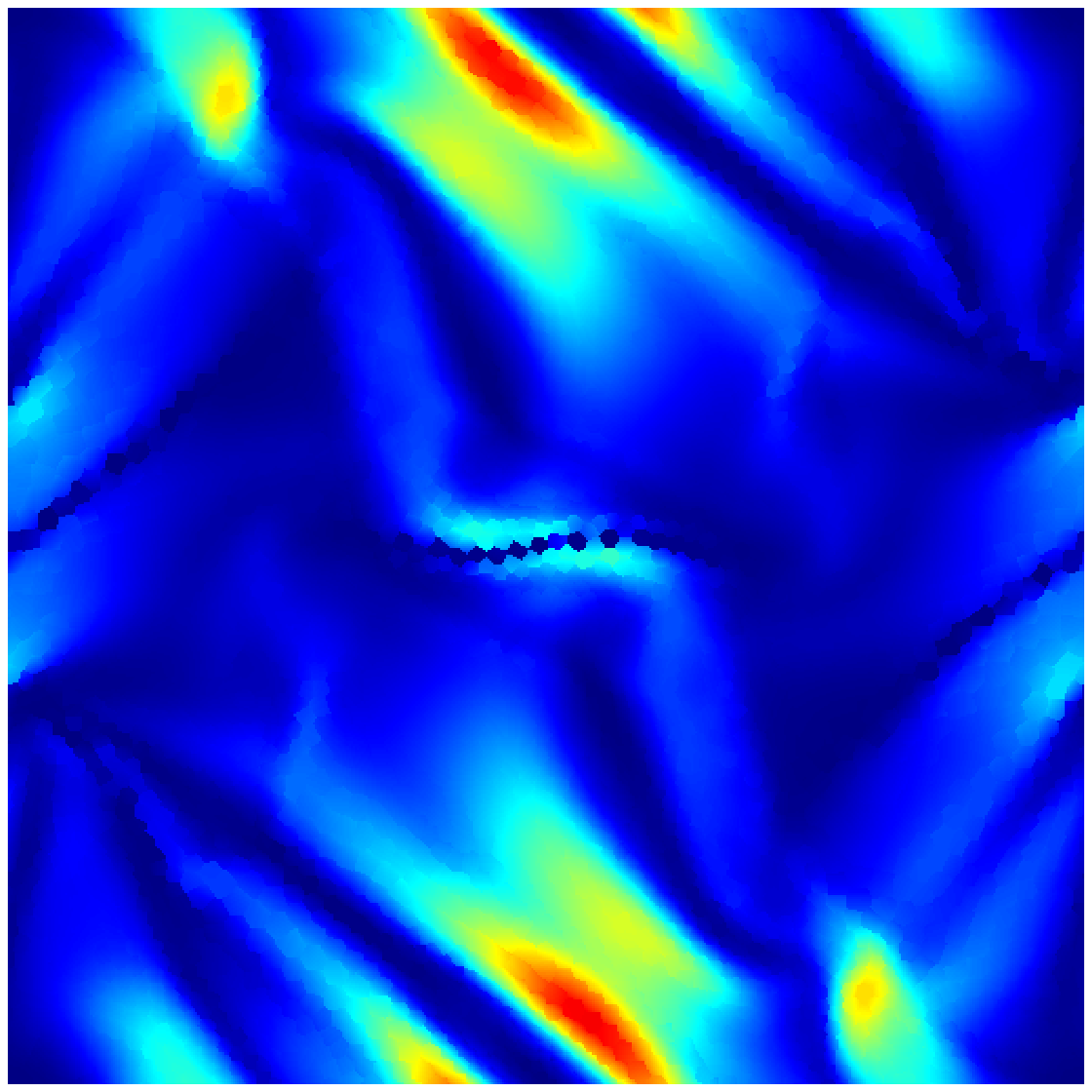} &
\includegraphics[width=0.22\textwidth]{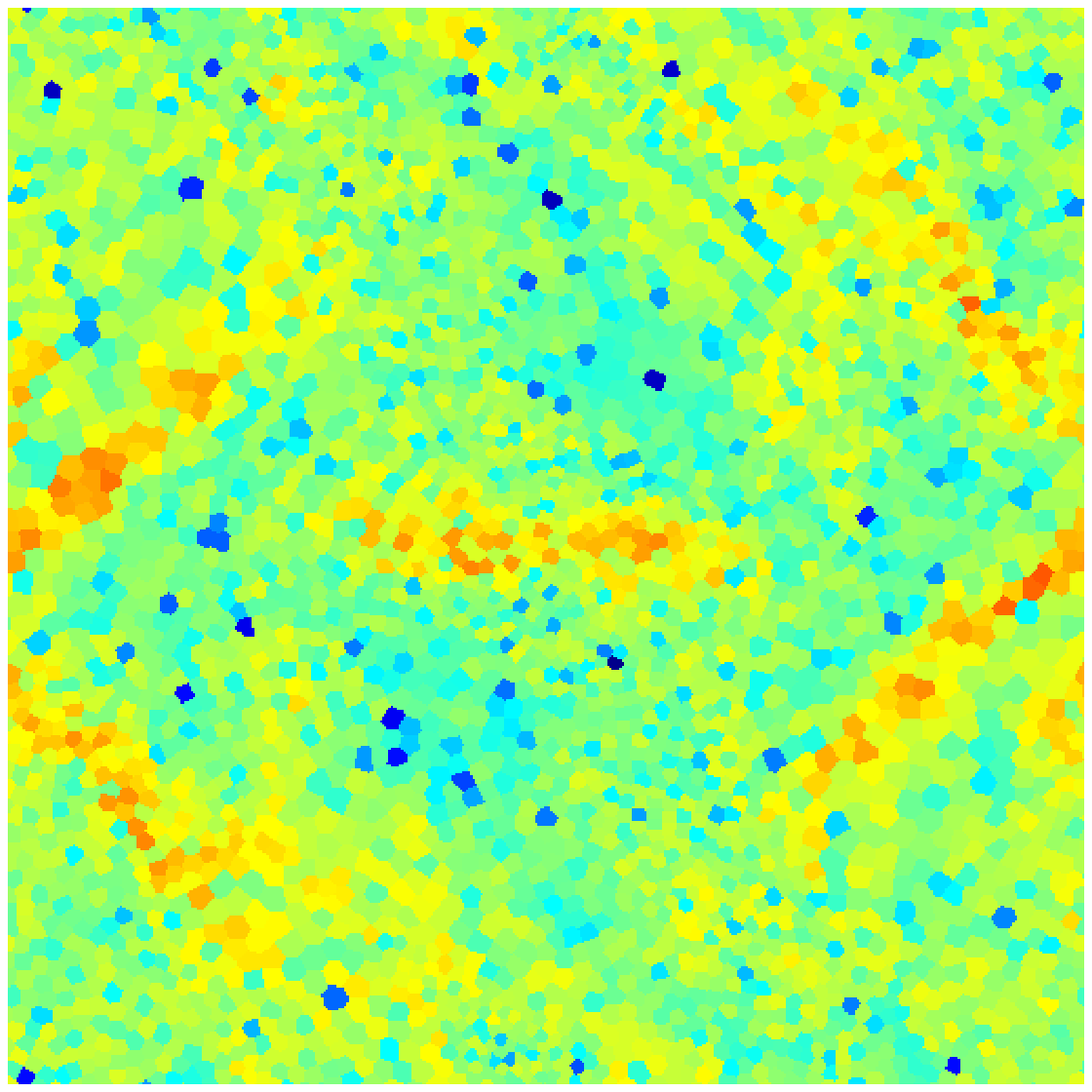} \\
\end{tabular}
\caption{Same as Fig.~\ref{fig:otmain}, except at $t=0.5$. }
\label{fig:otmain2}
\end{figure*}

\section{Numerical Tests}\label{sec:tests}

\subsection{Orszag-Tang vortex}\label{sec:ot}

We demonstrate that our numerical method works successfully by solving the Orszag-Tang vortex problem \citep{1979JFM....90..129O}, which is an excellent test of supersonic MHD turbulence and generally one of the more difficult tests for MHD solvers to handle. We use the initial conditions as described by \cite{1991PhFlB...3...29P}:
\begin{equation}
\rho = \frac{\gamma^2}{4\pi} ,
\end{equation}
\begin{equation}
p = \frac{\gamma}{4\pi} ,
\end{equation}
\begin{equation}
\mathbf{v} = (-\sin(2\pi y), \sin(2\pi x)) ,
\end{equation}
\begin{equation}
\mathbf{B} = (-\sin(2\pi y), \sin(4\pi x)).
\end{equation}
The domain is a box of side length $1$ with periodic boundaries. The gas has adiabatic index $\gamma = 5/3$. 

\begin{figure}
\centering
\includegraphics[width=0.47\textwidth]{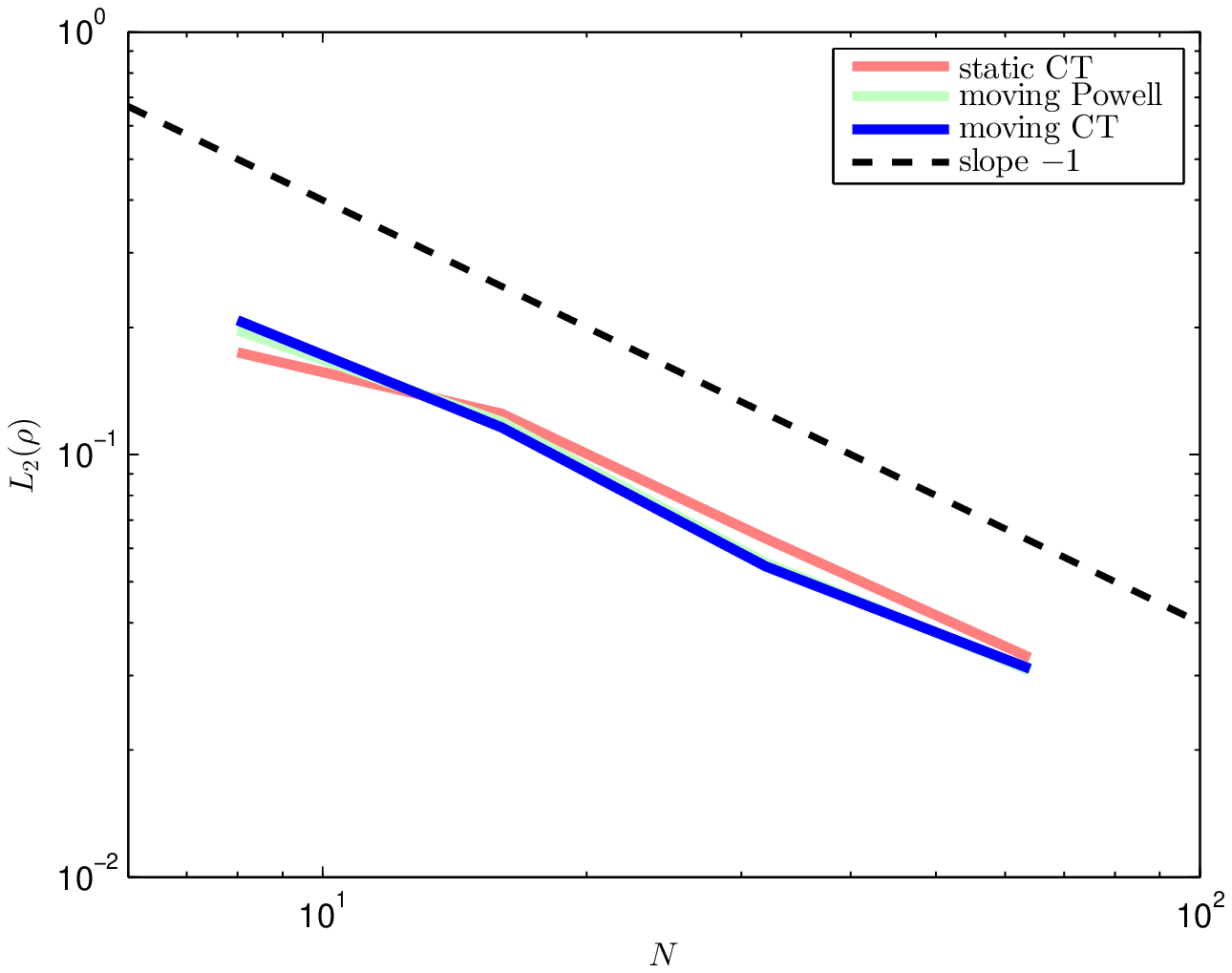}
\caption{The convergence rate of the moving and static mesh CT methods and the moving Powell method for the Orszag-Tang test at $t=0.2$. Both show first order convergence, as expected due to the presence of shocks which limits some of the slopes and reduces the second order method to first order. The moving mesh approach shows slightly smaller errors.}
\label{fig:otconv}
\end{figure}

We show the results of the moving CT method at resolution $64^2$, and compare it with a static CT approach on a hexagonal grid, and the moving Powell cleaning scheme. Plots of the density and magnetic energy density are shown in Figs.~\ref{fig:otmain} and \ref{fig:otmain2}\footnote{animations of the simulations are available at \url{https://www.cfa.harvard.edu/~pmocz/research.html}}. All the methods produce qualitatively similar results. Fig.~\ref{fig:otconv} shows that the moving CT method converges to the solution of a high resolution ($512^2$) static CT technique at the same rate as the static CT method (first order convergence is observed due to the presence of shocks, which results in gradients that are slope-limited). The errors in the moving CT method are slightly less than the static CT approach, which can be attributed to the moving method's better control of advection.

We quantify the strength of the magnetic field divergence error in two ways. First, is the relative divergence error compared to the magnetic field of the cell, used in \cite{2011ApJS..197...15D,2011MNRAS.418.1392P,2012ApJ...758..103G,2013MNRAS.432..176P}.
This is calculated for each cell $i$ as:
\begin{equation}
\frac{|\nabla\cdot\mathbf{B}_i|\cdot |R_i|}{|\mathbf{B}_i|}
\end{equation}
However, an alternate measure is to normalize by an ``effective'' magnetic field quantity, namely, the square root of twice the total pressure, as follows:
\begin{equation}
\frac{|\nabla\cdot\mathbf{B}_i|\cdot |R_i|}{\sqrt{2p_i}}
\end{equation}
This arguably provides a better indication of when the divergence error becomes dynamically important, since by the first measure the error can be large in the case when the magnetic fields are small but it is likely that the largest errors in the dynamics are incurred
at those points. The second diagnostic also includes the importance of gas pressure in the normalization. We plot this second measure of divergence in Figs.~\ref{fig:otmain} and \ref{fig:otmain2} for the Orszag-Tang test. As expected, the CT algorithms show no error.

It is worth pointing out that the moving CT scheme that uses the HLLD solver produces less smooth solutions at late times once strong shocks break out in the simulation ($t=0.5$, see Fig.~\ref{fig:otmain2}) in the density field (but not any of the other fluid variable fields) compared to the other methods. This is attributable to the moving mesh nature of the code and using an approximate Riemann solver: small errors in the density field advect with the flow and do not diffuse. More diffusive approaches, such as using a Rusanov flux solver, or a Powell cleaning scheme with HLLD, or a static CT HLLD solver, produce a smoother density field. 

\begin{figure}
\centering
\includegraphics[width=0.47\textwidth]{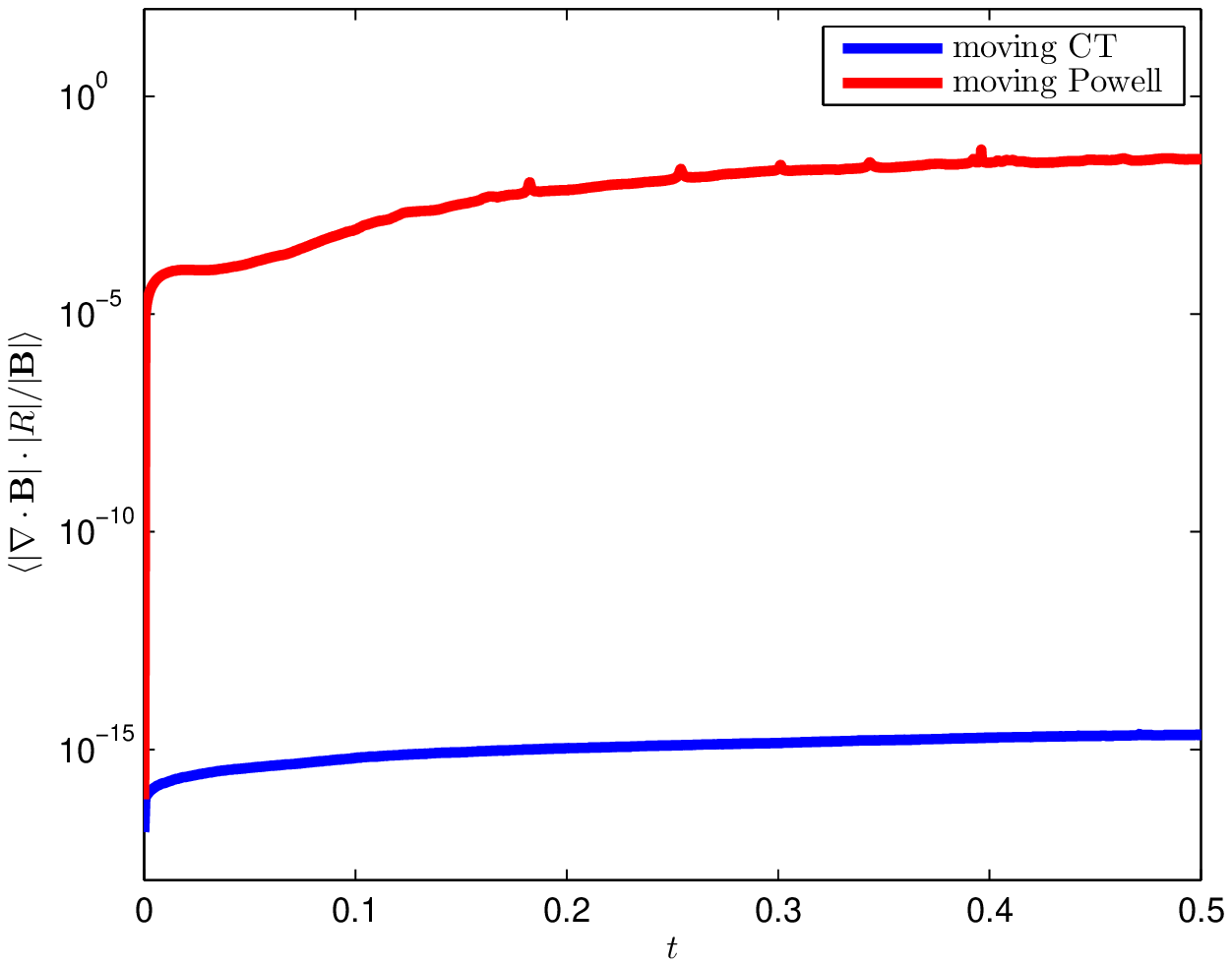}
\caption{The average relative magnetic field divergence errors in the moving CT approach are kept to $0$ at the level of machine precision, while the Powell cleaning scheme shows an average error of $10^{-3}$ for the Orszag-Tang test. The Powell cleaning scheme may even demonstrate relative errors of the order of unity in a few individual cells at a given time (Fig.~\ref{fig:otmain2}). }
\label{fig:otdivb}
\end{figure}

We verify that the average relative magnetic field divergence errors of a cell are zero to the level of machine precision in Fig.~\ref{fig:otdivb}. The average relative divergence errors of a moving mesh approach with the Powell cleaning scheme is on the order of $10^{-3}$, and a few individual cells can have errors greater than order unity.

\subsubsection{Galilean Invariance}\label{sec:gi}

\begin{figure*}
\centering
\begin{tabular}{cccc}
moving CT -- boosted & static CT -- boosted & moving Powell -- boosted  & static CT -- no boost \\
\includegraphics[width=0.23\textwidth]{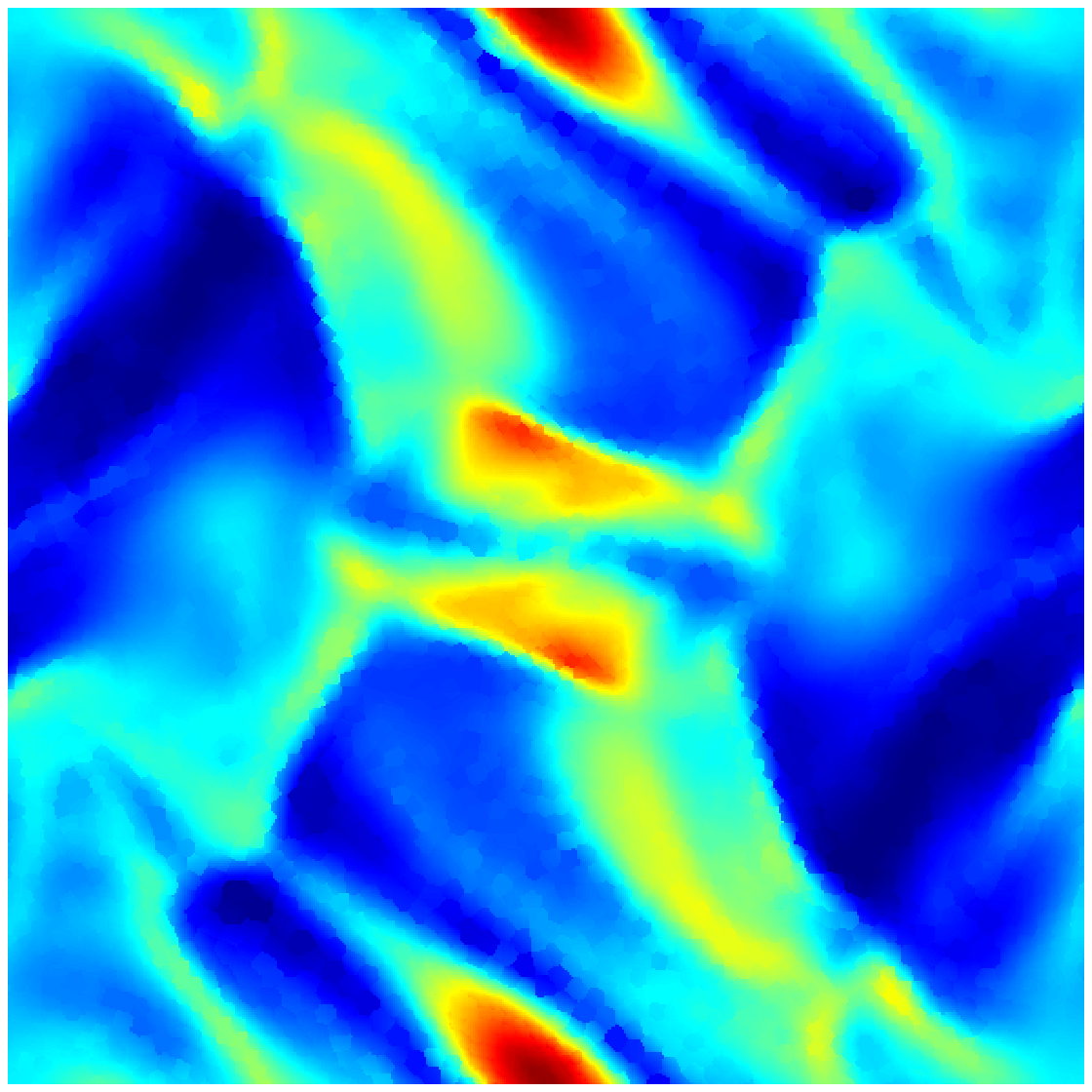} &
\includegraphics[width=0.23\textwidth]{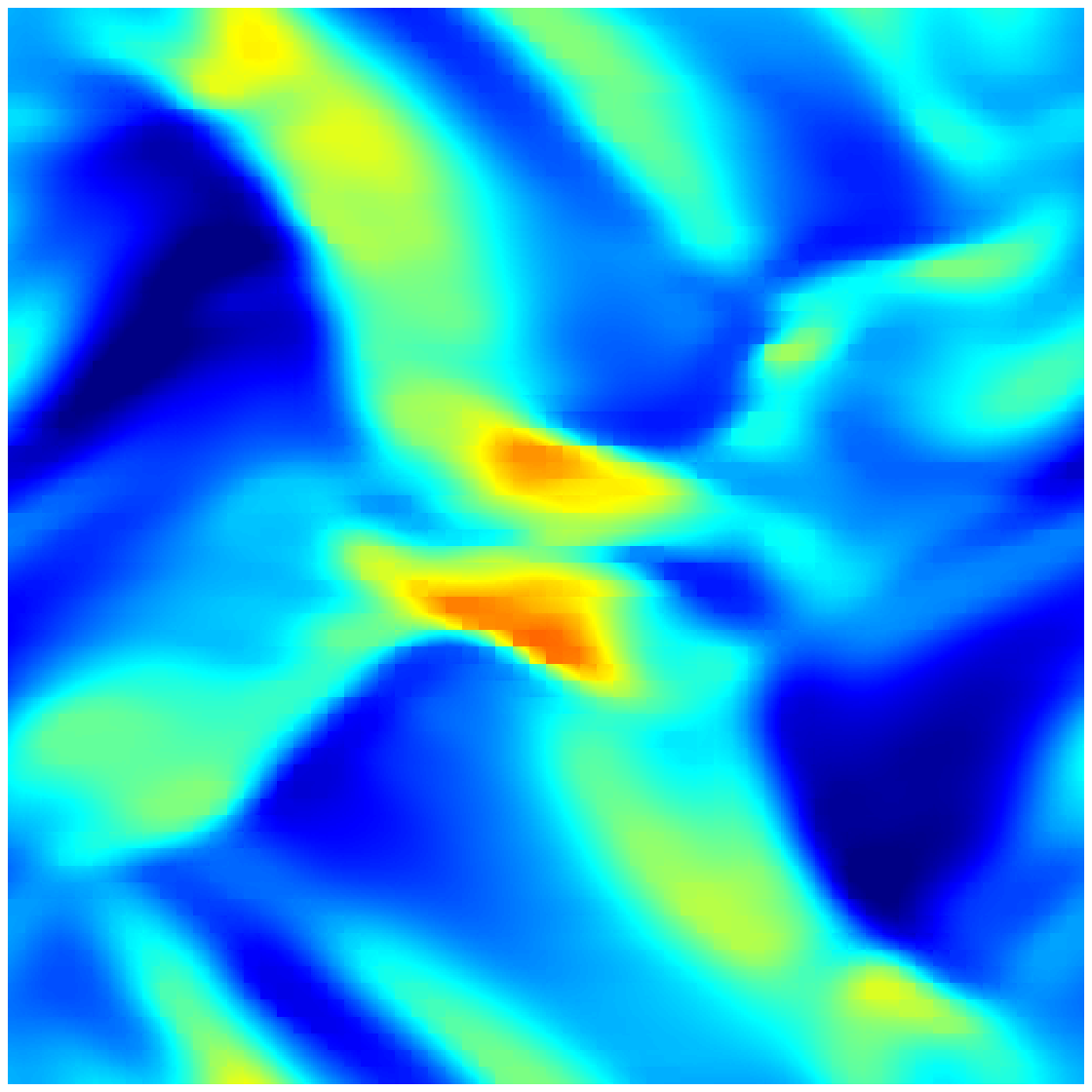} &
\includegraphics[width=0.23\textwidth]{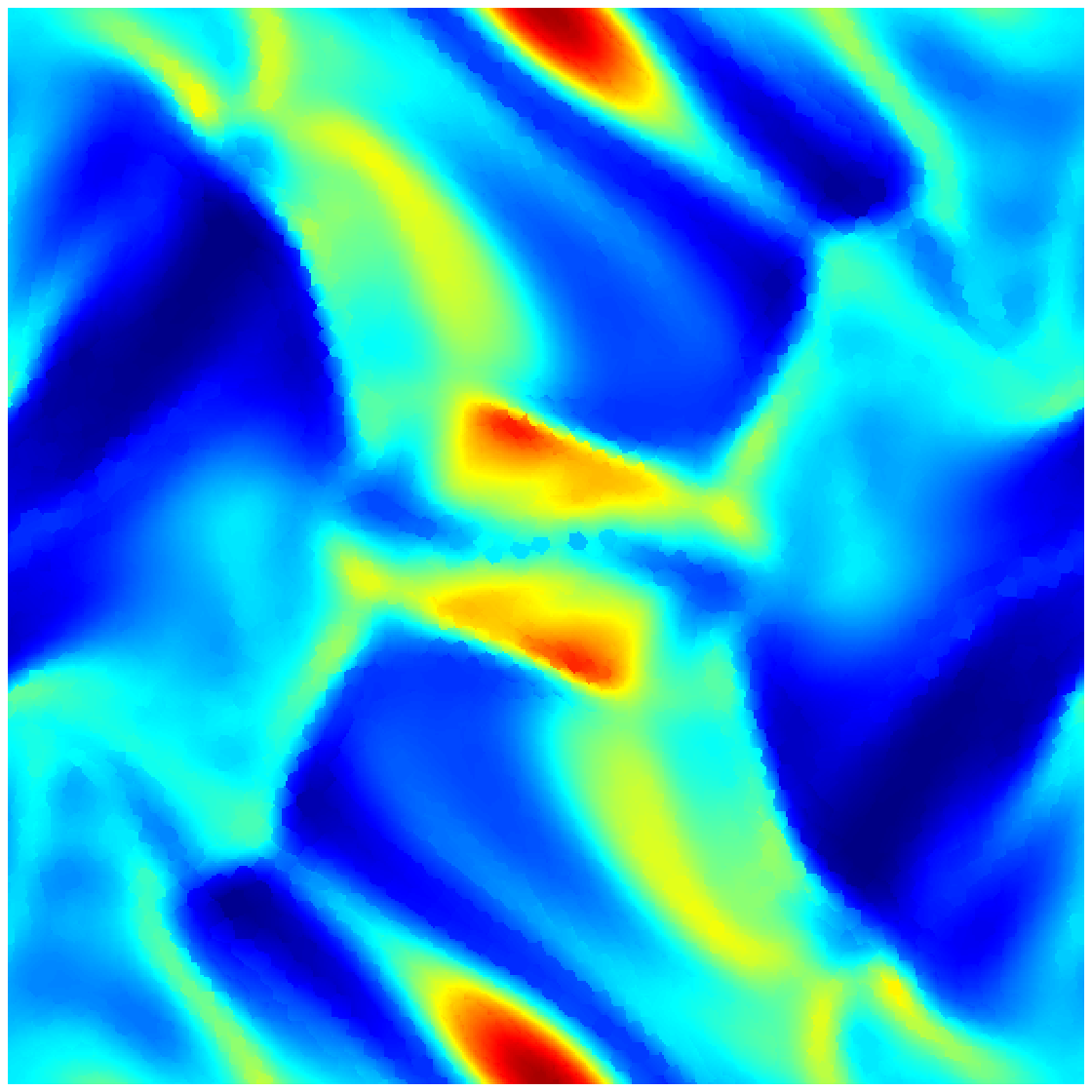} &
\includegraphics[width=0.23\textwidth]{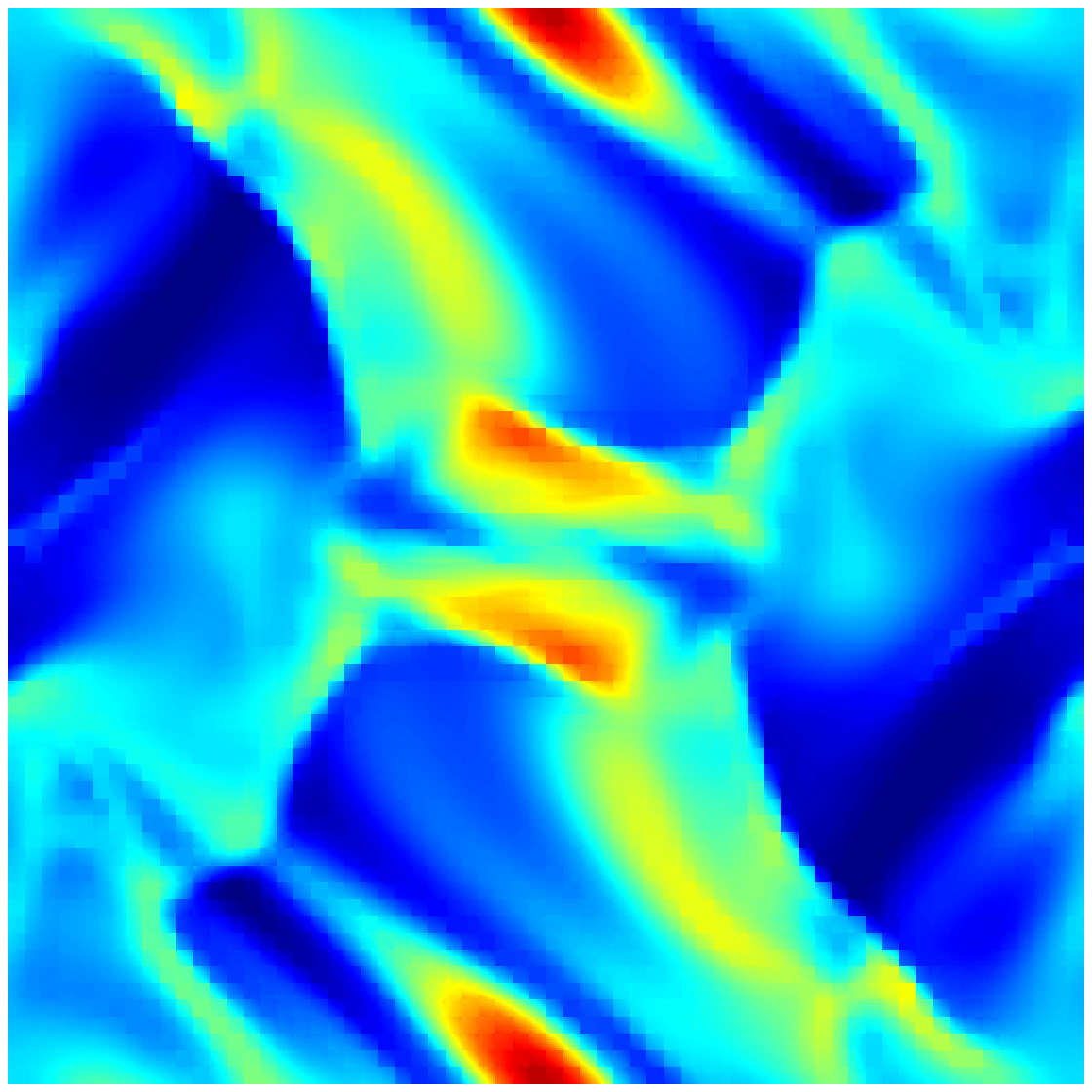} \\
\end{tabular}
\caption{The density field of the Orszag-Tang test at $t=0.5$ with initial conditions boosted by a Mach number of $10$ solved using the moving CT approach (left), the fixed grid CT approach (second from left), and the moving Powell cleaning scheme (third from left), compared to the solution obtained with a fixed grid CT solver and no boost in the initial conditions (right). The moving mesh approach shows less diffusion and more symmetry with the boost applied than the fixed grid approach, due to its Lagrangian nature.}
\label{fig:otgal}
\end{figure*}

We demonstrate that our method is less susceptible to being dominated by truncation errors under a Galilean boost than static mesh approaches. Our technique also requires less strict timestep criteria, due to the Lagrangian nature of our method (fluxes are always solved in the rest frames of faces/edges). We boost the $x$-direction velocities in the initial conditions of the Orszag-Tang test by a velocity of $10$ (corresponding to a Mach number of $10$), and compare the moving and static CT approaches at $t=0.5$. The results are shown in Fig.~\ref{fig:otgal}. The moving mesh approach maintains symmetry to a greater degree and has less diffusion (in the static CT approach, the overdense regions near at $x=0.5$, $y=0.1,0.9$ are diffused when the Galilean boost is applied). Such unwanted numerical artefacts could lead to potential inaccuracies in, for example, the study of density distribution functions to understand supersonic turbulence and the collapse of high density material in the formation of stars. We note that the errors in the fixed grid CT approach can be reduced with sufficiently high resolution, however, this may not always be attainable with limitations on computational performance. We also show the results of the moving Powell scheme, whose solution is also largely unaffected by the velocity boost, due to its Lagrangian nature (although, as we show in \S~\ref{sec:loop}, the Powell scheme loses some of its Galilean-invariant properties due to its source terms).

\subsection{Advection of a magnetic loop}\label{sec:loop}

The advection of a magnetic field loop is a common and important test of an MHD solver's ability to maintain the divergence of $\mathbf{B}$ and also demonstrates the scheme's diffusivity. It is often used as one of a series of tests to validate a numerical technique \citep{2008ApJS..178..137S,2011ApJS..197...15D}, and many MHD solving techniques face difficulty when evolving this simple setup. In this test problem, a magnetic field loop with dynamically unimportant magnetic fields is advected by a constant velocity field. We simulate the advection in a periodic box of size $1$, adiabatic index $\gamma=5/3$, and initial conditions given by:
\begin{equation}
\rho = 1 
\end{equation}
\begin{equation}
p = 1
\end{equation}
\begin{equation}
\mathbf{v} = (\sin(\pi/3),\cos(\pi/3),0)
\end{equation}
\begin{equation}
\mathbf{A} = (0,{\rm max}(0.001\cdot(0.3-r),0))
\end{equation}
where $r$ is the radial distance to the centre of the loop and $\mathbf{A}$ is the vector potential of the magnetic field.

\begin{figure}
\centering
\begin{tabular}{ccc}
initial condition (translated) & moving CT \\
\includegraphics[width=0.14\textwidth]{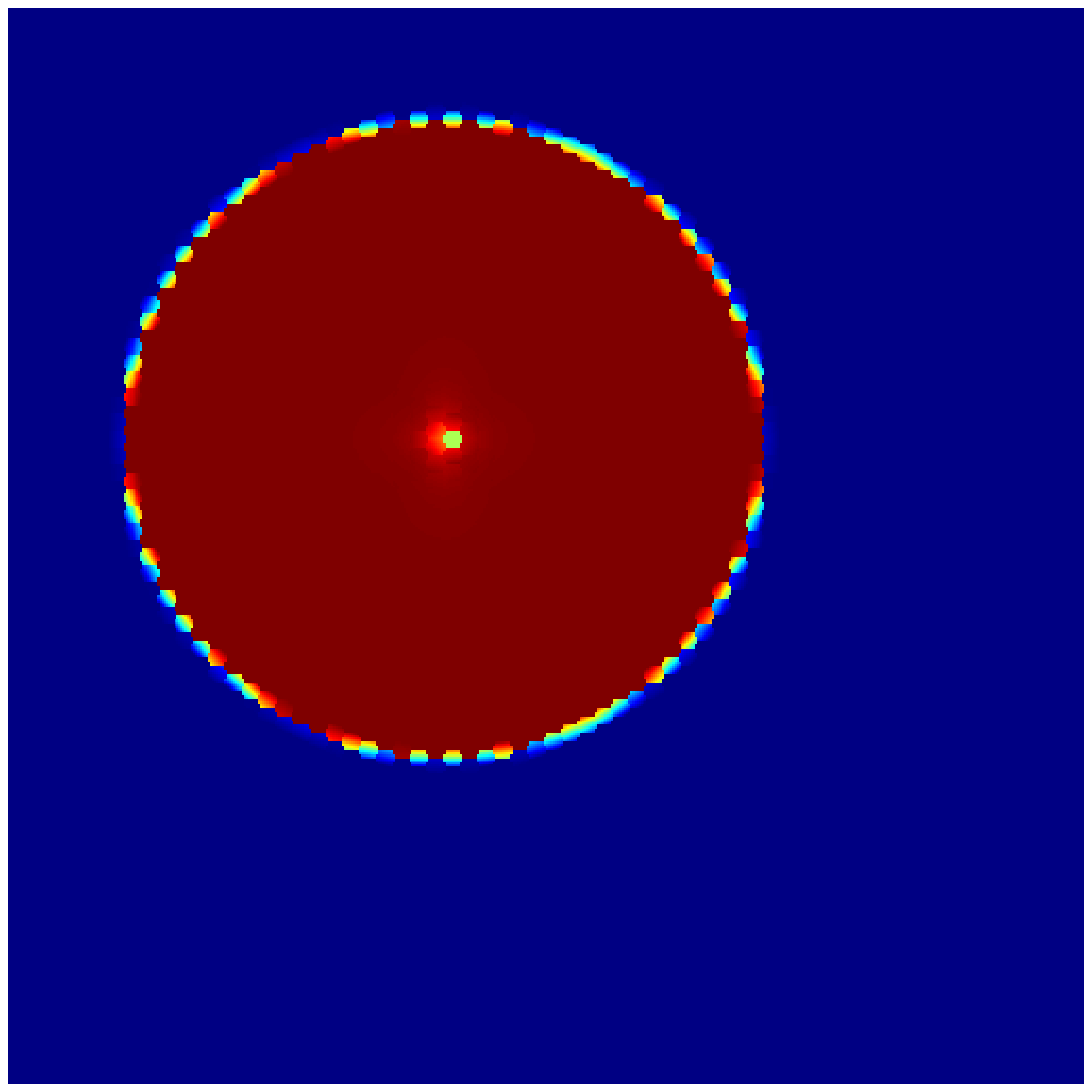} &
\includegraphics[width=0.14\textwidth]{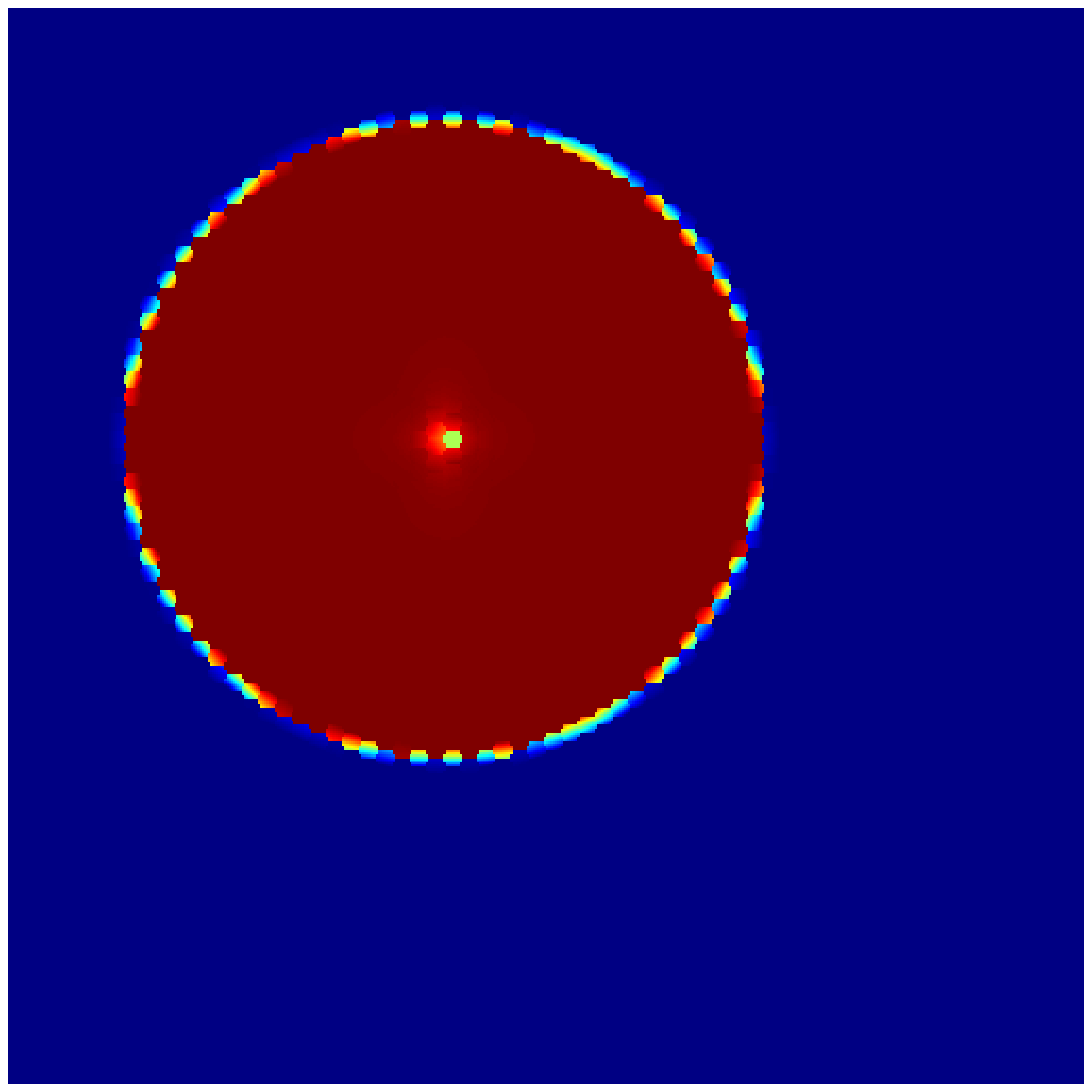} \\
static CT & moving Powell \\
\includegraphics[width=0.14\textwidth]{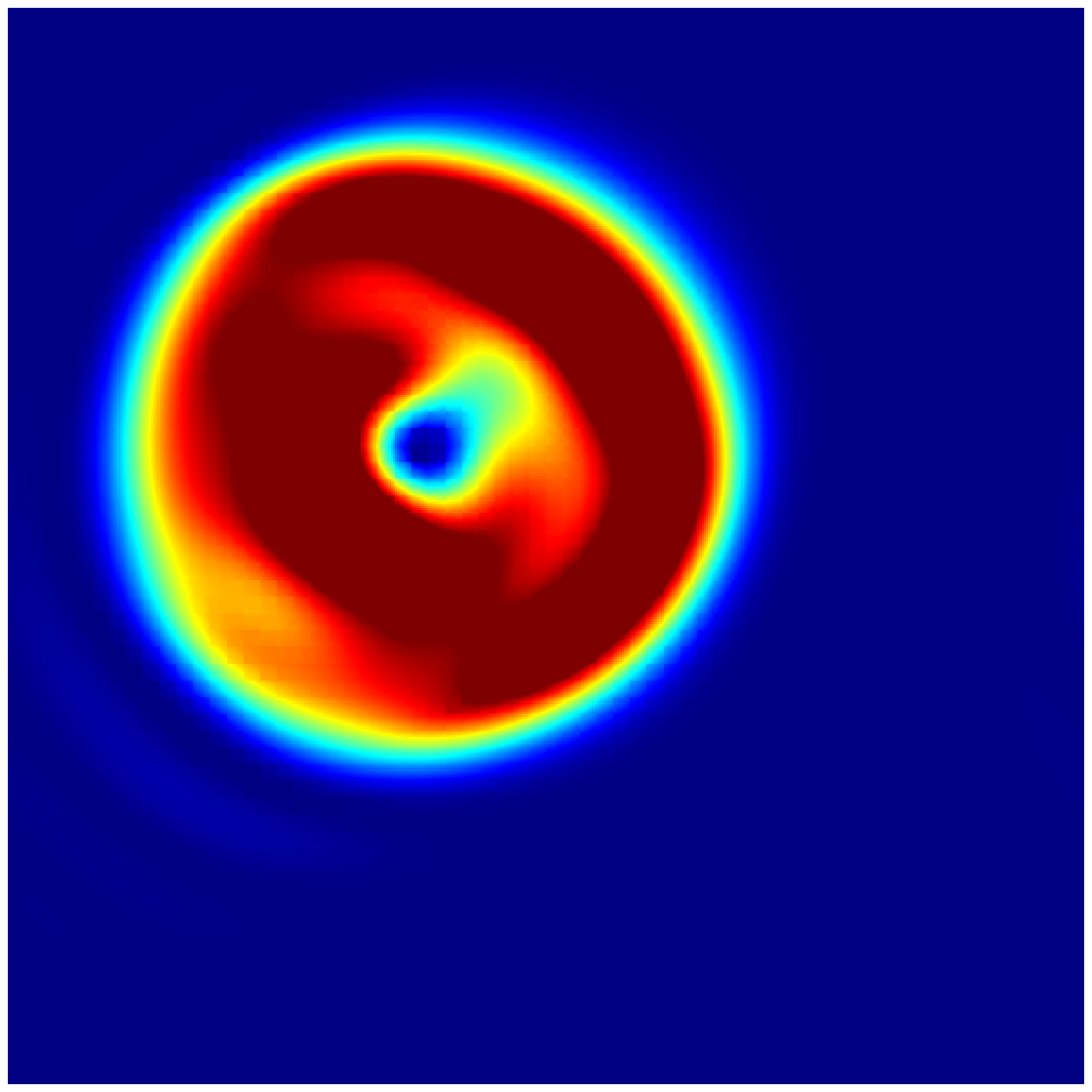} &
\includegraphics[width=0.14\textwidth]{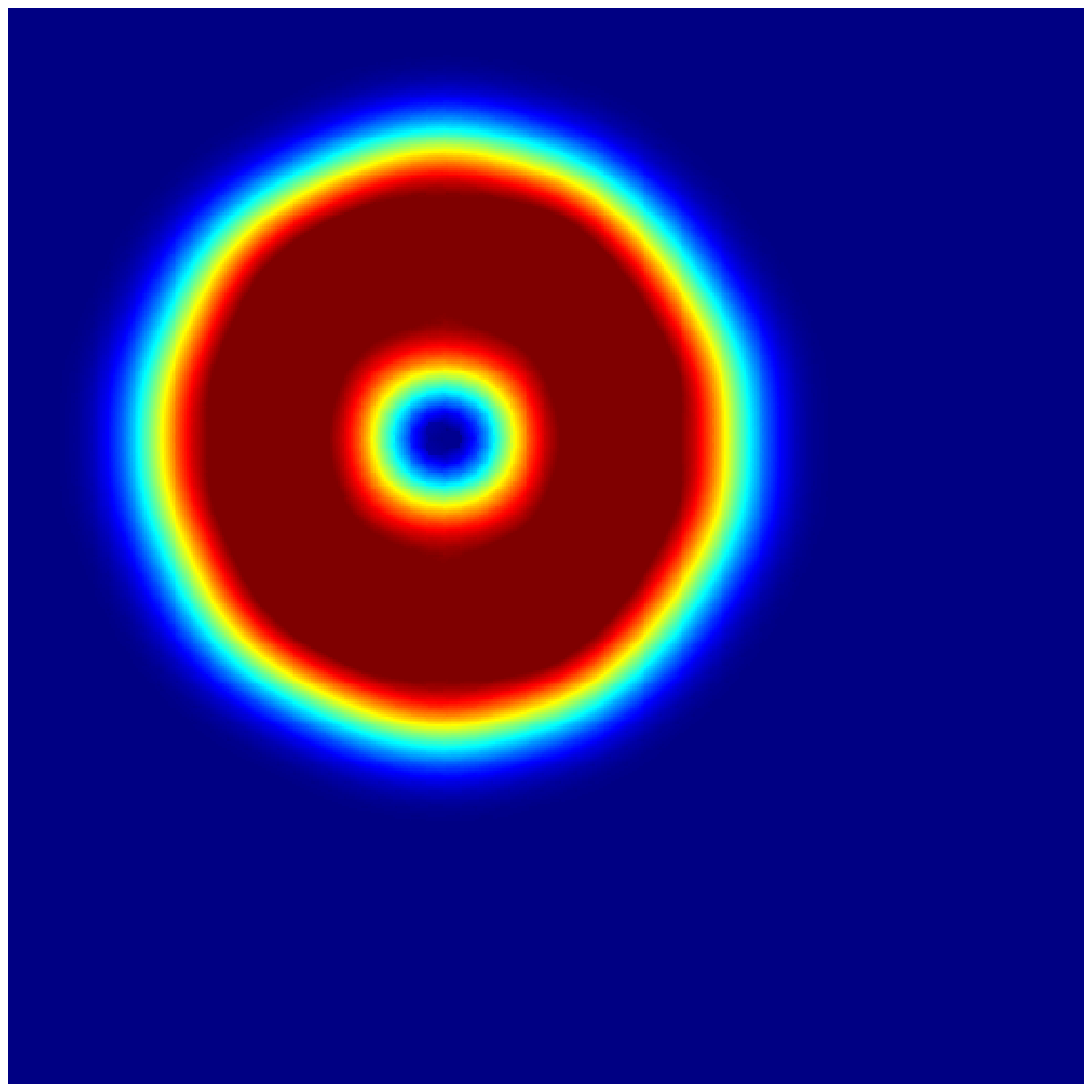} \\
\end{tabular}
\includegraphics[width=0.35\textwidth]{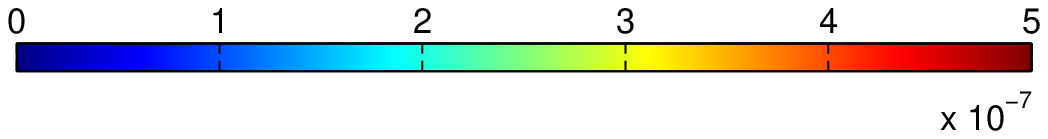} \\
\includegraphics[width=0.47\textwidth]{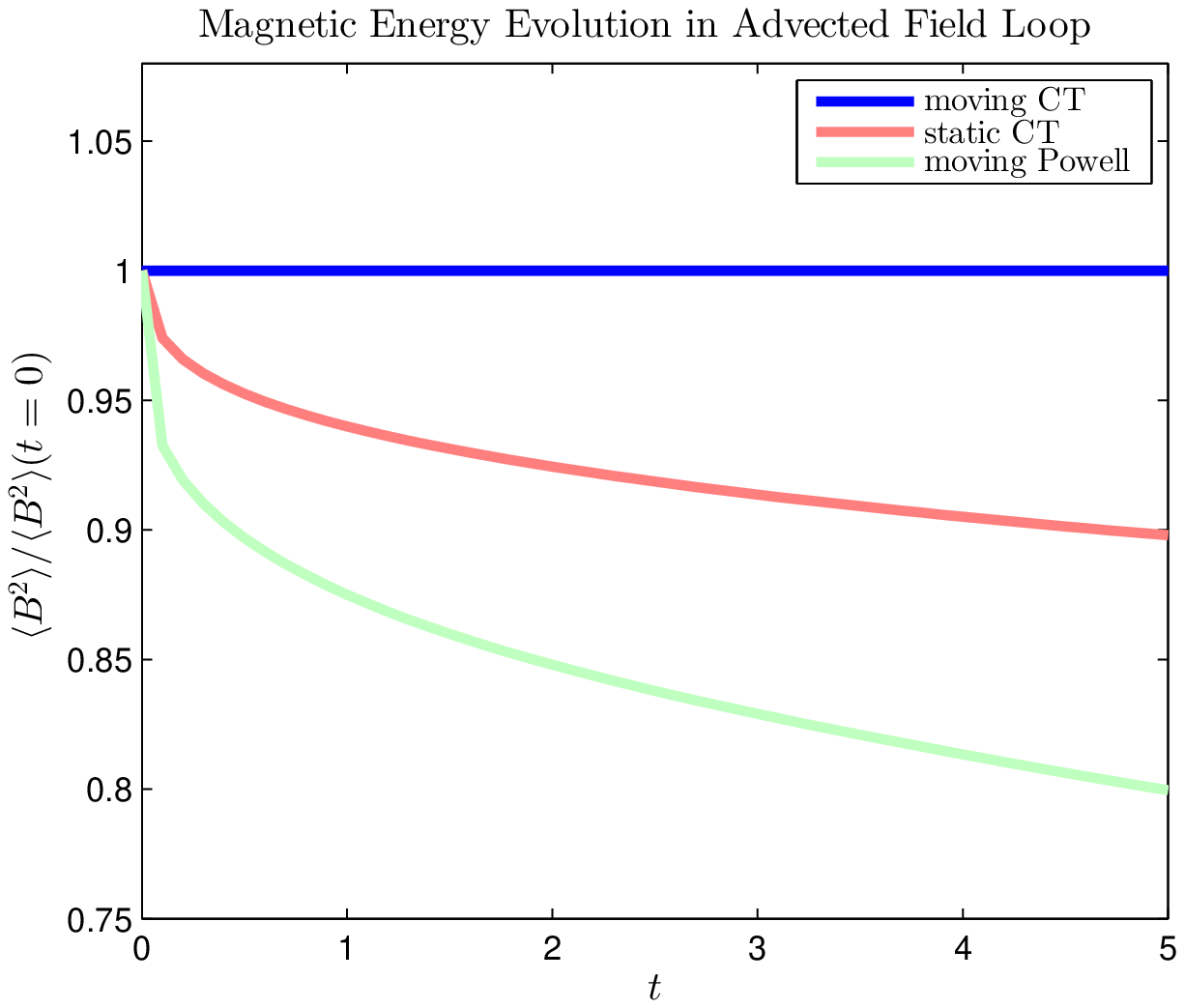} \\
\caption{Plots of the magnetic energy density in the advection of a field loop test at the initial condition $t=0$ (top left) and the advected solution at $t=2.2$ for a moving CT approach (top right), a static CT approach (bottom left) and a moving Powell approach (bottom right). The moving CT approach advects the initial conditions to machine precision. The fixed grid CT approach on the other hand shows diffusion and develops asymmetries and unphysical structures. The moving Powell scheme also shows diffusivity due to the presence of source terms. The average magnetic energy density stays constant in the moving CT approach (agreeing with the exact solution) while it decays with the static CT and moving Powell approaches (bottom).
}
\label{fig:advect}
\end{figure}

We compare the moving CT method to a fixed grid flux-interpolated CT method and the moving Powell method. For the moving mesh, we use a hexagonal grid of resolution $64^2$ and for the fixed grid we use a Cartesian grid with resolution of $64^2$. A plot of the evolution of the average magnetic energy density $\frac{1}{2}\mathbf{B}^2$ is shown in Fig.~\ref{fig:advect}, along with the magnetic energy density at time $t=2.2$. The moving mesh CT approach does incredibly well. It preserves the advection of the solution to machine precision, due to its Lagrangian properties. The magnetic energy density does not decay at all with time but rather stays constant. This is independent of the resolution used to simulate the advecting loop. In comparison, the fixed grid approach shows diffusivity and unphysical structures and asymmetries develop in the energy density. The magnetic energy density decays with time. These errors can be lessened by increasing the resolution of the run, but never fully eliminated in the fixed grid approach. The moving Powell cleaning scheme also shows diffusivity since it is not fully Galilean-invariant due to the addition of source terms. It does, however, maintain symmetry since the problem is solved in the rest frame of the motion of the loop.

\subsection{Strong Shock}\label{sec:s}

We simulate a 1D strong MHD shock on a 2D domain. The initial left state is $(\rho, v_{||}, v_\perp, p, B_{||}, B_\perp) = (1,10,0,20,5/\sqrt{4\pi},5/\sqrt{4\pi})$ and the initial right state is $(\rho, v_{||}, v_\perp, p, B_{||}, B_\perp) = (1,-10,0,1,5/\sqrt{4\pi},5/\sqrt{4\pi})$ for this Riemann problem, with adiabatic index $\gamma = 5/3$, and is found in \cite{Toth:2000:DBC:349920.349997}. The shock is set up to travel at an angle $30^\circ$ with respect to a line of symmetry of the mesh. This Riemann problem has the exact solution where $B_{||}$ stays constant throughout the different shock regions during the evolution of the shock. 

\begin{figure}
\centering
\includegraphics[width=0.47\textwidth]{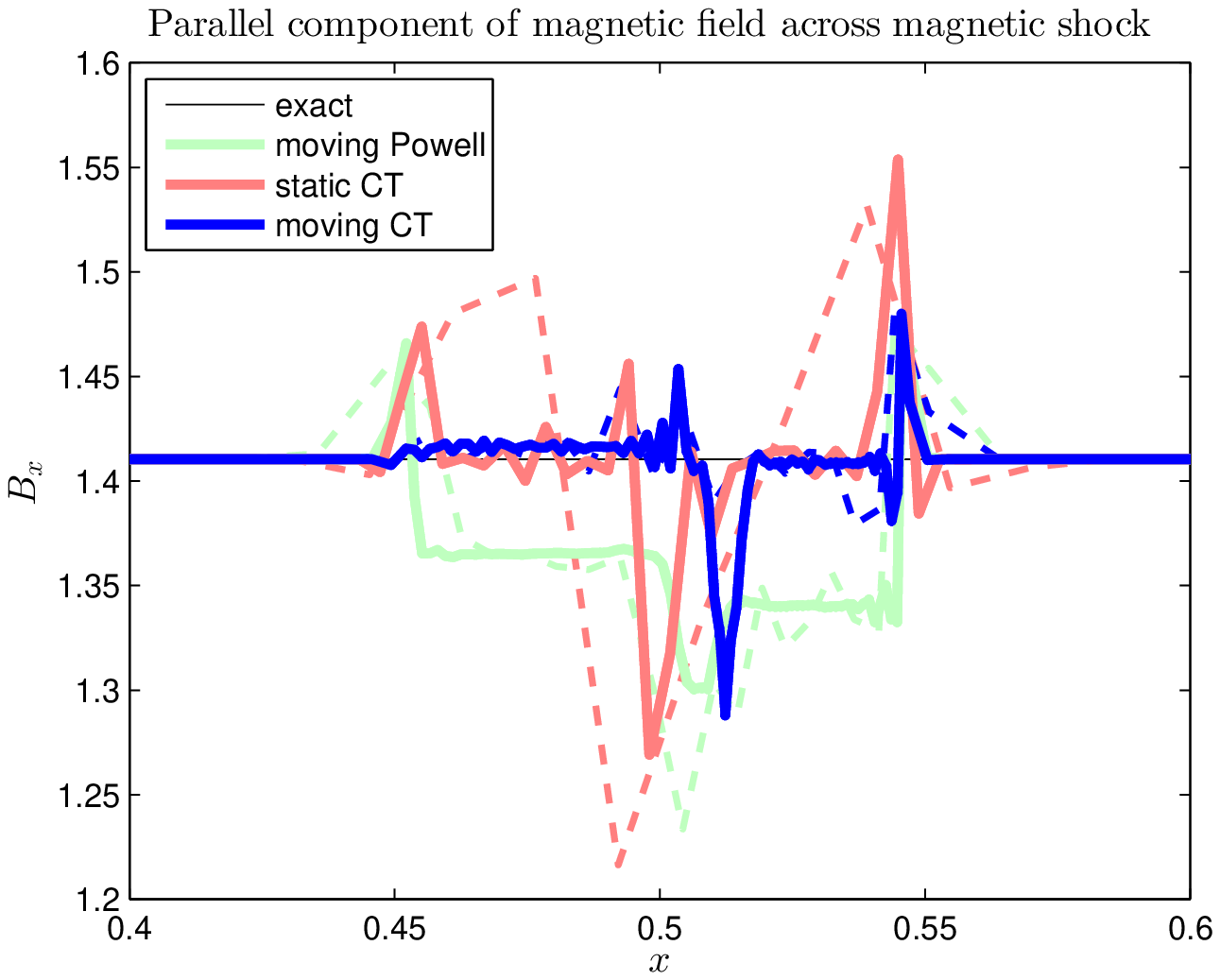}
\caption{The parallel component of the magnetic field across the shock for our moving CT, static CT, and moving Powell schemes at $t=0.01$. Thick solid and thin dashed lines correspond to resolutions of $256$ and $64$ respectively. Both CT approaches recover the exact solution with some spiked error at the discontinuity, as well as some oscillating errors in the rest of the domain which can be reduced with increased resolution. The moving Powell approach, however, produces incorrect jump conditions due to its non-conservative formulation, which expand with the shock and do not disappear with increased resolution.
}
\label{fig:shock}
\end{figure}

The shock is simulated with the moving CT, static CT, and moving Powell approaches, and the results of the evolved parallel component of the magnetic field are shown in Fig.~\ref{fig:shock}, for resolutions of $64$ and $256$ cells along the direction of the shock in a domain of length $1$. The results show that the non-conservative Powell approach performs the worst, producing incorrect jump conditions across discontinuities due to the cleaning source-terms. The depth of the deviations from the exact solution does not disappear with increased resolution. The CT schemes do much better. They also have errors across discontinuities, but they are smaller and oscillate around the exact solution, and therefore can be more reliably used to simulate strong shocks.

\section{Discussion}\label{sec:disc}

\subsection{Importance of $\nabla\cdot\mathbf{B}=0$}\label{sec:divB}

A divergence-free (solenoidal) magnetic field means that there are no field lines that meet at monopolar singularities. By Stokes' theorem, there is no net magnetic flux out of any enclosed surface. The use of Stokes' theorem to rewrite the $\nabla\cdot\mathbf{B}=0$ condition is the key to CT. Numerical methods which do not ensure the divergence of $B$ is kept small violate these properties which can lead to errors in the flow or even severe numerical instabilities in the solution. A plain finite volume scheme for the MHD equation can be unstable. Cleaning schemes do keep the divergence small enough so that the methods remain stable (at the cost of conserving the fluid variables, which may lead to convergence to the wrong answer). CT schemes offer a better discretization than finite volume approaches that maintains $\nabla\cdot\mathbf{B}=0$ without any modification to the MHD equations or loss of conservation.

If $\nabla\cdot\mathbf{B}=0$ is not preserved, one effectively litters the computational domain with magnetic monopole numerical artefacts. The Lorentz force no longer remains orthogonal to the magnetic field. To see this, consider the terms in the momentum equation in conservative form which arise from the magnetic field, which can be rewritten as:
\begin{equation}
\nabla \cdot \left( \frac{1}{2}\mathbf{B}^2 - \mathbf{B}\mathbf{B}^T \right) = -\left(\nabla\times\mathbf{B}\right)\times\mathbf{B} - \mathbf{B}\left(\nabla\cdot\mathbf{B}\right)
\end{equation}
which reduces to the Lorentz force acting per unit volume only in the case that $\nabla\cdot\mathbf{B}=0$. Any finite divergence in the discretized representation of the magnetic fields will bias the Lorentz force.

Large divergences in the magnetic fields make a simple MHD solver unstable, which is why CT or cleaning schemes need to be applied. \cite{Toth:2000:DBC:349920.349997} carries out an extensive comparison of CT and divergence cleaning schemes on fixed grids. The largest problem identified with the Powell cleaning scheme is that the method is non-conservative, which may cause incorrect jump-conditions occasionally. CT schemes on the other hand are conservative, divergence-free, robust, and more accurate.

\subsection{Variations and extensions of the method}\label{sec:var}

There are several ways to modify the CT scheme described above, which we sketch here. For example, the halving of the timestep in the cases where two or more neighbouring faces disappear in the same timestep may be avoided if one remaps all the edges of the connected disappearing faces to a single degenerate point and redistributes the fluxes of the vanishing faces to the neighbouring faces that do not disappear (this requires implementing a more involved remapping scheme). Alternatively, one could also invent a scheme where only faces that maintain connectivity throughout a timestep and share no edge with a face that disappears are updated at the end of the timestep, and a reconstruction scheme is used at the end of the timestep to create second-order estimates of the magnetic flux through faces that have appeared.

There are different ways one could obtain the estimate of the electric field at the edges of the cells. One could easily use a field-interpolated approach instead of the flux interpolated approach taken here.

The estimation of the cell-average magnetic fields from the face average fluxes is also free to be modified, and perhaps more accurate ones can be developed (our choice of mapping works well for regularized meshes). The gradient estimation of the magnetic fields in the reconstruction step may also be improved. For example, one may use some sort of local projection scheme on to a divergence-free basis.

Additionally, one could use a second-order accurate estimate for the velocity of the edges in the mesh that appear at the end of the timestep rather than calculating it exactly, which would mean that the geometrical information of the mesh at the beginning and end of the timestep would not have to be kept in memory at the same time.

We have highlighted some of the various modifications possible to the basic framework which can be explored in future work to optimize the moving mesh CT strategy. The goal of our current paper is to lay out the basic theoretical framework for the method. The basic strategy for maintaining $\nabla\cdot\mathbf{B}=0$ is straightforward: update magnetic fluxes across face cells using any second order accurate of the electric field in the rest frames of the edges, and remap faces that disappear across a timestep by shrinking them to a degenerate point.

\subsubsection{Adaptive Timestepping}\label{sec:ats}

Extending the moving CT method presented above to have adaptive timestepping through an efficient, local method is a goal for future research. We have not yet implemented such a scheme, but describe here general approaches and also additional challenges not found in the case of static meshes. An effective  adaptive timestepping method for a moving mesh would likely use a power-of-two hierarchical time-binning procedure \citep{2010MNRAS.401..791S}, which accounts for Voronoi cells changing their volume throughout the simulation and maintains the stability and conservation-laws of the fluid solver. The basic idea is simple: place the cells in a nested hierarchy of cells with partial synchronization. Active cells in a timestep use their current fluid quantities in the interpolation step and inactive ones are simply advected and use their most recently calculated fluid quantities in the interpolation step so that the Riemann problem may be solved across the cells. The flux is always added to both cells to maintain the conserved quantities. So an inactive cell $i$ has flux added to it across faces shared with neighbouring cells in smaller timestep bins. Advecting inactive cells maintains self-consistency of the connectivity of the Voronoi diagram as it is reconstructed at various timestep hierarchy levels.

The idea can be extended to the moving CT approach. The electric fields at an edge need to be estimated at the smallest timestep that any of the faces that join the edge fall into. The change in magnetic flux: $\Delta t\,E$ (with the appropriate sign due to orientation) then needs to be applied to all the faces (including inactive ones) that join the electric field in order to maintain the divergence free condition. This idea works straightforwardly in the cases that no changes in mesh connectivity take place across timesteps, and even in the cases where all the cells which experience a change in connectivity fall into in the same timestep bin. However, special care has to be taken to resolve a change in connectivity that occurs between cells that fall into different timestep bins, since faces that are surrounded by inactive cells may appear and their magnetic flux through the face has to be accurately calculated and stored with the inactive cells, and additionally inactive cells may lose a face and thus a remap involving inactive cells is required. Thus each timestep bin would have to be adaptive to accommodate inactive cells that change the number of faces. Our current choice of using rare time-step halving in cases where the connectivity changes too much also complicates the issue, and thus a more involved remapper (Section~\ref{sec:var}) is preferred. The development of the details of an adaptive timestepping scheme is left for future work.

\section{Concluding remarks}\label{sec:conclusion}

We have presented a new, stable, accurate, and robust CT scheme for solving the MHD equations on a moving unstructured mesh which is conservative and preserves the divergence of the magnetic field to zero at the level of machine precision. Such a CT scheme for a moving mesh has been claimed to be difficult to construct, maybe even impossible, by other authors \citep{2011ApJS..197...15D,2011MNRAS.418.1392P,2013MNRAS.432..176P}, but we have demonstrated that the scheme can be achieved. The CT method significantly improves the other current methods used to evolve the MHD equations on unstructured meshes and moving meshes, which are not necessarily conservative and can show large divergence errors and incorrect shock jump conditions. The new numerical method also has significant advantages over CT approaches on a fixed grid. Namely, our method is a quasi-Lagrangian scheme and greatly reduces advection errors. In pure advection flows, it preserves the solution at the level of machine-precision, which is a uniquely powerful feature of the method. Due to the moving mesh formulation, the method is also automatically adaptive in its resolution. Additionally, Galilean boosts/large velocities in the flow affect the truncation errors to a significantly smaller extent than in fixed grid codes due to the Galilean invariant properties of the moving mesh code. It is vital to construct a moving mesh code that preserves $\nabla\cdot \mathbf{B}=0$ to machine precision, otherwise if differences between moving and fixed grid MHD codes are observed it becomes difficult to tell whether it is due to the advantages of moving mesh codes (such as better treatment of advection) or to the non-zero divergence errors that exist in the solution.

The new method offers new exciting possibilities of simulating MHD flows more accurately than before. One particular application it is well-suited for is the study of supersonic MHD turbulence. Here the flow has high Mach number velocities so fixed grid codes experience very restrictive timestep criteria, which can be lessened on a moving mesh. Additionally, advection errors would be reduced to a significant degree in moving mesh CT simulations. An analytic theory for supersonic MHD turbulence is virtually non-existent and numerical simulations are considered the best framework for studying this problem, necessitating numerical methods that minimize any unphysical effects of numerical errors on the solution. The moving mesh CT approach is a very viable candidate for this task.

The moving mesh CT method is also well-suited for studying astrophysical problems where advection is important. Most simulations, from accretion disc simulations to cosmological simulations, have regions that are only marginally resolved due to limitations in computational resources. As we see from our low-resolution simulations of advection, the solution in static CT methods and moving mesh methods with divergence cleaning schemes can have large diffusion errors which can be reduced only with higher resolution simulations. The moving CT approach, however, is able to accurately advect the solution to machine precision, which is one of the main benefits of the method.

\section*{Acknowledgements}
This material is based upon work supported by the National Science Foundation Graduate Research Fellowship under grant no. DGE-1144152. LH acknowledges support from NASA grant NNX12AC67G and NSF award AST-1312095. PM would like to thank Paul Duffell for insightful discussions on the manuscript.

\bibliography{mybib}{}

\bsp
\label{lastpage}
\end{document}